%% file: main.tex
\documentclass[runningheads]{llncs}
\usepackage{graphicx}

\usepackage{tikz}
\usepackage{comment}
\usepackage{amsmath,amssymb} 
\usepackage{color}
\usepackage{makecell}
\usepackage{bbm} 
\usepackage{enumitem}

\usepackage{times}
\usepackage{epsfig}
\usepackage{graphicx}
\usepackage{amsmath}
\usepackage{amssymb}
\usepackage{soul}
\usepackage{multirow}
\usepackage{booktabs}
\usepackage{wrapfig}
\usepackage{footnote}
\usepackage{subcaption}
\usepackage[font=small]{caption}
\usepackage[symbol]{footmisc}

\usepackage[accsupp]{axessibility}  

\begin{document}
\input{macros.tex}
\pagestyle{headings}
\mainmatter
\def\ECCVSubNumber{7704}  

\newcommand{\methodname}{The Shape Part Slot Machine}
\title
{\methodname: Contact-based Reasoning for Generating 3D Shapes from Parts}

\titlerunning{SPSM: Contact-based Reasoning for Generating 3D Shapes from Parts} 

\author{Kai Wang\inst{1}$^*$ \and
Paul Guerrero\inst{2} \and
Vladimir Kim\inst{2} \and
Siddhartha Chaudhuri\inst{2} \and \\
Minhyuk Sung\inst{2,3} \and
Daniel Ritchie\inst{1}}
\authorrunning{K. Wang et al.}
\institute{Brown University \and
Adobe Research \and 
KAIST
}

\renewcommand{\thefootnote}{\fnsymbol{footnote}}
\footnotetext[1]{This work was done partially during Kai Wang's internship at Adobe Research}
\maketitle

\begin{figure}[h]
    \centering
    \includegraphics[width=\linewidth]{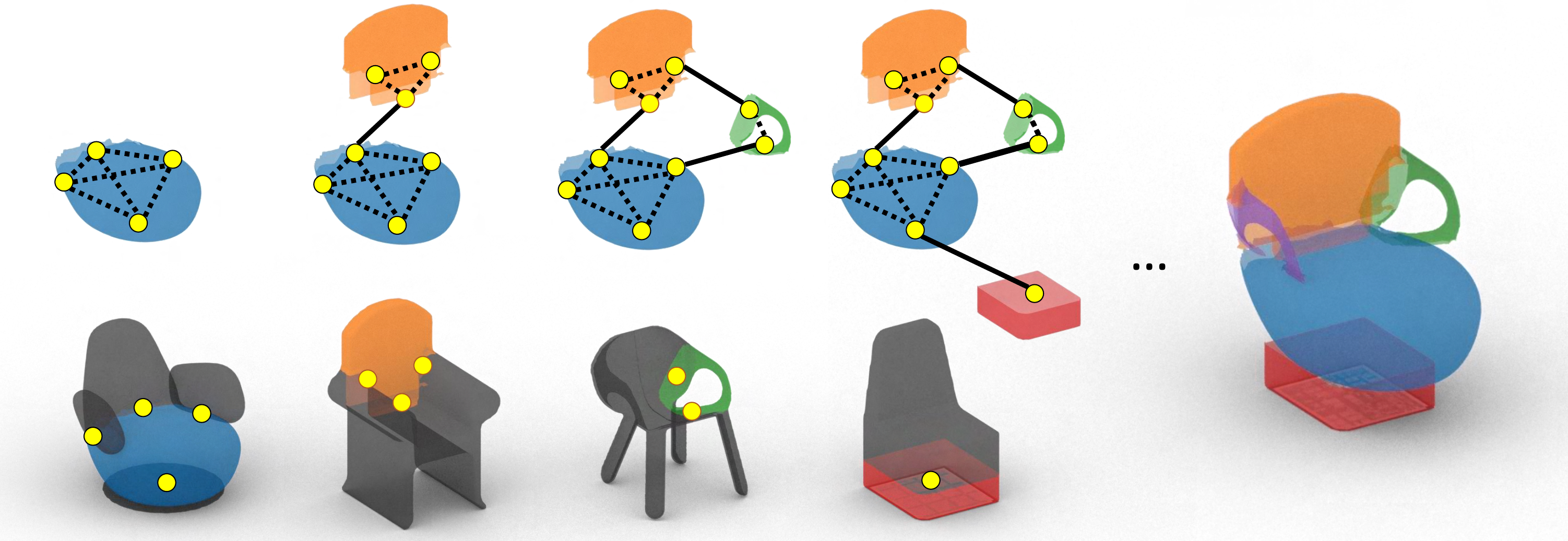}
    \caption{
    Our system synthesizes novel 3D shapes by assembling them from parts.
    Internally, it represents shapes as a graph of the regions where parts connect to one another (which we call slots).
    It generates such a graph by retrieving part subgraphs from different shapes in a dataset.
    Once a full graph has been generated, the system then optimizes for affine part transformations to produce a final output shape.
    }
    \label{fig:teaser}
    \vspace{-3em} 
\end{figure}

\begin{abstract}
\input{sections/0-abstract.tex}
\end{abstract}

\input{sections/1-intro.tex}
\input{sections/2-relatedwork.tex}
\input{sections/3-overview.tex}

\input{sections/4-representation.tex}
\input{sections/5-retrieval.tex}
\input{sections/6-optimization.tex}
\input{sections/7-results.tex}
\input{sections/8-conclusion.tex}
\input{sections/9-acknowledgements.tex}

\clearpage
\bibliographystyle{splncs04}
\bibliography{main}

\renewcommand\thesection{\Alph{section}}
\setcounter{section}{0}

\input{supp/1-data.tex}
\input{supp/2-retrieval.tex}
\input{supp/3-testtime.tex}
\input{supp/4-implementation.tex}
\input{supp/5-baselines.tex}

\input{supp/6-results.tex}

\end{document}

%% file: macros.tex
\newcommand{\revision}[1]{#1}
\newcommand{\revisionnew}[1]{{\revision{#1}}}

\newcommand{\reals}{\ensuremath{\mathbb{R}}}

\newcommand{\indicator}{\ensuremath{\mathbbm{1}}}

\newcommand{\norm}[1]{\left\lVert#1\right\rVert}

\newcommand{\unit}[1]{\ensuremath{\,\mathrm{#1}}}

\newcommand{\denselist}{\itemsep 0pt\parsep=0pt\partopsep 0pt\vspace{-\topsep}}

\newcommand*{\vcent}[1]{\vcenter{\hbox{#1}}}

\newenvironment{packed_itemize}
{\begin{itemize}
    \vspace{-\topsep}
    \setlength{\itemsep}{1pt}
    \setlength{\parskip}{0pt}
    \setlength{\parsep}{0pt}
}{\end{itemize}}

\newcommand{\nolistbottomspace}{\vspace{-\topsep}}

\newcommand{\parahead}[1]{\noindent\textbf{#1}:\ }

%% file: sections/0-abstract.tex
We present the Shape Part Slot Machine, a new method for assembling novel 3D shapes from existing parts by performing contact-based reasoning.
Our method represents each shape as a graph of ``slots,'' where each slot is a region of contact between two shape parts.
Based on this representation, we design a graph-neural-network-based  model for generating new slot graphs and retrieving compatible parts, as well as a gradient-descent-based optimization scheme for assembling the retrieved parts into a complete shape that respects the generated slot graph.
This approach does not require any semantic part labels; interestingly, it also does not require complete part geometries---reasoning about the slots proves sufficient to generate novel, high-quality 3D shapes.
We demonstrate that our method generates shapes that outperform existing modeling-by-assembly approaches regarding quality, diversity, and structural complexity.

%% file: sections/1-intro.tex
\section{Introduction}
\label{sec:intro}

There is increasing demand for high-quality 3D object models across multiple fields: gaming and virtual reality; advertising and e-commerce; synthetic training data for computer vision and robotics; and more.
The traditional practice of manual 3D modeling is time-consuming and labor-intensive and is not well-suited to scaling to this demand.
Thus, visual computing researchers have pursued data-driven methods which can augment human creative capabilities and accelerate the modeling process.

One promising technology in this space are generative models of 3D shapes.
Such generative models could suggest new, never-before seen shapes, freeing users from tedious and time-consuming low-level geometric manipulations to focus on high-level creative decisions.
Recent work in this space has focused on deep generative models of shapes in the form of volumetric occupancy grids, point clouds, or implicit fields.
While these methods demonstrate impressive abilities to synthesize the bulk shape of novel objects, the local geometry they produce often exhibits noticeable artifacts: oversmoothing, bumpiness, extraneous holes, etc.
\revision{
At present, none of these generative models has achieved geometric output quality resembling the shapes they are trained on. 
}
An alternative approach would be to avoid synthesizing novel geometry altogether and instead learn how to re-combine existing high-quality geometries created by skilled modeling artists.
This paradigm is known in the computer graphics literature as \emph{modeling by assembly}, where it once received considerable attention.
Since the deep learning revolution, however, the focus of most shape generation research has shifted to novel geometry synthesis.
The few post-deep-learning methods for modeling by assembly have shown promise but have not quite lived up to it: handling only coarse-grained assemblies of large parts, as well as placing parts by directly predicting their world-space poses (leading to `floating part' artifacts).

In this paper, we present a new generative model for shape synthesis by part assembly which addresses these issues.
Our key idea is to use a representation which focuses on the connectivity structure of parts.
This choice is inspired by several recent models for novel geometry synthesis which achieve better structural coherence in their outputs by adopting a part-connectivity-based representation~\cite{StructureNet,jones2020shapeAssembly,yang2020dsmnet}.
In our model, the first-class entities are the regions where one part connects to another.
We call these regions \emph{slots} and our model the \emph{Shape Part Slot Machine}.

In our model, a shape is represented by a graph in which slots are nodes and edges denote connections between them.
We define shape synthesis as iteratively constructing such a graph by retrieving parts and connecting their slots together.
We propose an autoregressive generative framework for solving this problem, composed of several neural network modules tasked with retrieving compatible parts and determining their slot connections.
Throughout the iterative assembly process, the partial shape is represented only by its slot graph: it is not necessary to assemble the retrieved parts together until the process is complete, at which point we use a gradient-descent-based optimization scheme to find poses and scales for the retrieved parts which are consistent with the generated slot graph.

We compare the Shape Part Slot Machine to other modeling-by-assembly and part-connectivity-based generative models.
We find that our approach consistently outperforms the alternatives in its ability to generate visually and physically plausible shapes.

In summary, our contributions are:
\begin{packed_itemize}
    \item The \emph{Slot graph} representation for \revision{reasoning about part structure of} shapes.
    \item An autoregressive generative model for slot graphs by iterative part retrieval and assembly.
    \item A demonstration that local part connectivity structure is enough to synthesize globally-plausible shapes: neither full part geometries nor their poses are required.
\end{packed_itemize}

%% file: sections/2-relatedwork.tex
\section{Related Work}
\label{sec:relatedwork}

\parahead{Modeling by Part Assembly}
The Modeling By Example system pioneered the paradigm of modeling-by-assembly with interactive system for replacing parts of an object by searching in database~\cite{ModelingByExample}.
The Shuffler system added semantic part labels, enabling automatic `shuffling' of corresponding parts~\cite{Shuffler}.
Later work handled more complex shapes by taking symmetry and hierarchy into account~\cite{PartBasedRecombination}.
Other modes of user interaction include guiding exploration via sketches~\cite{sketch2design_cgf13} or abstract shape templates~\cite{akzm_shapeSynth_eg14}, searching for parts by semantic attributes~\cite{AttribIt}, or having the user play the role of fitness function in an evolutionary algorithm~\cite{FitAndDiverse}.
Probabilistic graphical models have been effective for suggesting parts~\cite{SidVangelisAssembly} or synthesizing entire shapes automatically~\cite{SidVangelisSynthesis}.
Part-based assembly has also been used for reconstructing shapes from images~\cite{PartBasedStructureRecovery}.

Our work is most closely related to ComplementMe, which trains deep networks to suggest and place unlabeled parts to extend a partial shape~\cite{ComplementMe}.
Our model is different in that we use a novel, part-contacts-only representation of shapes, which we show enables handling of more structurally complex shapes.

\parahead{Deep Generative Models of Part-based Shapes}
Our work is also related to deep generative models which synthesize part-based shapes.
One option is to make voxel-grid generative models part-aware~\cite{GlobalToLocal,SAGnet19}.
Many models have been proposed which generate sets of cuboids representing shape parts~\cite{3DPRNN}; some fill the cuboids with generated geometry in the form of voxel grids~\cite{GRASS} or point clouds~\cite{StructureNet,jones2020shapeAssembly,jones2021shapeMOD}.
Other part-based generative models skip cuboid proxies and generate part geometries directly, as point clouds~\cite{CompoNet}, implicit fields~\cite{PQNet}, or deformed genus zero meshes~\cite{gao2019sdmnet,yang2020dsmnet}. 
All of these models synthesize part geometry.
In contrast, our model synthesizes shapes by retrieving and assembling existing high-quality part meshes.

\parahead{Estimating Poses for 3D Parts}
Many part-based shape generative models must pose the generated parts.
Some prior work looks at this problem on its own: given a set of parts, how to assemble them together?
One method predicts a 6DOF pose for each part such that they become assembled~\cite{HuangZhan2020PartAssembly}; another predicts per-part translations and scales and also synthesizes geometry to make the transformed parts connect seamlessly~\cite{yin2020coalesce}.
Rather than predict part poses directly, we solve for per-part poses and scales that satisfies contact constraints encoded in a slot graph.
This approach has its root in
\revision{
early work in modeling by assembly~\cite{SidVangelisSynthesis} but without the need for part labels and separate steps computing how parts should attach.
}
It is also similar in spirit to
that of ShapeAssembly~\cite{jones2020shapeAssembly}, working with part meshes rather than cuboid abstractions.

%% file: sections/3-overview.tex
\section{Overview}
\label{sec:overview}

Assembling novel shapes from parts requires solving two sub-problems: finding a set of compatible parts, and computing the proper transforms to assemble the parts.
These tasks depend on each other, e.g. replacing a small chair seat with a large one will shift the chair legs further away from the center.
Instead of solving these sub-problems separately, we propose a system for solving them jointly.
Specifically, our system synthesizes shapes by iteratively constructing a representation of the contacts between parts.
The assembly transformations for each part can then be computed directly from this representation.

In our system, a shape is represented as a \textit{slot graph}: each node corresponds to a ``slot" (part-to-part contact region) on a part; each edge is either a \textit{part edge} connecting slots of the same part or a \textit{contact edge} connecting slots on two touching parts.
Section~\ref{sec:representation} defines this graph structure and describes how we extract them from available data.

This reduces the task of assembling novel shapes to a graph generation problem: retrieving sub-graphs representing parts from different shapes and combining them into new graphs.
We solve this problem autoregressively, assembling part sub-graphs one-by-one into a complete slot graph. 
At each iteration, given a partial slot graph, our system inserts a new part using three neural network modules: the first determines \textit{where} a part be should connect to the current partial graph, the second decides \textit{what} part to connect, and third determines \textit{how} to connect the part. 
We describe this generation process in Section~\ref{sec:graphgen}.

Finally, given a complete contact graph, the system runs a gradient-based optimization process that assembles parts into shapes by solving for poses and scales of the individual parts such that the contacts implied by the generated slot graph are satisfied.
We describe the process in Section~\ref{sec:optimization}.

%% file: sections/4-representation.tex
\section{Representing Shapes with Slot Graphs}
\label{sec:representation}

In this section, we define slot graphs, describe how we extract them from segmented shapes, and how we encode them with neural networks.

\subsection{Slot-based Graph Representation of Shapes}
\label{sec:graph_representation}

\begin{figure}[t!]
    \centering
    \includegraphics[width=0.4\linewidth]{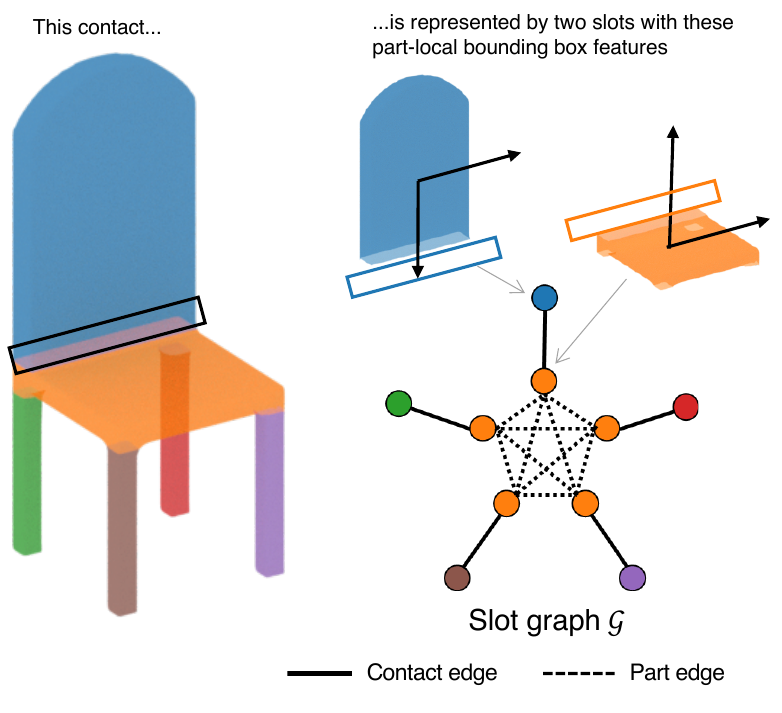}
    \vspace{-1em}
    \caption{
    A slot graph. Nodes are part-to-part contact regions called \emph{slots} and describe the contact geometry with bounding boxes.
    \emph{Contact edges} connect two slots on two adjacent parts, while \emph{part edges} connect all slots of the same part.
    }
    \label{fig:representation}
    \vspace{-1em}
\end{figure}

A good shape representation that models how parts connect allows the generative model to reason independently about part connectivity and part geometry.
Given a shape $S$ and its part decomposition $\{P_1 \ldots P_N\}$, we call regions where parts connect ``slots", and use them as nodes in a graph $\mathcal{G} = (V, E_c, E_p)$, as illustrated in Figure~\ref{fig:representation}. Each pair of parts may be connected with multiple slots, and each slot $\mathbf{u}_{ij} \in V$ on part $P_i$ that connects to $P_j$ has a corresponding slot $\mathbf{u}_{ji}$ on part $P_j$ that connects back to $P_i$.
Each node $\mathbf{u}_{ij}$ stores the following properties:
\begin{packed_itemize}
    \item The axis-aligned bounding box (AABB) of the slot, in a coordinate frame centered on $P_i$.
    \item The same AABB, but normalized such that the bounding box of the entire part is a unit cube. This provides a scale-invariant view of how parts connect.
\nolistbottomspace
\end{packed_itemize}
A slot graph $\mathcal{G}$ has two types of edges:
\textit{contact edges} $\mathbf{e}^c_{ij} \in E_c$ connect every pair of contacting slots $\textbf{u}_{ij}, \textbf{u}_{ji}$ and
\textit{part edges} $\mathbf{e}^p_{ijk} \in E_p$ connect every pair of slots $\textbf{u}_{ij}, \textbf{u}_{ik}$ in the same part $P_i$.

This representation encodes neither the geometry nor the pose of each part.
Omitting this information encourages generalization: the choice of parts will be based only on the compatibility of their attachment regions and connectivity structure; it will not be biased by a part's world-space position in its original shape nor its complete geometry.

This representation also does not encode part symmetries; nevertheless, we demonstrate in Section~\ref{sec:results} that our model often generates appropriately symmetrical shapes.
We can also optionally include logic that enforces symmetries at test time (see Section~\ref{sec:graphgen}).

\subsection{Extracting Slot Graphs from Data}
Given a set of part-segmented shapes, we follow StructureNet~\cite{StructureNet} to extract part adjacencies and symmetries.
We use adjacencies to define slots, and define connecting regions as points within distance $\tau$ of the adjacent part. Additionally, we ensure that symmetrical parts have symmetrical slots and prune excess slots where multiple parts overlap at the same region.
See supplemental for details.

\subsection{Encoding Slots Graphs with Neural Networks}

We encode a slot graph
into a graph feature $h_\mathcal{G}$ and per-slot features $h_\mathbf{u}$ using messaging passing networks~\cite{NeuralMessagePassing}.

\parahead{Initializing Node Embeddings}
We initialize slot features $h_\mathbf{u}$ using a learned encoding of the slot properties $x_\mathbf{u}$ (two six-dimensional AABBs) with a three-layer MLP $f_\text{init}$: $h_\mathbf{u}^0 = f_\text{init}(x_\mathbf{u})$.
As we discuss later, some of our generative model's modules also include an additional one-hot feature which is set to one for particular nodes that are relevant to their task (effectively `highlighting' them for the network).

\parahead{Graph Encoder}
The node embeddings are then updated with our message passing network using an even number of message passing rounds. In each round, node embeddings are updated by gathering messages from adjacent nodes.
We alternate the edge sets $E$ during each round, using only part edges $E = E_p$ for odd rounds ($t=1,3,5\ldots$) and only contact edges $E = E_c$ for even rounds ($t=2,4,6\ldots$):
\[
h_\mathbf{t} = f_\text{update}^t\Big(h_\mathbf{u}^{t-1}, \sum_{\mathbf{uv} \in E} f_\text{msg}^t (h_\mathbf{u}^{t-1}, h_\mathbf{v}^{t-1}, h_\mathbf{u}^0, h_\mathbf{v}^0) \Big)
\]
where $f_{\text{msg}}$ is a multi-layer perceptron (MLP) that computes a message for each pair of adjacent nodes, 
and $f_\text{update}$ is a MLP that updates the node embedding from the summed messages. 
We also include skip connections to the initial node embeddings $h_\mathbf{u}^0$.
All MLPs have separate weights for each round of message passing.

\parahead{Gathering Information from the Graph}
To obtain the final node features $h_\mathbf{u}$, we concatenate its initial embedding with as its embeddings after every even round of message passing (i.e. those using contact edges) and feed them into an MLP $f_\text{node}$:
\[
h_\mathbf{u} = f_\text{node}(h_\mathbf{u}^0, h_\mathbf{u}^2 \ldots h_\mathbf{u}^T)
\]
To obtain the feature $h_\mathcal{G}$ of an entire graph, we first perform a graph readout over the embeddings at round $t$: 
\[
h_\mathcal{G}^t =\sum_{\mathbf{u} \in V} (f_\text{project}(h_\mathbf{u}^t) \cdot f_\text{gate}(h_\mathbf{u}^t))
\]
where $f_\text{project}$ projects node features into the latent space of graph features and $f_\text{gate}$ assigns a weight for each of the mapped features.
We then compute the final graph feature $h_\mathcal{G}$ similar to the way we compute the node features:
\[
h_\mathcal{G} = f_\text{graph}(h_\mathcal{G}^0, h_\mathcal{G}^2 \ldots h_\mathcal{G}^T)
\]
In cases where we need a feature for a subset of nodes $V' \subset V$ in the graph, we simply perform the readout over $V'$ instead of $V$.

\begin{figure*}[t!]
    \centering
    \includegraphics[width=\linewidth]{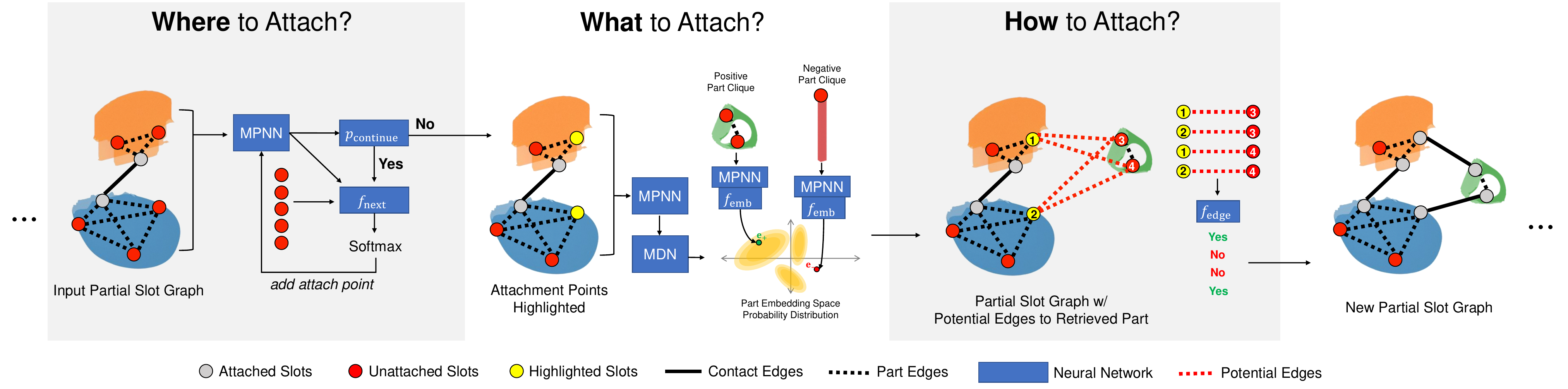}
    \vspace{-2em}
    \caption{
    Our slot graph generative model uses three neural network modules to build a graph step by step.
    \emph{Where to Attach?}: Predicts which slots on the current partial shape the next-retrieved part should be attached to.
    \emph{What to Attach?}: Learns an embedding space for part slot graphs and predicts a probability distribution over this space in which parts which are compatible with the highlighted slots have high probability.
    \emph{How to attach?}: Determines which slots on the retrieved part should connect to which slots on the current partial shape.
    }
    \label{fig:architecture}
\end{figure*}

%% file: sections/5-retrieval.tex
\section{Generating Slot Graphs}
\label{sec:graphgen}

Our system casts shape synthesis by part assembly as a problem of generating novel slot graphs.
To do so, we first extract a graph representation of each part:
for every shape $S$ in a dataset of shapes, and for every part $P_i \in S$ represented by a slot graph $\mathcal{G} = (V, E_c, E_p)$, we create a part clique $C_{P_i} \subseteq \mathcal{G}$ representing this part by taking all the slots $\mathbf{u}_{ij} \in V$ associated with this part, as well as all the part edges $\mathbf{e}_{ijk} \in E_p$. 
We remove the all contact edges $\mathbf{e}_{ij} \in E_c$ that connects $P_i$ to other parts in the shape. 
Our goal, then, is find a set of part cliques $\mathbf{C}$ that can be connected together into a novel slot graph $\mathcal{G'} = (V', E_c', E_p')$, where $V' = \{\mathbf{u} \in \mathbf{C}\}$, $E_p' = \{\mathbf{e} \in \mathbf{C}\}$, and $E_c'$ is the set of contact edges that need to be added to make all the slots attached i.e. connected to exactly one other slot via a contact edge.

There can be thousands of parts in available shape datasets, each containing multiple slots that can be attached in different ways.
Thus, it is infeasible to search this combinatorial space exhaustively.
Instead, we \emph{learn} how to build novel slot graphs autoregressively, attaching one part clique at a time to a partial slot graph, until it is complete (i.e. all slots are attached).
We learn this autoregressive process with teacher forcing: 
for every training sample, we take a random, connected partial slot graph $\mathcal{G}'$ consisting of one or more part cliques from a graph $\mathcal{G} = (V, E_c, E_P)$ extracted from a dataset shape $S$.
We then select a random part clique $C_{P_j}|P_j \in S$ (referred to as $C_\text{target}$ in the following sections) that is attached to $\mathcal{G}'$ on one or more slots $V_\text{target} = \{\mathbf{u}_{ij} \mid \mathbf{u}_{ij} \in \mathcal{G}', \mathbf{u}_{ji} \in C_{P_j}\}$ via set of contact edges $E_\text{target} = \{\mathbf{e}^c_{ij} \mid \mathbf{u}_{ij} \in V_\text{target}\}$.
The goal of a single generation step, then, is to maximize
\[
    p(V_\text{target}, C_\text{target}, E_\text{target} \mid \mathcal{G'})
\]
Rather than learn this complex joint distribution directly, we instead factor it into three steps using the chain rule:
\begin{itemize}[noitemsep,topsep=0pt,parsep=0pt,partopsep=0pt]
\item
\textbf{Where} to attach: maximizing $p(V_\text{target} \mid \mathcal{G'})$
\item
\textbf{What} to attach: maximizing $p(C_\text{target} \mid \mathcal{G'}, V_\text{target})$
\item
\textbf{How} to attach: \mbox{maximizing $p(E_\text{target} \mid \mathcal{G'}, V_\text{target}, C_\text{target})$
}
\end{itemize}
In the remainder of this section, we detail the formulation for the networks we use for each of the three steps, as well as how we use them during test time.
Figure~\ref{fig:architecture} visually illustrates these steps.

\parahead{Where to Attach?}
Given a partial slot graph $\mathcal{G}'$, we first identify the slots $V_\text{target}$ to which the next-retrieved part clique should attach.
We predict each element of $V_\text{target}$ autoregressively (in any order), where each step $i$ takes as input $\mathcal{G}'$ and the already-sampled slots $V^i_\text{target} = \{V_{\text{target},0} \ldots  V_{\text{target},i-1} \}$ (highlighted in $\mathcal{G}'$ with a one-hot node feature).
We first use a MLP $f_\text{continue}$ to predict the probability $p_\text{continue}$ that another slot should be added ($p_\text{continue} = 0$ if $V^i_\text{target} = V_\text{target}$ and $1$ otherwise).
If more slots should be included, then we use an MLP $f_\text{next}$ to predict a logit for each of the unattached slots $\tilde{V}$ in $\mathcal{G}'$ that are not already in $V^i_\text{target}$.
These logits are then fed into a softmax to obtain a probability distribution over possible slots:
\begin{align*}
    p(V_\text{target} | \mathcal{G}') = \prod_{i=1}^{|V_\text{target}|} p^i_\text{cont} \cdot p^i_\text{next}[V_{\text{target},i}]
    \\
    p^i_\text{cont} = \begin{cases}
        p_\text{continue}(h_\mathcal{G'} | V^i_\text{target}) & i < |V_\text{target}| \\
        1 - p_\text{continue}(h_\mathcal{G'} | V^i_\text{target}) & i = |V_\text{target}|
    \end{cases}
    \\
    p^i_\text{next} = \text{softmax}(\big[ f_\text{next}(h_\mathbf{u} | V^i_\text{target}) | \mathbf{u} \in \tilde{V} / V^i_\text{target} \big])
\end{align*}

\parahead{What to Attach?}
Having selected the slots $V_\text{target}$ to attach to, we then retrieve part cliques compatible with the partial graph $\mathcal{G'}$ and the selected slots.
Similar to prior work~\cite{ComplementMe}, we take a contrastive learning approach to this problem: the probability of the ground truth part clique should be greater than that of randomly sampled other part cliques (i.e. negative examples) by some margin $m$:
\[
p(C_\text{target} \mid \mathcal{G}', V_\text{target}) > p(C_\text{negative} \mid \mathcal{G}', V_\text{target}) + m
\]
We use two neural networks to enforce this property.
The first maps part cliques $C$ into an embedding space $\mathbb{R}_\text{emb}$.
\[
X_C = f_\text{emb} (h_{C})
\]
where $f_\text{emb}$ is the embedding MLP and $h_{C}$ is the graph feature computed from $C$ alone.
The second network is a mixture density network (MDN) that outputs a probability distribution over this embedding space:
\[
P(X | \mathcal{G}', V_\text{target}, X \in \mathbb{R}_\text{emb})
= \text{MDN}(h_\mathcal{G'}, h_\mathcal{G'_\text{target}})
\]
Where $V_\text{target}$ are highlighted in the input node features and $h_\mathcal{G'_\text{target}}$ is obtained by computing graph features using $V_\text{target}$ only.

\begin{figure*}[t!]
    \centering
    \setlength{\tabcolsep}{0pt}
    \renewcommand{\arraystretch}{0}
    \begin{tabular}{cccccccc}
        Input & Input + GT & GT & Best & $2$nd & $5$th & $25$th & $50\%$
        \\ [-1em]
        \includegraphics[width=.11\linewidth]{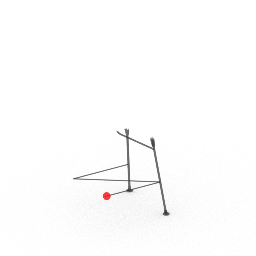} &
        \includegraphics[width=.11\linewidth]{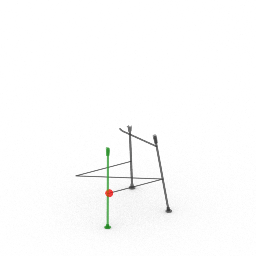} &
        \includegraphics[width=.11\linewidth]{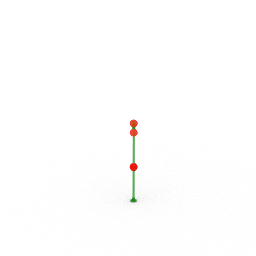} &
        \includegraphics[width=.11\linewidth]{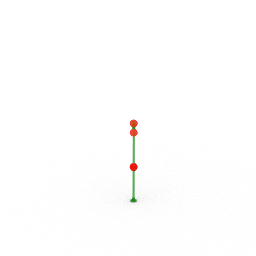} &
        \includegraphics[width=.11\linewidth]{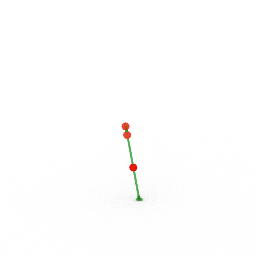} &
        \includegraphics[width=.11\linewidth]{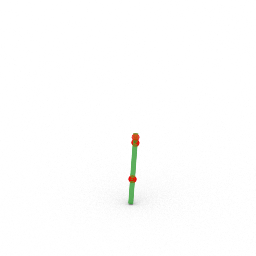} &
        \includegraphics[width=.11\linewidth]{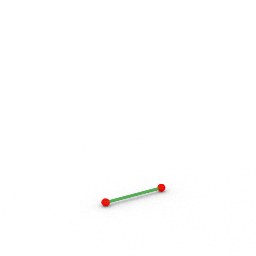} &
        \includegraphics[width=.11\linewidth]{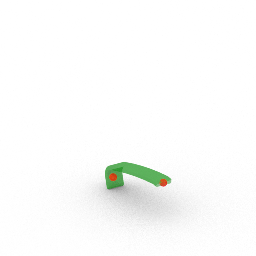}
        \\ [-0.5em]
        \includegraphics[width=.11\linewidth]{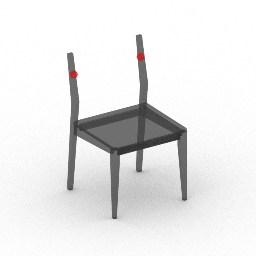} &
        \includegraphics[width=.11\linewidth]{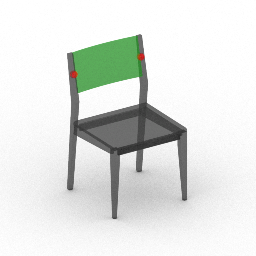} &
        \includegraphics[width=.11\linewidth]{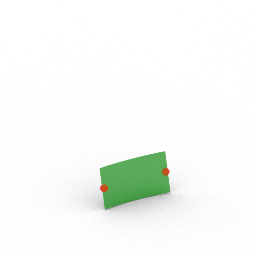} &
        \includegraphics[width=.11\linewidth]{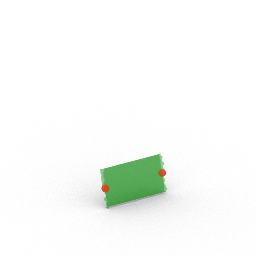} &
        \includegraphics[width=.11\linewidth]{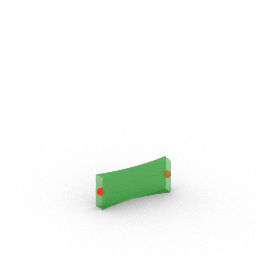} &
        \includegraphics[width=.11\linewidth]{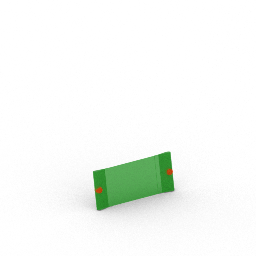} &
        \includegraphics[width=.11\linewidth]{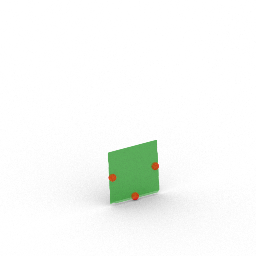} &
        \includegraphics[width=.11\linewidth]{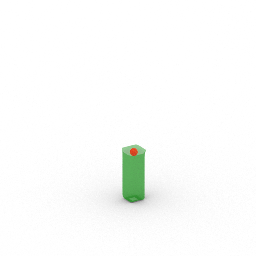}
        \\ [-1.5em]
        \includegraphics[width=.11\linewidth]{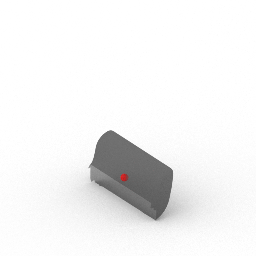} &
        \includegraphics[width=.11\linewidth]{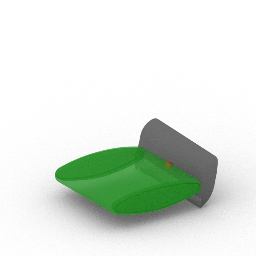} &
        \includegraphics[width=.11\linewidth]{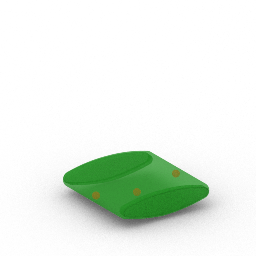} &
        \includegraphics[width=.11\linewidth]{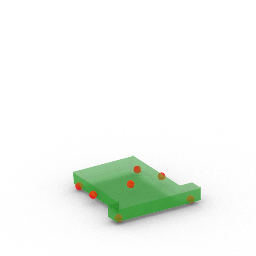} &
        \includegraphics[width=.11\linewidth]{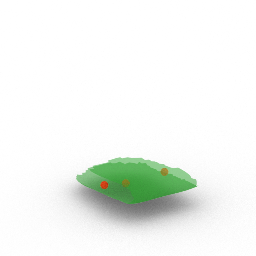} &
        \includegraphics[width=.11\linewidth]{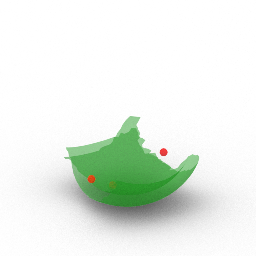} &
        \includegraphics[width=.11\linewidth]{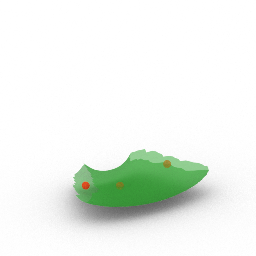} &
        \includegraphics[width=.11\linewidth]{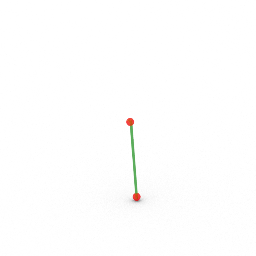}
        \\ [-0.25em]
        \includegraphics[width=.11\linewidth]{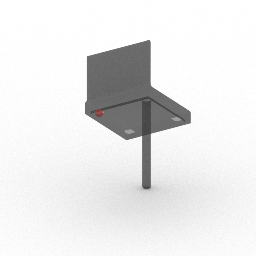} &
        \includegraphics[width=.11\linewidth]{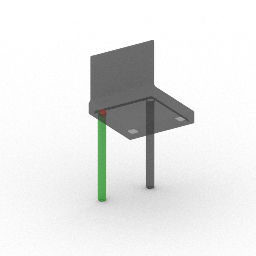} &
        \includegraphics[width=.11\linewidth]{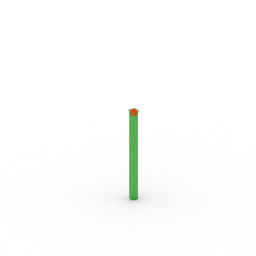} &
        \includegraphics[width=.11\linewidth]{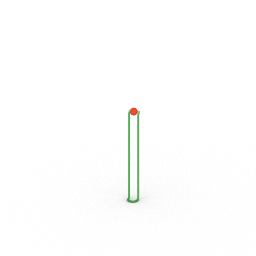} &
        \includegraphics[width=.11\linewidth]{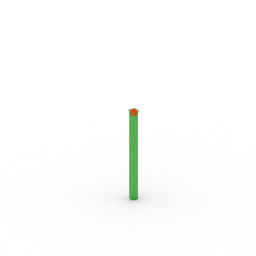} &
        \includegraphics[width=.11\linewidth]{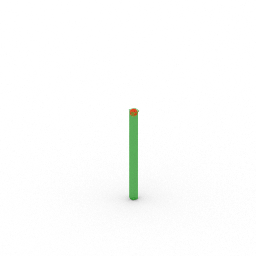} &
        \includegraphics[width=.11\linewidth]{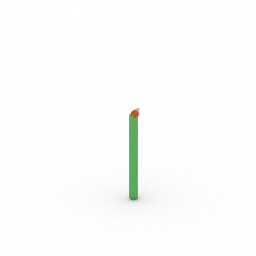} &
        \includegraphics[width=.11\linewidth]{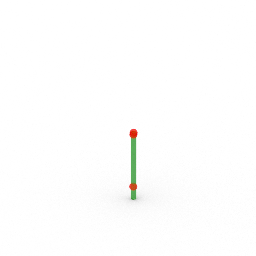}
    \end{tabular}
    \vspace{-0.75em}
    \caption{
    \revision{
    Example outputs of the \textbf{What to Attach?} module.
    We visualize the input partial slot graph within the parts that contain them (grey) and the center of the selected slots (red), as well as the ground truth part (green, 2nd column). The parts and slots are in their ground truth world-space pose, which is \emph{not} available to the neural network.
    We then visualize, individually, the ground truth part and the retrieved candidates ranked $1$st, $2$nd, $5$th, $25$th, and at the $50$th percentile, respectively, along with all of their slots (red).
    }
    } 
    \label{fig:retrieved_parts}
    \vspace{-2em}
\end{figure*}

\revision{
We visualize the behavior of this module trained on Chair in Figure~\ref{fig:retrieved_parts}.
When the input demands a very specific type of structure (first $2$ rows), our module can retrieve the part cliques that match such structure.
When the input has fewer constraints ($3$rd row), our module retrieves a wide variety of partial cliques that can be attached.
In the $4$th row, our module retrieves chair legs that are {\em structurally} compatible. The legs are not necessarily {\em geometrically} compatible, as geometry information is not available to the module. 
}

\parahead{How to Attach?}
The last module learns to connect the retrieved part clique $C_\text{target}$ to the partial slot graph $\mathcal{G'}$.
It predicts a probability for every pair of slots $\mathbf{u}_{ij} \in V_\text{target}, \mathbf{u}_{ji} \in C_\text{target}$ that could be connected via a contact edge:
\[
p(\mathbf{e}^c_{ij} \mid \mathcal{G}, V_\text{target}, C_\text{target}) = f_\text{edge}(h_{\mathbf{u}_{ij}}, h'_{\mathbf{u}_{ji}})
\]
Where $V_\text{target}$ are highlighted in input node features, $f_\text{edge}$ is a MLP and $h_{\mathbf{u}_{ij}}$ and $h'_{\mathbf{u}_{ji}}$ are computed with two neural networks, one over $\mathcal{G'}$ and one over $\mathcal{C_\text{target}}$. $p(\mathbf{e}^c_{ij}) = 1$ if $\mathbf{e}^c_{ij} \in E_\text{target}$ and $0$ otherwise.
If both $V_\text{target}$ and $C_\text{target}$ contain one slot, then these slots must be connected, and this module can be skipped.
To encourage the networks to learn more general representations, we augment $V_\text{target}$ with random unattached slots in $\mathcal{G}'$.

\parahead{Generating New Slot Graphs at Test Time}
At test time, we generate new slot graphs by iteratively querying the three modules defined above.
Although the modules we learn are probabilistic and support random sampling, we find MAP inference sufficient to produce a diverse range of shapes.
We terminate the generation process when the slot graph is complete i.e. when all slots in the graph are attached to exactly one slot from a different part.

\begin{figure}[t!]
    \centering
    \vspace{-1em}
    \setlength{\tabcolsep}{0pt}
    \renewcommand{\arraystretch}{0}
    \begin{tabular}{cccc}
        \\
        \includegraphics[width=.11\linewidth]{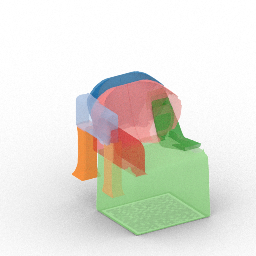} &
        \includegraphics[width=.11\linewidth]{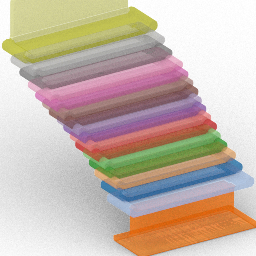} &
        \includegraphics[width=.11\linewidth]{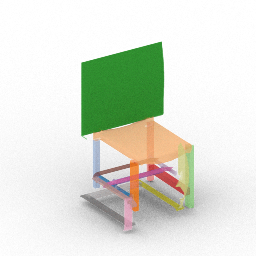} &
        \includegraphics[width=.11\linewidth]{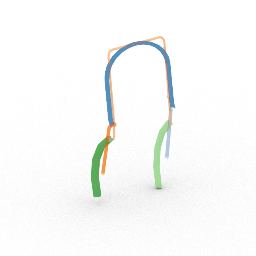}
    \end{tabular}
    \vspace{-0.5em}
    \caption{
    Typical structural outliers detected at test time.
    From left to right: redundant component (chair back), repetition of structures, inability to resolve local connections (chair base), not enough slots to finish structure.
    }
    \label{fig:outliers}
    \vspace{-1.2em}
\end{figure}

This stopping criterion, while simple, is not robust to errors: a random set of part cliques can occasionally form a complete slot graph by chance.
To combat these errors, we reject any partial shapes for which the ``how'' module assigns low probabilities to all candidate slot pairs.
We also use one-class Support Vector Machines to reject other structural outliers (see Figure~\ref{fig:outliers} for examples).

Finally, we also include logic to  enforce part symmetries.
When retrieving a part to connect with slots that were part of a symmetry group in their original shape, we alter the rank order of parts returned by our ``what'' module to prioritize (a) parts that occurred in symmetry groups in the dataset, followed by (b) parts that are approximately symmetrical according to chamfer distance.
See the supplemental for more details about these test-time sampling strategies.

%% file: sections/6-optimization.tex
\section{Assembling Shapes From Slot Graphs}
\label{sec:optimization}

A generated slot graph defines connectivity between shape parts but does not explicitly give part poses.
Thus, our system has an additional step to find world-space poses for all the retrieved parts.
In this section, we describe a simple gradient-descent-based optimization procedure for doing so, which takes a generated slot graph $\mathcal{G} = (V, E_c, E_p)$ that describes $N$ parts $P_1 \ldots P_N$, and predicts an affine transformation matrix $T_i$ for each part $P_i$.

\parahead{Objective Function}
To assemble a shape from its slot graph, we want each slot to be connected in the same way as it was in its original shape, which we approximate by enforcing that the distance from any point on a slot to the contacting slot should stay the same in the assembled shape.
Formally, for each slot $\mathbf{u}_{ij} \in V$, we select the set of points $S_{ij}$ that the slot contains (from the point sample of $P_i$ from Section~\ref{sec:representation}).
For each point $p \in S_{ij}$, we compute its distance $d_o(p)$ to the closest point on the slot that was originally connected to $\mathbf{u}_{ij}$ in the dataset.
We then optimize for $T_1 \ldots T_N$ via the following objective: for every slot $\mathbf{u}_{ij} \in V$, every point sample $p \in S_{ij}$ should have the same distance to the connecting slot $\mathbf{u}_{ji}$ as the original distance $d_o(p)$:
\[
f(T_1 \ldots T_N) =
\sum_{\mathbf{u}_{ij} \in V}
\sum_{p \in S_{ij}}
\Big(
\big(\min_{q \in S_{ji}} d(T_i p,T_j q)\big) - d_o(p)
\Big)^2
\]

\begin{figure}[t!]
    \centering
    \small
    \setlength{\tabcolsep}{1pt}
    \begin{tabular}{cc}
         Translation & Trans + scale
         \\
        \includegraphics[width=.13\linewidth]{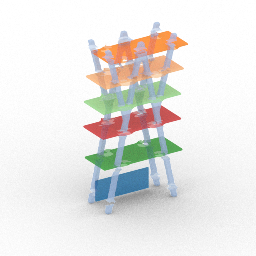} &
        \includegraphics[width=.13\linewidth]{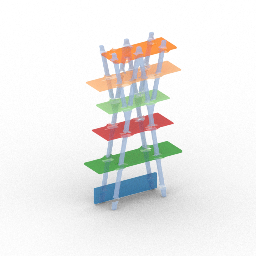}
    \end{tabular}
    \vspace{-1em}
    \caption{
    Optimizing part affine transformations to satisfy a slot graph.
    We show the output of the initial translation-only phase of optimization \& the final output with both translation and scale.
    }
    \label{fig:pose_opt}
    \vspace{-1.2em}
\end{figure}

\parahead{Optimization Process}
We minimize this objective using gradient descent.
Rather than full affine transformations, we optimize only translation and anisotropic scale.
This prevents certain part re-uses from happening (e.g. re-using a horizontal crossbar as a vertical crossbar), but we find that the space of possible outputs is nonetheless expressive.
To minimize unnecessary distortion, we prefer translation over scaling whenever possible: we optimize for translation only for the first $1000$ iterations, and then alternate between translation and scaling every $50$ iterations for the next $500$ iterations.
Optimizing for scales is essential in cases where translation alone cannot satisfy the slot graph. We show one such example in Figure~\ref{fig:pose_opt}, where the shelf layers are scaled horizontally to match the V shape of the frame.

%% file: sections/7-results.tex
\section{Results \& Evaluation}
\label{sec:results}

In this section, we evaluate our method's ability to synthesize novel shapes.
We use the PartNet~\cite{PartNet} dataset, segmenting shapes with the finest-grained PartNet hierarchy level and filtering out shapes with inconsistent segmentations and/or disconnected parts.
More implementation details are in the supplemental.

\begin{figure*}[t!]
    \centering
    \setlength{\tabcolsep}{1pt}
    \begin{tabular}{cccccccccc}
        \multirow{2}{*}{\raisebox{8em}{\rotatebox{90}{Chair}}} &
        \includegraphics[width=.105\linewidth]{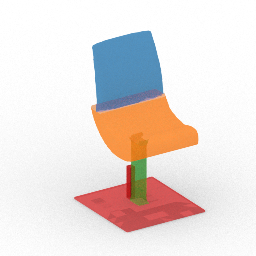} &
        \includegraphics[width=.105\linewidth]{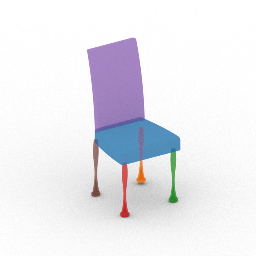} &
        \includegraphics[width=.105\linewidth]{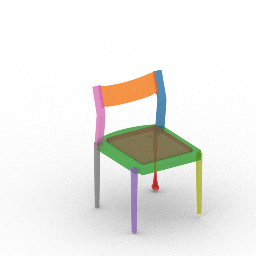} &
        \includegraphics[width=.105\linewidth]{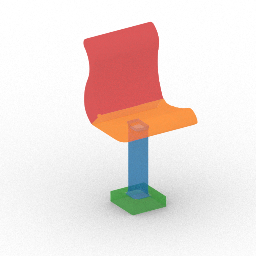} &
        \includegraphics[width=.105\linewidth]{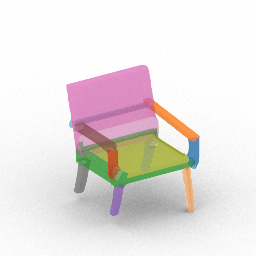} &
        \includegraphics[width=.105\linewidth]{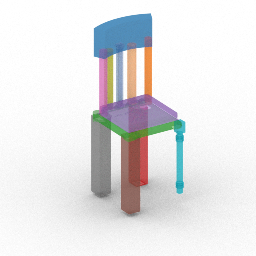} &
        \includegraphics[width=.105\linewidth]{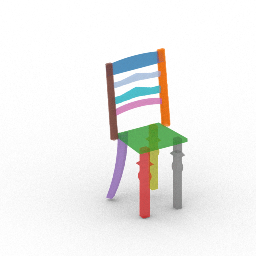} &
        \includegraphics[width=.105\linewidth]{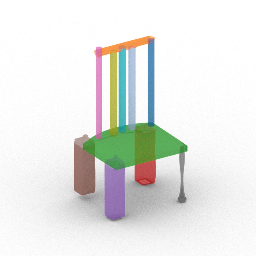} &
        \includegraphics[width=.105\linewidth]{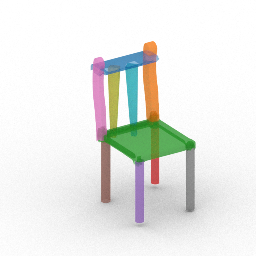}
        \\ [-0.5em]
        &
        \includegraphics[width=.105\linewidth]{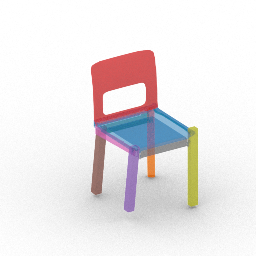} &
        \includegraphics[width=.105\linewidth]{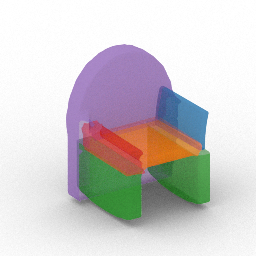} &
        \includegraphics[width=.105\linewidth]{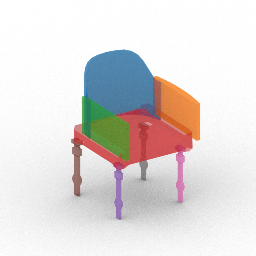} &
        \includegraphics[width=.105\linewidth]{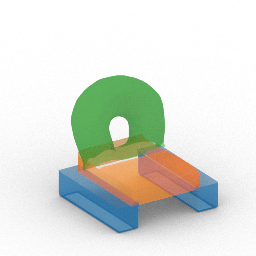} &
        \includegraphics[width=.105\linewidth]{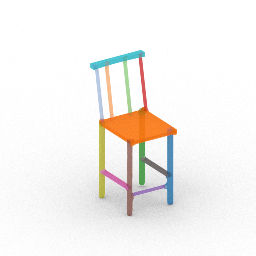} &
        \includegraphics[width=.105\linewidth]{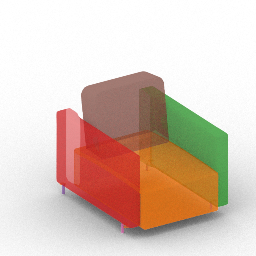} &
        \includegraphics[width=.105\linewidth]{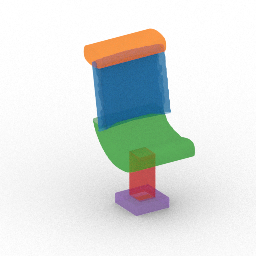} &
        \includegraphics[width=.105\linewidth]{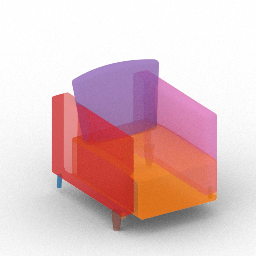} &
        \includegraphics[width=.105\linewidth]{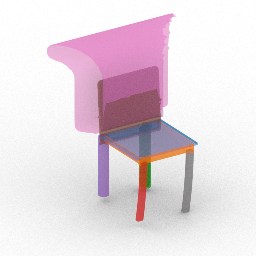}
        \\ [-0.5em]
        \multirow{2}{*}{\raisebox{8em}{\rotatebox{90}{Table}}} &
        \includegraphics[width=.105\linewidth]{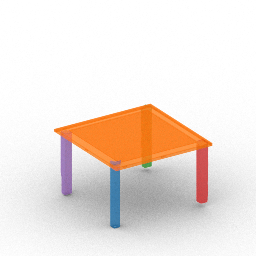} &
        \includegraphics[width=.105\linewidth]{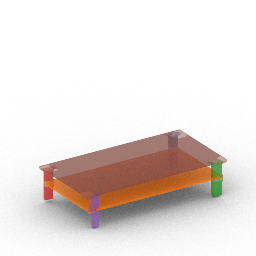} &
        \includegraphics[width=.105\linewidth]{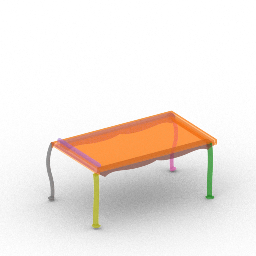} &
        \includegraphics[width=.105\linewidth]{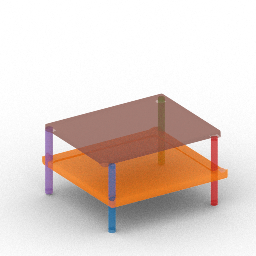} &
        \includegraphics[width=.105\linewidth]{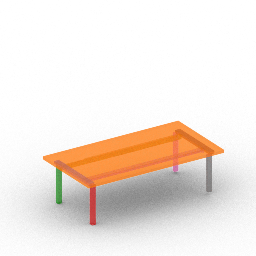} &
        \includegraphics[width=.105\linewidth]{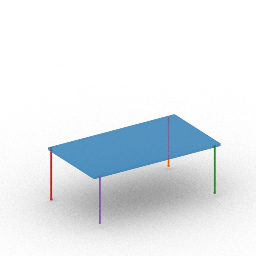} &
        \includegraphics[width=.105\linewidth]{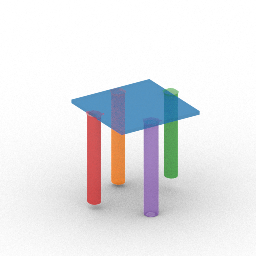} &
        \includegraphics[width=.105\linewidth]{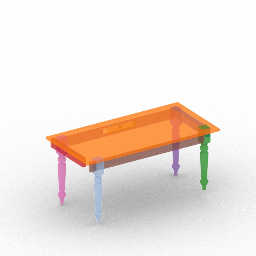} &
        \includegraphics[width=.105\linewidth]{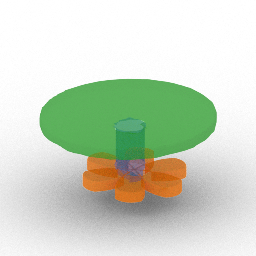}
        \\ [-0.5em]
        &
        \includegraphics[width=.105\linewidth]{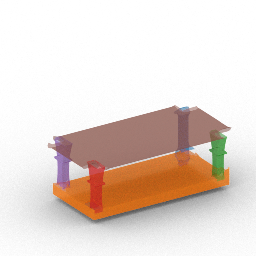} &
        \includegraphics[width=.105\linewidth]{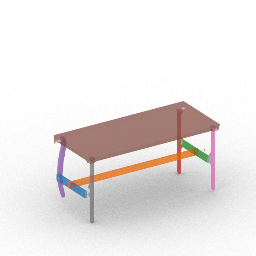} &
        \includegraphics[width=.105\linewidth]{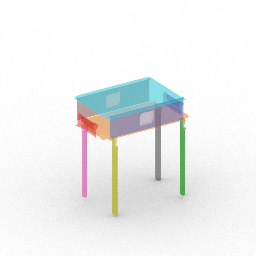} &
        \includegraphics[width=.105\linewidth]{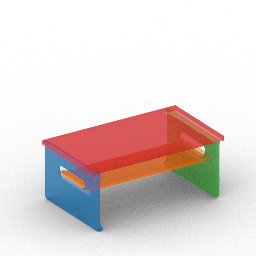} &
        \includegraphics[width=.105\linewidth]{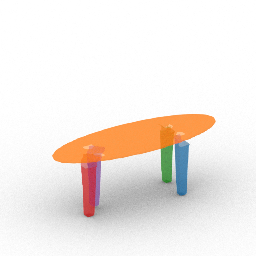} &
        \includegraphics[width=.105\linewidth]{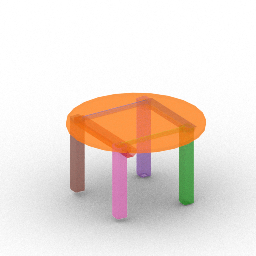} &
        \includegraphics[width=.105\linewidth]{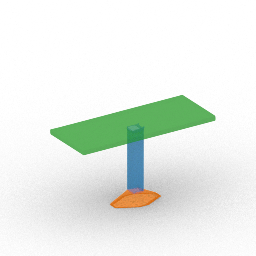} &
        \includegraphics[width=.105\linewidth]{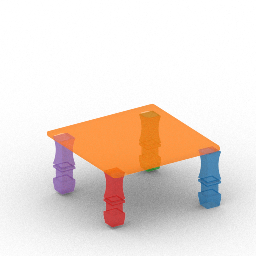} &
        \includegraphics[width=.105\linewidth]{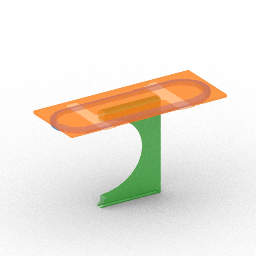}
        \\ [-0.5em]
        \multirow{2}{*}{\raisebox{8em}{\rotatebox{90}{Storage}}} &
        \includegraphics[width=.105\linewidth]{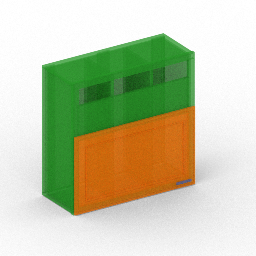} &
        \includegraphics[width=.105\linewidth]{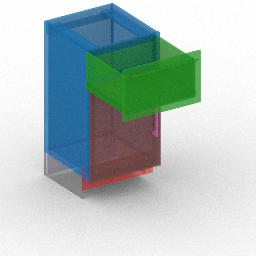} &
        \includegraphics[width=.105\linewidth]{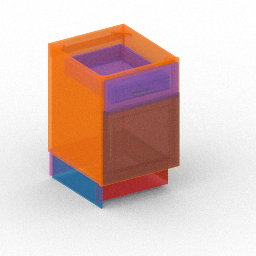} &
        \includegraphics[width=.105\linewidth]{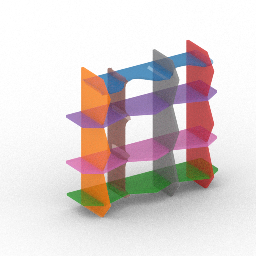} &
        \includegraphics[width=.105\linewidth]{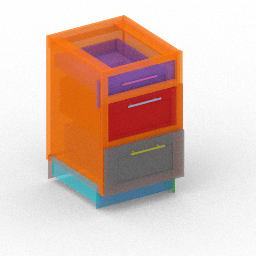} &
        \includegraphics[width=.105\linewidth]{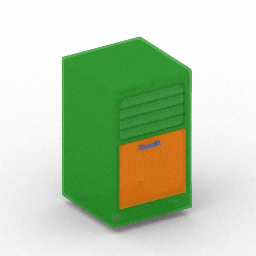} &
        \includegraphics[width=.105\linewidth]{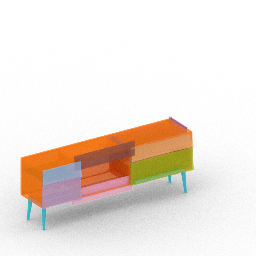} &
        \includegraphics[width=.105\linewidth]{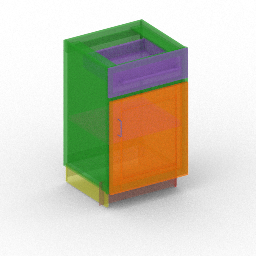} &
        \includegraphics[width=.105\linewidth]{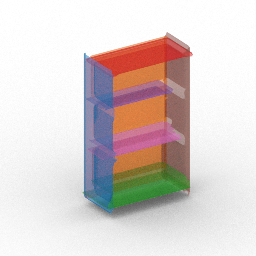}
        \\ [-0.5em]
        &
        \includegraphics[width=.105\linewidth]{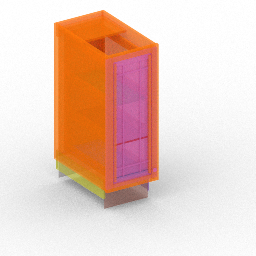} &
        \includegraphics[width=.105\linewidth]{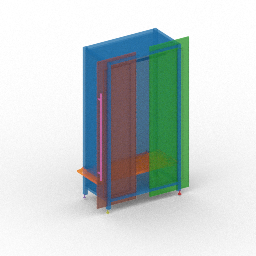} &
        \includegraphics[width=.105\linewidth]{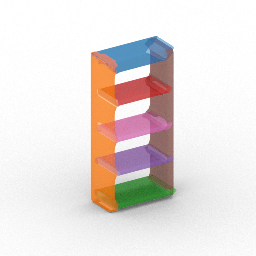} &
        \includegraphics[width=.105\linewidth]{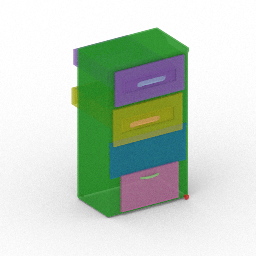} &
        \includegraphics[width=.105\linewidth]{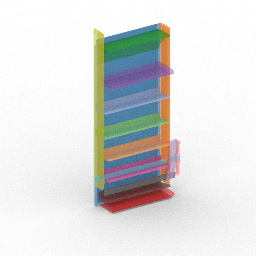} &
        \includegraphics[width=.105\linewidth]{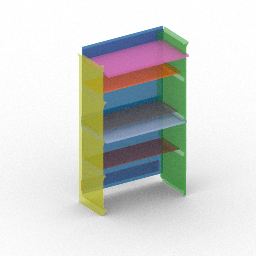} &
        \includegraphics[width=.105\linewidth]{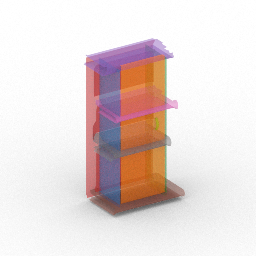} &
        \includegraphics[width=.105\linewidth]{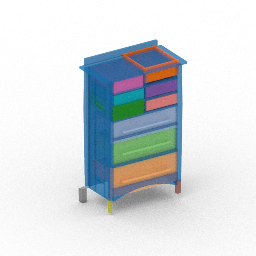} &
        \includegraphics[width=.105\linewidth]{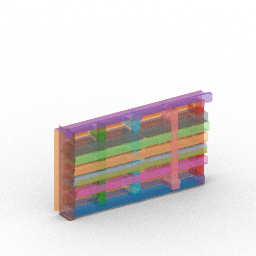}
        \\ [-0.5em]
        \multirow{2}{*}{\raisebox{8em}{\rotatebox{90}{Lamp}}} &
        \includegraphics[width=.105\linewidth]{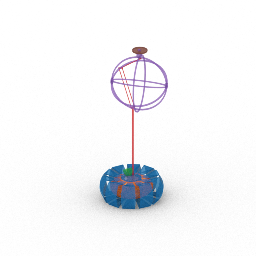} &
        \includegraphics[width=.105\linewidth]{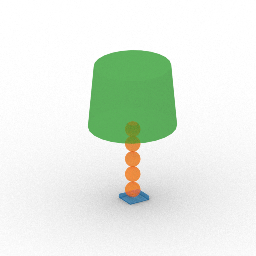} &
        \includegraphics[width=.105\linewidth]{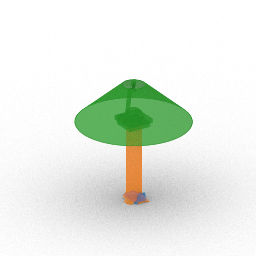} &
        \includegraphics[width=.105\linewidth]{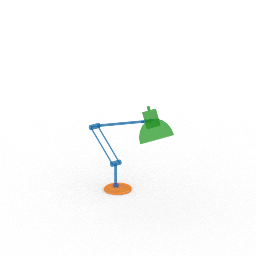} &
        \includegraphics[width=.105\linewidth]{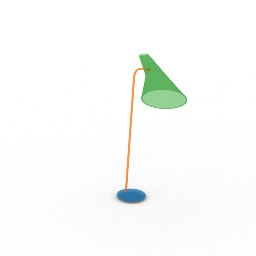} &
        \includegraphics[width=.105\linewidth]{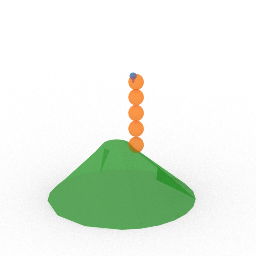} &
        \includegraphics[width=.105\linewidth]{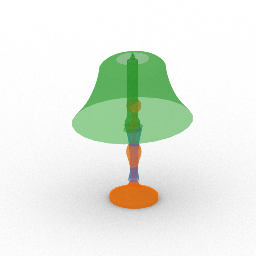} &
        \includegraphics[width=.105\linewidth]{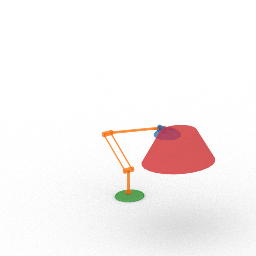} &
        \includegraphics[width=.105\linewidth]{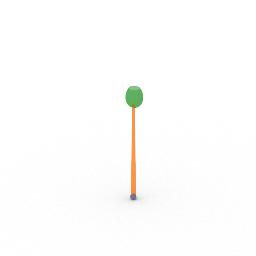}
        \\ [-0.5em]
        &
        \includegraphics[width=.105\linewidth]{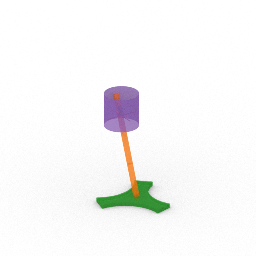} &
        \includegraphics[width=.105\linewidth]{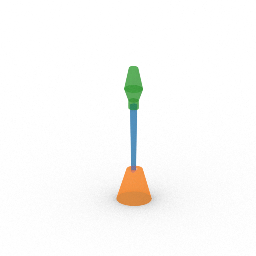} &
        \includegraphics[width=.105\linewidth]{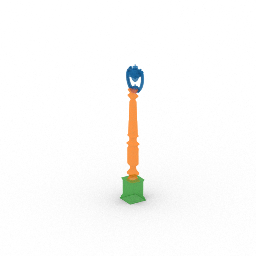} &
        \includegraphics[width=.105\linewidth]{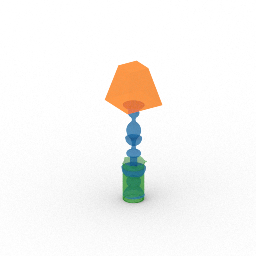} &
        \includegraphics[width=.105\linewidth]{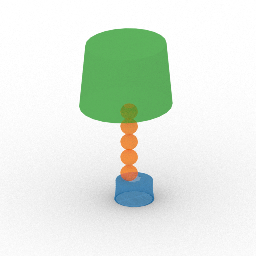} &
        \includegraphics[width=.105\linewidth]{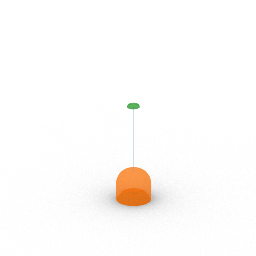} &
        \includegraphics[width=.105\linewidth]{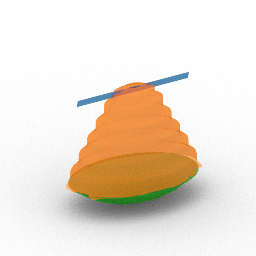} &
        \includegraphics[width=.105\linewidth]{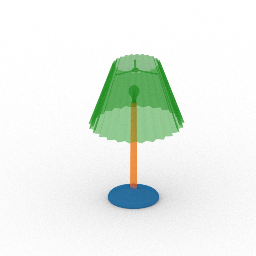} &
        \includegraphics[width=.105\linewidth]{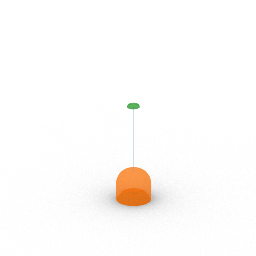}
        \\
    \end{tabular}
    \vspace{-1.5em}
    \caption{Examples of range of shapes our method is able to generate. Each part has a different color that correlates with the order they are inserted. \revisionnew{The blue part is used for initialization. See the supplementary material for more details about the color palette.}}
    \label{fig:lots_of_shapes}
\end{figure*}

\begin{table}[t!]
    \centering
    \footnotesize
    \setlength{\tabcolsep}{2pt}
    \caption{Comparing our system to baselines and ablations on generating visually and physically plausible shapes.}
    \begin{tabular}{@{}llccccc@{}}
        \toprule
        \textbf{Category} & \textbf{Method} & \textbf{Root$\uparrow$ } & \textbf{Stab$\uparrow$ } & \textbf{Fool$\uparrow$ } & \textbf{FD$\downarrow$ } & \textbf{Parts}\\
        \midrule
        \multirow{8}{*}{\emph{Chair}} 
        & Ours & $98.1$ & $70.1$ & $6.4$ & $61.1$ & $7.8$ \\
        & ComplementMe & $90.2$ & $41.1$ & $6.6$ & $83.0$ & $5.7$ \\
        & StructureNet & $81.0$ & $61.3$ & $4.0$ & $37.5$ & $12.1$ \\
        & Oracle & $94.5$ & $83.4$ & $25.8$ & $13.3$ & $-$ \\
        \cmidrule(lr){2-2}
        & ComplementMe (w/sym) & $88.3$ & $79.1$ & $21.9$ & $21.6$ & $4.3$ \\
        & Ground Truth & $100.0$ & $100.0$ & $-$ & $-$ & $11.1$\\ 
        \cmidrule(lr){2-2}
        & Ours (no symmetry) & $98.2$ & $68.0$ & $8.1$ & $58.9$ & $7.8$ \\
        & Ours (no duplicate) & $97.5$ & $67.4$ & $13.8$ & $61.2$ & $7.7$ \\
        \midrule
        \multirow{3}{*}{\emph{Table}} 
        & Ours & $98.2$ & $82.8$ & $10.6$ & $61.6$ & $6.8$\\
        & ComplementMe & $90.2$ & $62.0$ & $7.7$ & $93.5$ & $4.7$ \\
        & StructureNet & $82.8$ & $78.5$ & $2.3$ & $85.2$ & $7.8$ \\
        \cmidrule(lr){2-2}
        & ComplementMe (w/sym) & $87.1$ & $84.0$ & $35.8$ & $18.7$ & $3.3$ \\
        & Ground Truth & $100.0$ & $100.0$ & $-$ & $-$ & $9.3$ \\
        \midrule
        \multirow{3}{*}{\emph{Storage}} 
        & Ours & $99.4$ & $90.8$ & $15.5$ & $42.6$ & $6.9$ \\
        & ComplementMe & $91.4$ & $72.4$ & $8.9$ & $89.8$ & $3.4$ \\
        & StructureNet & $89.6$ & $82.2$ & $6.8$ & $105.5$ & $8.3$ \\
        \cmidrule(lr){2-2}
        & ComplementMe (w/sym) & $85.3$ & $70.6$ & $11.5$ & $71.4$ & $3.3$\\
        & Ground Truth & $100.0$ & $99.4$ & $-$ & $-$ & $13.6$\\
        \midrule
        \multirow{3}{*}{\emph{Lamp}} 
        & Ours & $89.6$ & $-$ & $21.5$ & $42.4$ & $3.4$\\
        & ComplementMe & $62.0$ & $-$ & $35.8$ & $26.0$ & $3.4$\\
        & Ground Truth & $92.6$ & $-$ & $-$ & $-$ & $4.2$\\
        \bottomrule
    \end{tabular}
    \label{tab:quality}
\end{table}

\begin{figure}[t!]
    \centering
    \setlength{\tabcolsep}{1pt}
    \begin{tabular}{cccccccccc}
        \raisebox{0.8em}{\rotatebox{0}{(a)}} &
        \includegraphics[width=.105\linewidth]{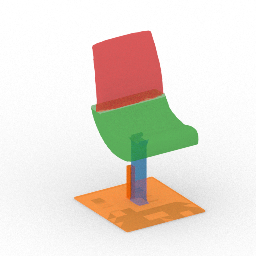} &
        \includegraphics[width=.105\linewidth]{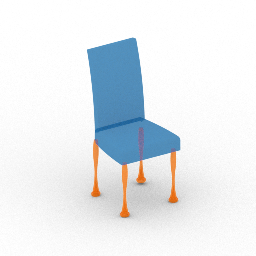} &
        \includegraphics[width=.105\linewidth]{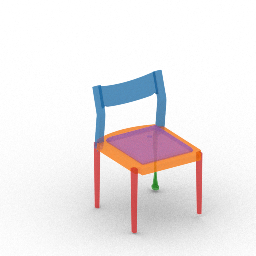} &
        \includegraphics[width=.105\linewidth]{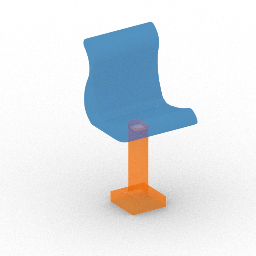} &
        \includegraphics[width=.105\linewidth]{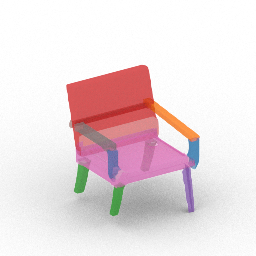} &
        \includegraphics[width=.105\linewidth]{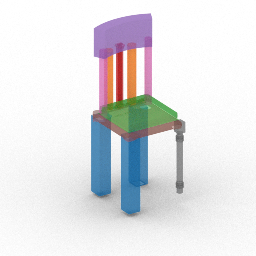} &
        \includegraphics[width=.105\linewidth]{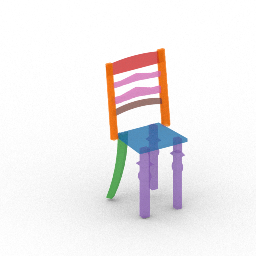} &
        \includegraphics[width=.105\linewidth]{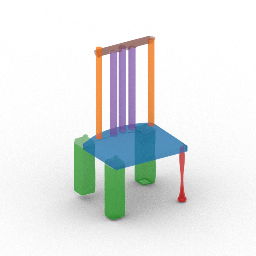} &
        \includegraphics[width=.105\linewidth]{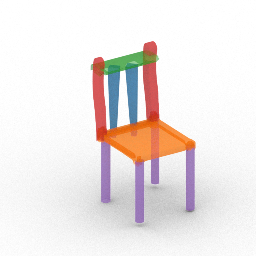}
        \\ [-0.5em]
        \raisebox{0.8em}{\rotatebox{0}{(b)}} &
        \includegraphics[width=.105\linewidth]{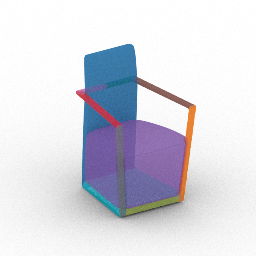} &
        \includegraphics[width=.105\linewidth]{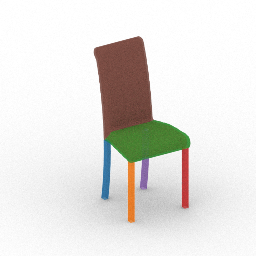} &
        \includegraphics[width=.105\linewidth]{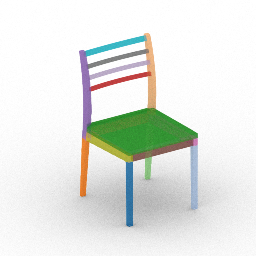} &
        \includegraphics[width=.105\linewidth]{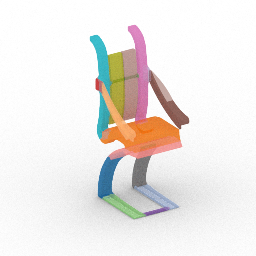} &
        \includegraphics[width=.105\linewidth]{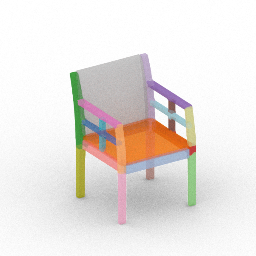} &
        \includegraphics[width=.105\linewidth]{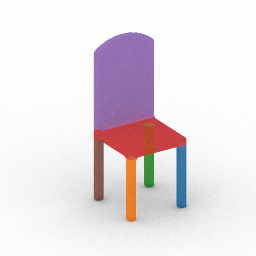} &
        \includegraphics[width=.105\linewidth]{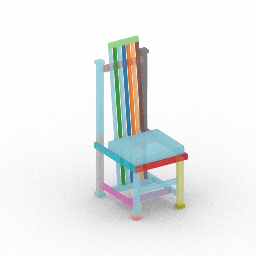} &
        \includegraphics[width=.105\linewidth]{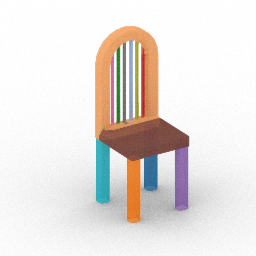} &
        \includegraphics[width=.105\linewidth]{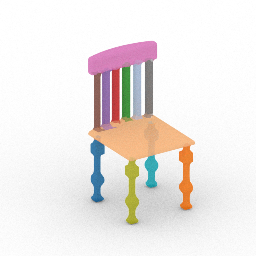}
        \\ [-0.5em]
        \raisebox{0.8em}{\rotatebox{0}{(c)}} &
        \includegraphics[width=.105\linewidth]{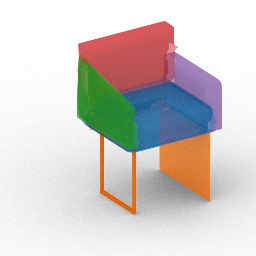} &
        \includegraphics[width=.105\linewidth]{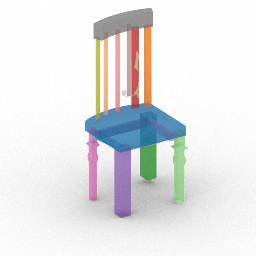} &
        \includegraphics[width=.105\linewidth]{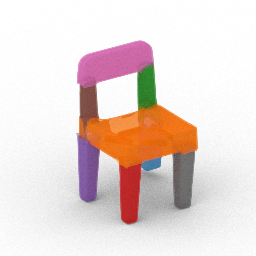} &
        \includegraphics[width=.105\linewidth]{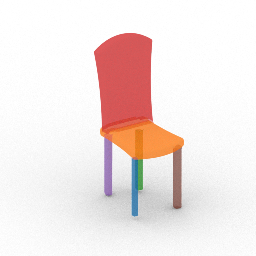} &
        \includegraphics[width=.105\linewidth]{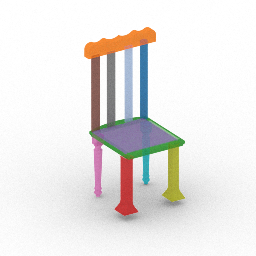} &
        \includegraphics[width=.105\linewidth]{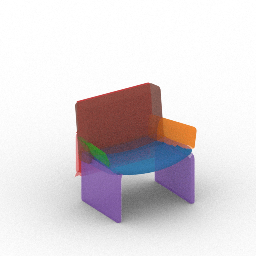} &
        \includegraphics[width=.105\linewidth]{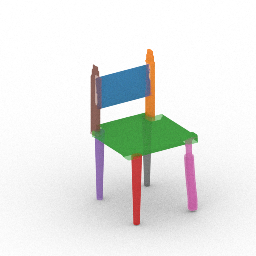} &
        \includegraphics[width=.105\linewidth]{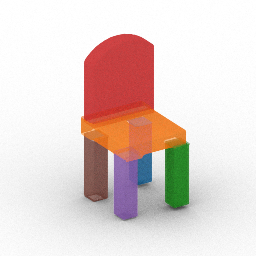} &
        \includegraphics[width=.105\linewidth]{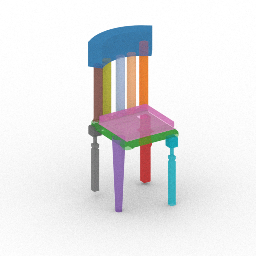}
        \\ [-0.5em]
        \raisebox{0.8em}{\rotatebox{0}{(d)}} &
        \includegraphics[width=.105\linewidth]{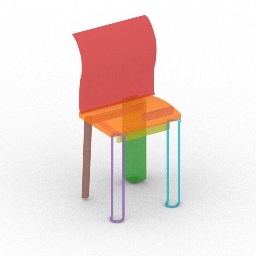} &
        \includegraphics[width=.105\linewidth]{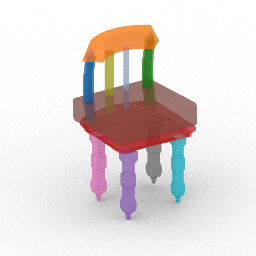} &
        \includegraphics[width=.105\linewidth]{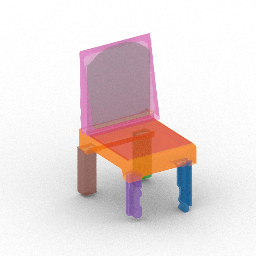} &
        \includegraphics[width=.105\linewidth]{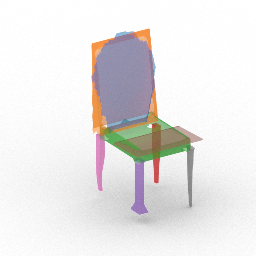} &
        \includegraphics[width=.105\linewidth]{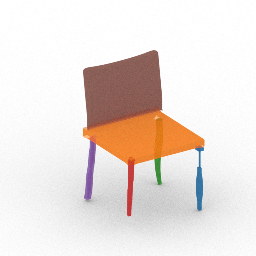} &
        \includegraphics[width=.105\linewidth]{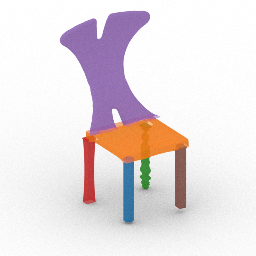} &
        \includegraphics[width=.105\linewidth]{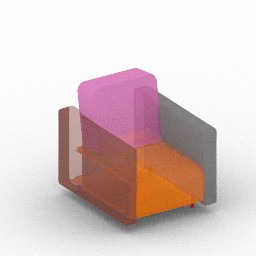} &
        \includegraphics[width=.105\linewidth]{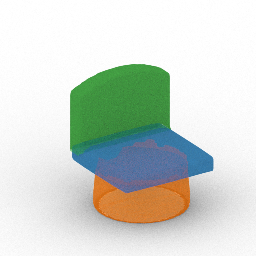} &
        \includegraphics[width=.105\linewidth]{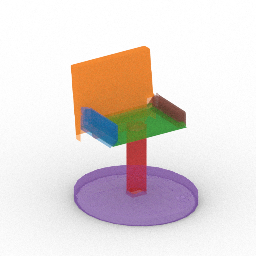}
    \end{tabular}
    \vspace{-1em}
    \caption{
    \revision{
    (a): Chairs in the first row of Figure~\ref{fig:lots_of_shapes}, where parts coming from the same source shape now have the same color.
    (b): Geometric nearest neighbor of the the same chairs in the training set.
    (c): Chairs generated without enforcing part symmetries.
    (d): Chairs generated with the explicit rule that no parts coming from the same source shape can be attached together.
    Our method uses parts from different shapes to generate novel shapes. It can generate approximately symmetric shapes without explicit rules, and can connect parts from different shapes together plausibly. 
    }
    },
    \label{fig:generalization}
    \vspace{-3em}
\end{figure}

\parahead{Novel Shape Synthesis}
Figure~\ref{fig:lots_of_shapes} shows examples of shapes our method is capable of assembling.
Our model learns to generate shapes with complex structures specific to each shape categories.
Though it does not use global part position information during graph generation, the resulting slot graphs lead to consistent global positions for the individual parts once optimized.
Although we choose not to encode full geometries, our model is still able to generate shapes with interesting variations both structurally and geometrically.
We further quantitatively compare the results generated by our model against against these alternatives: 
\begin{itemize}[noitemsep,topsep=0pt,parsep=0pt,partopsep=0pt]
\item
\textbf{ComplementMe}~\cite{ComplementMe} is the previous state-of-art for modeling by part assembly.
It retrieves compatible parts and places them together by predicting per-part translation vectors.
ComplementMe also does not implement a stopping criteria for generation, so we train a network that takes a partial shape point cloud as input and predicts whether generation should stop.
We also stop the generation early if the output of the part retrieval network does not change from one step to the next.
\revision{Finally, ComplementMe relies on a part discovery process where most groups of symmetrical parts are treated as a single part (e.g. four chair legs). 
We notice that, when trained on our data, ComplementMe suffers from a significant performance decrease on Chair and Table, and \revisionnew{struggles to generate more complex Storage, as is evident from the average number of parts} (See Table~\ref{tab:quality}). Therefore, for these categories, we also include results where parts are grouped by symmetry (w/sym) for reference. We stress that, under this condition, both retrieving and assembling parts are significantly easier, thus the results are not directly comparable.
}
\item
\textbf{StructureNet}~\cite{StructureNet} is an end-to-end generative model outputs a hierarchical shape structure, where each leaf node contains a latent code that can either be decoded into a cuboid or a point cloud. We modify it to output meshes by, for each leaf node, retrieving the part in the dataset whose StructureNet latent code is closest to the leaf node's latent code and then transforming the part to fit the cuboid for that leaf node.
\item
We also include an \textbf{Oracle} as an upper bound on retrieval-based shape generation. 
The oracle is an autoregressive model that takes as input at each step (a) the bounding boxes for all parts in a ground-truth shape and (b) point clouds for parts retrieved so far.
Retrieved parts are scaled so that they fit exactly to the bounding box to which they are assigned.
\nolistbottomspace
\end{itemize}
See supplemental for more details about these baselines.
We use an evaluation protocol similar to ShapeAssembly~\cite{jones2020shapeAssembly} which evaluates both the physical plausibility and quality of generated shapes: 
\begin{itemize}[noitemsep,topsep=0pt,parsep=0pt,partopsep=0pt]
\item
\textbf{Rootedness $\uparrow$ (Root)} measures the percentage of shapes for which there is a connected path between the ground to all parts; 
\item
\textbf{Stability $\uparrow$ (Stable)} measures the percentage of shapes that remains upright under gravity and a small force in physical simulation, \revisionnew{we do not report this for lamps because lamps such as chandeliers do not need to be stable}; 
\item
\textbf{Realism $\uparrow$ (Fool)} is the percentage of test set shapes classified as ``generated” by a PointNet trained to distinguish between dataset and generated shapes; 
\item
\textbf{Freschet Distance $\downarrow$ (FD)}~\cite{FrechetInceptionDistance} measures distributional similarity between generated and dataset shapes in the feature space of a pre-trained PointNet classifier.
\item
\textbf{Parts} is the mean number of parts in generated shapes.
\end{itemize}
Table~\ref{tab:quality} summarizes the results.
By using a contact-based representation, our model is able to generate shapes that are more physically plausible (rooted and stable) than the baselines.
while being comparable in terms of the overall shape quality, measured by Frechet distance and classifier fool percentage.
Our model performs particularly well for storage furniture; we hypothesize rich connectivity information of this shape category allows our model to pick parts that match particularly well.
Our model fares less well on lamps, where connectivity structure is simple and the geometric variability of parts (which out model does not encode) is highly variable.
\revision{
ComplementMe works well on lamps, thanks to its focus on part geometry. Its performance drops significantly on all other categories with more complicated shape structures.
}
We provide more details, as well as random samples for all methods, in the supplementary material.

\parahead{Generalization Capacity}
It is important that a generative model that follows the modeling by assembly paradigm learns to recombine parts from different sources into novel shapes.
We demonstrate our model's capacity for this in Figure~\ref{fig:generalization}: it is able to assemble parts from multiple source shapes together \revision{into novel shapes different from those seen during training}, with or without explicit restrictions \revision{whether parts from the same source shape can be connected to each other}.
We also see that while including symmetry reasoning improves geometric quality, our method is able to generate shapes that are roughly symmetrical without it. 
This is also reflected in Table~\ref{tab:quality}: removing symmetry or prohibiting using multiple parts from the same source shape has minimal impact on our metrics.
We provide more analysis of generalization in the supplemental.

\parahead{Performance of Individual Modules}
Finally, we evaluate each individual model module, using the following metrics:
\begin{itemize}[noitemsep,topsep=0pt,parsep=0pt,partopsep=0pt]
    \item \textbf{Attach Acc}: How often the "where" module correctly selects the slots to attach, given the first slot.
    \item \textbf{Average Rank}: Average percentile rank of ground truth next part according to the ``what'' module.
    \item \textbf{Edge Acc}: How often the ``how'' module recovers the the ground truth edge pairs.
\end{itemize}

Table~\ref{tab:component_ablation} summarizes the results. Modules perform very well, with some lower numbers caused by inherent multimodality of the data.
\begin{table}[t!]
    \centering
    \footnotesize
    \caption{
    Evaluating our neural network modules in isolation.
    }
    \begin{tabular}{@{}lccc@{}}
        \toprule & \textbf{Attach Acc} & \textbf{Avg Rank} & \textbf{Edge Acc}  \\
        \midrule
        Chair & $96.70$ & $99.05$ & $94.79$ \\
        Table & $92.32$ & $99.14$ & $92.82$ \\
        Storage & $87.46$ & $99.08$ & $85.38$ \\
        Lamp & $98.87$ & $91.36$ & $91.89$ \\
        \bottomrule
    \end{tabular}
    \label{tab:component_ablation}
\end{table}

\begin{figure}[t!]
    \centering
    \setlength{\tabcolsep}{0pt}
    \renewcommand{\arraystretch}{0}
    \begin{tabular}{cccc}
        \\
        \includegraphics[width=.13\linewidth]{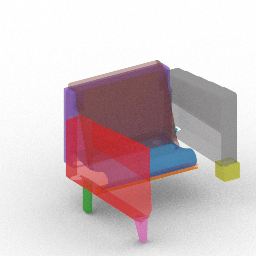} &
        \includegraphics[width=.13\linewidth]{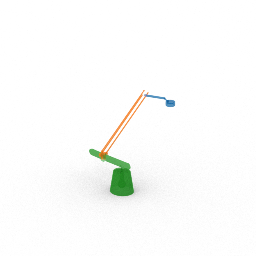} & 
        \includegraphics[width=.13\linewidth]{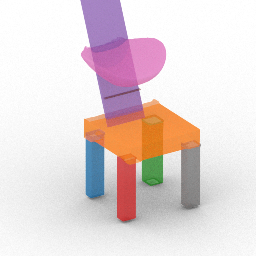} &
        \includegraphics[width=.13\linewidth]{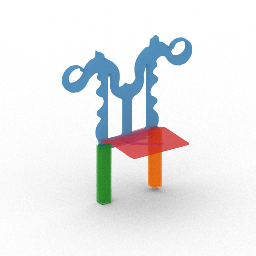} 
    \end{tabular}
    \vspace{-0.75em}
    \caption{
    Typical failure cases of our method.
    From left to right: a chair with a tiny seat, two opposite-facing lamps attached together awkwardly, a chair with a implausible back, a chair that misses seat and legs completely.
    }
    \label{fig:failure_cases}
    \vspace{-1.3em}
\end{figure}

\parahead{Limitations}
Even with outlier detection as mentioned in section~\ref{sec:graphgen}, poor-quality outputs can still occur.
Figure~\ref{fig:failure_cases} shows typical examples. Most are caused by our model's lack of focus on geometry: chairs with a tiny seat, lamps that face opposite directions, and chair backs that block the seat completely. 
Incorporating additional geometric features when appropriate could help.

%% file: sections/8-conclusion.tex
\section{Conclusion}
\label{sec:conclusion}

We presented the Shape Part Slot machine, a new modeling-by-part-assembly generative model for 3D shapes.
Our model synthesizes new shapes by generating slot graphs describing the contact structure between parts; it then assembles its retrieved parts by optimizing per-part affine transforms to be consistent with this structure.
The slot graph encodes surprisingly little information, yet we demonstrated experimentally that our model outperforms multiple baselines and prior modeling-by-assembly systems on generating novel shapes from PartNet parts.

There are multiple directions for future work.
Parts could be repurposed in more diverse ways if we had a method to transfer slot graphs between geometrically- and contextually-similar parts (so e.g. a chair seat that had armrests originally does not have to have them in all synthesized results).
More variety could also be obtained by optimizing for part orientations (so e.g. a vertical slat could be used as a horizontal one).

%% file: sections/9-acknowledgements.tex
$\;\\$
$\;\\$
\noindent\textbf{Acknowledgements}
This work was funded in part by NSF award \#1907547 and a gift fund from Adobe.
Daniel Ritchie is an advisor to Geopipe and owns equity in the company. Geopipe is a start-up that is developing 3D technology to build immersive virtual copies of the real world with applications in various fields, including games and architecture.
Minhyuk Sung acknowledges the support of NRF grant (2022R1F1A1068681), NST grant (CRC 21011), and IITP grant (2022-0-00594) funded by the Korea government (MSIT), and grants from Adobe, KT, and Samsung Electronics corporations.
Part of this work was done when Kai Wang interned at Adobe.

%% file: supp/1-data.tex
\section{Data Preparation}
\label{sec:data}

We use the PartNet~\cite{PartNet} dataset for all our experiments, following the train/validation/test split provided in the original paper.

\subsection{Obtaining Part Level Geometry}
Each shape in the PartNet data comes with a semantic hierarchy that decomposes the shape into parts in a coarse-to-fine manner.
We use the finest-grained level of parts in this hierarchy.
We filter the data using the following criteria:
\begin{packed_itemize}
\item
We remove shapes that contain only $1$ part or more than $30$ parts.
\item
We detect inconsistent shapes with parts that do not equal the union of their children.
For the chairs and tables dataset, we reject these inconsistent shapes.
The furniture and lamps dataset are smaller, so in these datasets, we keep the inconsistent parts, but discard their children.
\item
We remove shapes with parts that contain floating geometry
due to annotation errors. 
We detect such cases by first clustering the part's point cloud with DBSCAN~\cite{DBSCAN}. 
If there exist any cluster that is significantly smaller than other clusters, we reject the entire shape.
(We cannot reject all parts that consist of multiple disconnected clusters, because some parts contain multiple symmetric disconnected components).
\item
We remove shapes that are disconnected based on the adjacency edges we detect.
\end{packed_itemize}

Table~\ref{tab:dataset} summarizes the size of the dataset before and after filtering.

\begin{table}[t!]
    \centering
    \footnotesize
    \setlength{\tabcolsep}{2pt}
    \caption{Dataset statistics before and after our filtering process}
    \begin{tabular}{@{}llcc@{}}
        \toprule
        \textbf{Category} & \textbf{Split} & \textbf{Before} & \textbf{After} \\
        \midrule
        \multirow{3}{*}{\emph{Chair}} 
        & Train & 4489 & 3315 \\
        & Val & 617 & 438 \\
        & Test & 1217 & 886 \\
        \midrule
        \multirow{3}{*}{\emph{Table}} 
        & Train & 5707 & 4254 \\
        & Val & 843 & 637 \\
        & Test & 1668 & 1257 \\
        \midrule
        \multirow{3}{*}{\emph{Storage}} 
        & Train & 1588 & 1123 \\
        & Val & 230 & 152 \\
        & Test & 451 & 290 \\
        \midrule
        \multirow{3}{*}{\emph{Lamp}} 
        & Train & 1554 & 1187 \\
        & Val & 234 & 181 \\
        & Test & 419 & 321 \\
        \bottomrule
    \end{tabular}
    \label{tab:dataset}
\end{table}

\subsection{Extracting Relationships Between Parts}
After obtaining the geometry of individual parts,
we sample the surface of each part uniformly to obtain a $3000$-point representation.
We then detect relationships between parts based on the protocol of StructureNet~\cite{StructureNet}:

\parahead{Detecting Symmetry} 
We detect symmetry based on the methods proposed by Wang et al.~\cite{wang2011symmetry}.
We restrict the symmetry types to translational symmetry, reflectional symmetry about planes parallel to the three coordinate planes, and 4-way rotational symmetry about the y(up)-axis.
We create an undirected graph for each of the symmetry type, where every edge is a detected symmetry between a pair of parts. We treat each connected component in these graphs as a symmetry group.

\parahead{Pruning Symmetry}
We then prune the detected symmetries by enforcing that each part belongs to at most one symmetry groups. We prioritize larger groups. If two groups are of the same size, then we favor the simpler explanation: translational $>$ rotational $>$ reflectional.

\parahead{Detecting Adjacency}
We regard two parts, $A$ and $B$, as adjacent if the smallest distance between their respective points clouds is less than $\tau=\theta\mathbf{r}$, where $\mathbf{r}$ is the average bounding sphere radius of the two parts.
We first detect symmetries using $\theta=0.05$ i.e. the orignal setting of StructureNet.
We then do a second pass of adjacency detection for parts involved in symmetry groups in order to recover any undetected adjancies to a common neighbor:
We set $\theta=0.1$ if $A$ belongs to a symmetry group (before pruning) and $B$ is adjacent (using $\theta=0.05$) to any other parts in the same symmetry group, vice versa;
we further increase $\theta$ to $0.3$ if the involved symmetry group has more than $3$ parts and at least $3$ other parts are adjacent to $B$, vice versa.
We use the same threshold $\tau$ for computing the points for each slot (Section 4.2).

\parahead{Pruning Adjacency}
We then attempt to identify the set of adjancency relationships that best describe the part structure. 
Note that this might not be necessary for a dataset where the connections between parts are more clearly defined.
We prune the adjacency edges using the following set of heuristics, applied in order. All heuristics are only applied if removing the edge does not disconnect the adjacency graph:

We first remove any edges between parts in the same symmetry group, prior to symmetry pruning.

We then identify all triplet of parts $A,B,C$ that overlap at the same area, and thus pairwise adjacent. For each triplet, we check if there's an edge that we can prune, using the following heuristics, without loss of generality, applied in order:
    \begin{packed_itemize}
    \item
    If $B$ and $C$ shares a common parent in the PartNet hierarchy and $A$ has a different parent, then we store either $AB$ or $AC$ for deletion if we can break ties between them: we store the edge for the part that is either significantly farther from $A$, smaller in surface area, or with less adjacent parts. We do not store any edges if the ties between $B$ and $C$ cannot be broken.
    \item
    We store $AC$ for deletion if the y(up)-coordinate of the centroid of $B$ is between those of $A$ and $C$, and is of at least a distance of $0.05$ away from each.
    \item
    We store $BC$ for deletion if the surface area of both part $B$ and $C$ is signifcantly smaller than that of part $A$, or if $B$ and $C$ has roughly the same area but $A$ does not.
    \end{packed_itemize}
We then sort all the candidate edges for deletion, prioritizing on those detected with heuristics mentioned earlier, and then those belonging to parts with smaller surface areas.

After finding all the candidates edges, we iterate over them and delete edges, while respecting the detected symmetries. For each candidate $AB$, we check if $A$ and/or $B$ belongs to any symmetry groups. If $A$ is in a symmetry group, then we include all other adjacency edges from any parts in that symmetry group to $B$. We do the same for $B$.
We proceed to remove all these edges if the following conditions are met:
    \begin{packed_itemize}
    \item
    Removing these edges does not disconnect the graph.
    \item
    Removing these edges does not disconnect a symmetry group from its most frequent neighbor i.e. the part that has the highest number of adjacency edges to parts in the symmetry group. If multiple such neighbors exist, we prefer to keep the edges to the neighbor that is not in any symmetry groups. If there are still multiple such neighbors, we keep only $1$ of them, and allow deleting edges to the rest. 
    A special case occurs when every neighbor to a symmetry group is adjacent to exactly one part in the group. This often occurs when a group of symmetrical parts are decomposed further into subparts (e.g. four symmetrical legs are decomposed into four legs and for leg wheels). In such cases, we regard every part in the adjacent symmetry group as a most frequent neighbor.
    \item
    Either $A$ and $B$ are still both connected to $C$ in the original triplet, or if there exists other parts in the region where $A$, $B$ and $C$ overlaps and a path can be found from $A$ to $B$ via those parts or vice versa.
    \end{packed_itemize}

%% file: supp/2-retrieval.tex
\section{More Details on the ``What" Module}
\label{sec:retrieval}
We provide additional details on the \textbf{What to Attach?} module (Section 5) here.

Given the graph features $h_\mathcal{G'}, h_\mathcal{G'_\text{target}}$, the mixture density network (MDN) represents the conditional probability distribution 
$P(X | \mathcal{G}', V_\text{target}, X \in \mathbb{R}_\text{emb})$ as a mixture of $N$ gaussians, with mixing coefficients $\pi_1 \ldots \pi_N$, means $\mu_1 \ldots \mu_N$ and standard deviations $\sigma_1 \ldots \sigma_N$ respectively.
The probability of any embedding $X_C$, then, can be expressed as
\[
p(X_C) = \sum_{k=1}^N \pi_k \cdot \mathcal{N} (X_C \;|\; \mu_k, \sigma_k^2)
\]
We omit the conditions ($\mathcal{G}', V_\text{target}$) for simplicity of notation.
In practice, we use negative log likelihood to setup the triplet loss:
\[
\ell(X_C) = -\log\sum_{k=1}^N \pi_k \cdot \mathcal{N} (X_C \;|\; \mu_k, \sigma_k^2)
\]
Given a positive example $C_\text{target}$ and a negative example $C_\text{negative}$, we then obtain the final triplet loss as 
\begin{equation*}
    \mathcal{L}(X_{C_\text{target}}, X_{C_\text{negative}})
    = \max\{m + \ell(X_{C_\text{target}}) - \ell(X_{C_\text{negative}}), 0\}
\end{equation*}
Where $m$ is a constant margin.
We select the negative examples $C_\text{negative}$ at training time by computing the triplet loss between the positive example and a set of randomly sampled negative examples, and choose one that gives a non-zero loss, whenever possible (i.e. using only semi-hard triplets).

\begin{figure*}[t!]
    \centering
    \setlength{\tabcolsep}{0pt}
    \renewcommand{\arraystretch}{0}
    \begin{tabular}{cccccccc}
        Input & Input + GT & GT & Best & $2$nd & $5$th & $25$th & $50\%$
        \\ [-0.25em]
        \includegraphics[width=.11\linewidth]{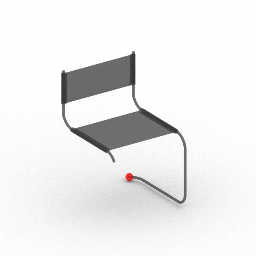} &
        \includegraphics[width=.11\linewidth]{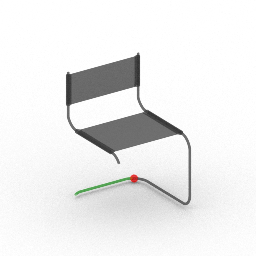} &
        \includegraphics[width=.11\linewidth]{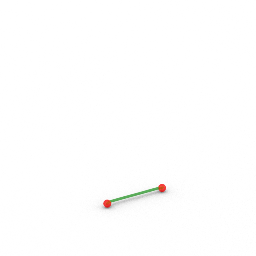} &
        \includegraphics[width=.11\linewidth]{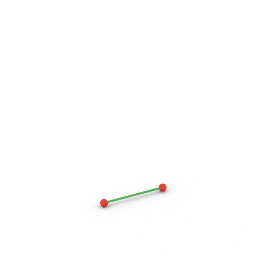} &
        \includegraphics[width=.11\linewidth]{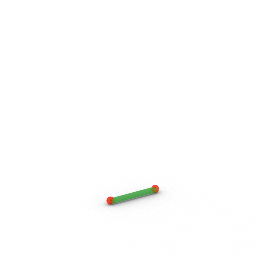} &
        \includegraphics[width=.11\linewidth]{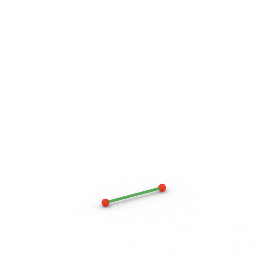} &
        \includegraphics[width=.11\linewidth]{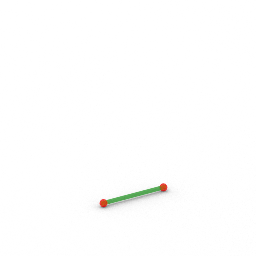} &
        \includegraphics[width=.11\linewidth]{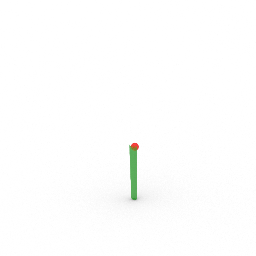}
        \\  [-0.25em]
        \includegraphics[width=.11\linewidth]{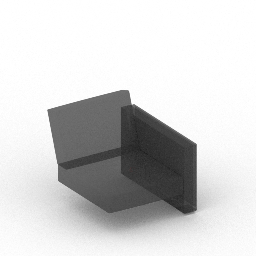} &
        \includegraphics[width=.11\linewidth]{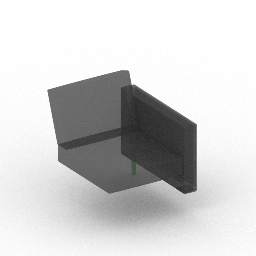} &
        \includegraphics[width=.11\linewidth]{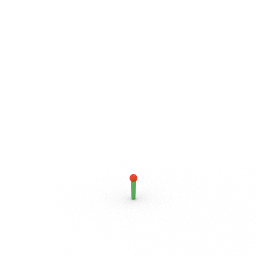} &
        \includegraphics[width=.11\linewidth]{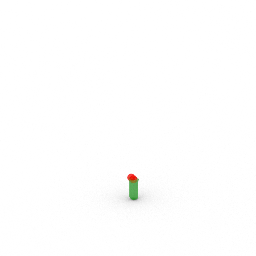} &
        \includegraphics[width=.11\linewidth]{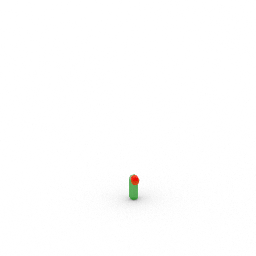} &
        \includegraphics[width=.11\linewidth]{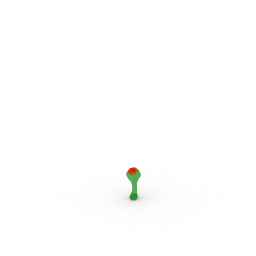} &
        \includegraphics[width=.11\linewidth]{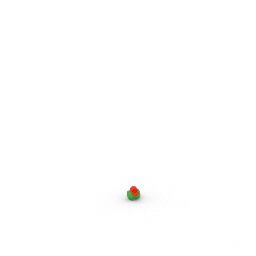} &
        \includegraphics[width=.11\linewidth]{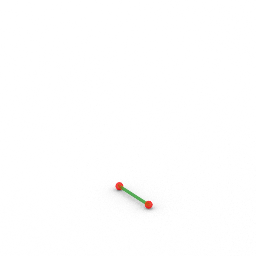}
        \\ [-0.25em]
        \includegraphics[width=.11\linewidth]{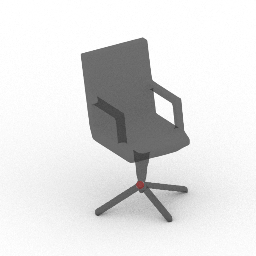} &
        \includegraphics[width=.11\linewidth]{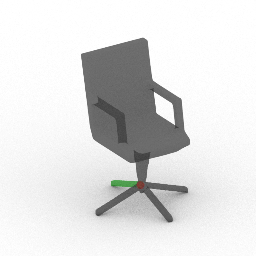} &
        \includegraphics[width=.11\linewidth]{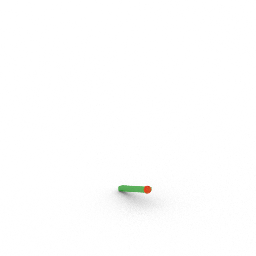} &
        \includegraphics[width=.11\linewidth]{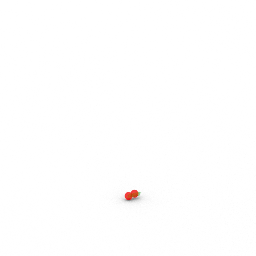} &
        \includegraphics[width=.11\linewidth]{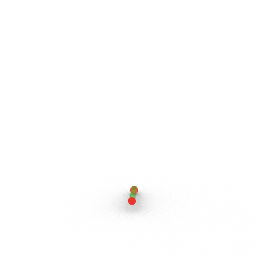} &
        \includegraphics[width=.11\linewidth]{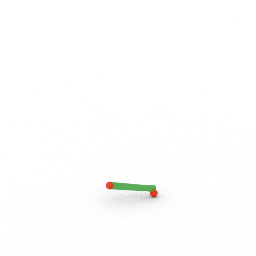} &
        \includegraphics[width=.11\linewidth]{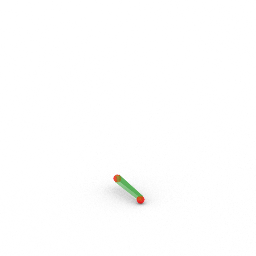} &
        \includegraphics[width=.11\linewidth]{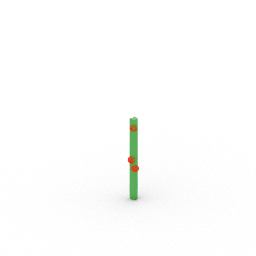}
        \\ [-0.25em]
        \includegraphics[width=.11\linewidth]{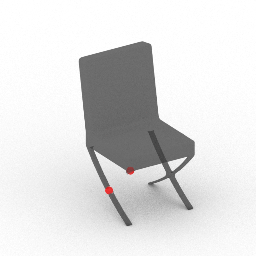} &
        \includegraphics[width=.11\linewidth]{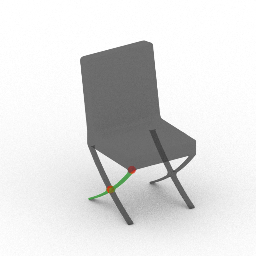} &
        \includegraphics[width=.11\linewidth]{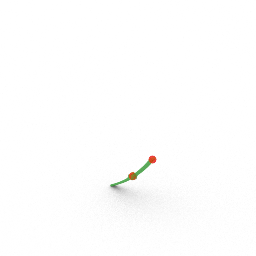} &
        \includegraphics[width=.11\linewidth]{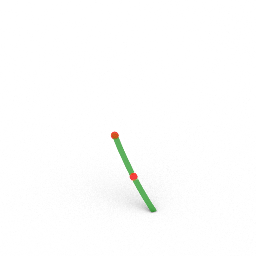} &
        \includegraphics[width=.11\linewidth]{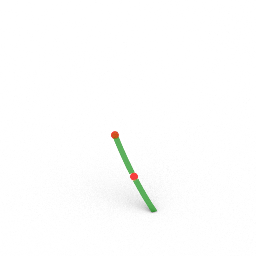} &
        \includegraphics[width=.11\linewidth]{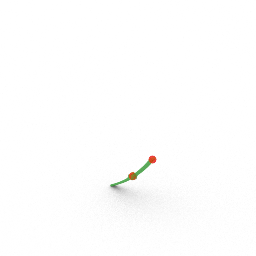} &
        \includegraphics[width=.11\linewidth]{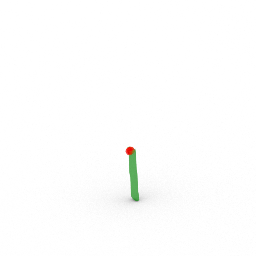} &
        \includegraphics[width=.11\linewidth]{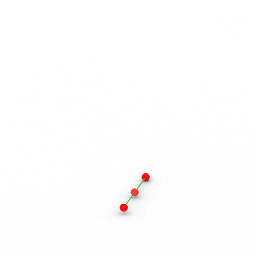}
    \end{tabular}
    \vspace{-0.75em}
    \caption{
    \revision{
    Additional outputs of the \textbf{What to Attach?} module.
    We visualize the input partial slot graph within the parts that contain them (grey) and the center of the selected slots (red), as well as the ground truth part (green, 2nd column). The parts and slots are in their ground truth world-space pose, which is \emph{not} available to the neural network.
    We then visualize, individually, the ground truth part and the retrieved candidates ranked $1$st, $2$nd, $5$th, $25$th, and at the $50$th percentile, respectively, along with all of their slots (red).
    }
    } 
    \label{fig:supp_retrieved_parts}
    \vspace{-2em}
\end{figure*}

In Figure~\ref{fig:supp_retrieved_parts}, we provide additional examples of the learned module on chairs (see also Figure 4 in the main paper). The first row shows another example of query that demands a very specific type of structure. The second row shows another example of a query that asks for chair legs. We show failure cases of our module in the last 2 rows, where it fails to reason about the exact spatial structure of shapes and retrieves parts that are oriented incorrectly.

%% file: supp/3-testtime.tex
\section{Generating New Slot Graphs at Test Time}
\label{sec:testtime}

Although trained on all shapes with less than or equal to $30$ parts, less than $5$ percent of the training shapes have more than $20$ parts, and each of those shapes have rather unique structures. 
Therefore, when generating new slot graphs, we only use parts from shapes with less than $20$ parts.
We start each shape by randomly selecting a part from the candidate parts.
We then iteratively query the three neural network modules, until the slot graph is complete (when all slots are attached). 
During this generation process, we use the output of the three neural modules to detect and reject partial slot graphs that are outliers:
\begin{packed_itemize}
\item
If the \textbf{Where} module gives a probability $p_\text{continue}$ of less than $0.5$ when there are no slots selected.
\item
Not all part cliques retrieved by the \text{What} module are good candidates. We reject a retrieved candidate $C_\text{target}$ if $|V_\text{target}| > |C_\text{target}|$, or if one of the edge predicted by the \textbf{Where} module has a probability less than $1/(\max(|V_\text{target}|+1, |C_\text{target}|)+0.5)$.
If all candidates within a margin of $60$ ($100$ for parts invovled in symmetry) from the highest scoring candidate are rejected, we reject the partial slot graph.
\end{packed_itemize}
We also reject the generated slot graph if any of the following conditions are met:
\begin{packed_itemize}
\item
The slot graph contains more than $20$ parts.
\item
The slot graph is detected as an outlier. We perform outlier detection using one-class Support Vector Machines (OCSVM). We fit one OCSVM for all graphs in the training set with the same number of parts. 
For each OCSVM that fits graphs with $N$ nodes, we use a feature size of $3(N-1)$, with the following features:
    \begin{packed_itemize}
    \item
    Number of parts with an (adjacency) degree of $1 \ldots N-1$.
    \item
    Number of parts where $1 \ldots N-1$ other parts are within a distance of $2$.
    \item
    Number of parts where the furthest part has a distance of $1 \ldots N-1$.
    \end{packed_itemize}
Other commonly used graph summary statistics, such as clustering coefficient, number of $n$-cycles, etc. are also possible candidates here, but we found the set of features we use to be sufficient for our purposes.
\end{packed_itemize}

%% file: supp/4-implementation.tex
\section{Implementation Details}
\label{sec:implementation}
We set the rounds of message passing, $T$, to $10$ for all our graph neural networks (GNN) operating on partial slot graphs. We set $T=4$ for GNNs operating on part cliques. Since no adjacency edges exists, this effectively leads to $2$ rounds of message passing.
We set the dimension of node embeddings to $64$ and the dimension of graph embeddings to $128$. 
All MLP we use have $2$ hidden layers and uses leaky ReLU as the activation function.
We use a mixture of $10$ gaussians for the MDN and a margin $m=20$ for the triplet loss.
We train all neural networks with the Adam~\cite{Adam} optimizer, and with a batch size of $32$. We select the negative examples for the \textbf{What} module from $32$ randomly selected slot graphs as well, for each training step.

%% file: supp/5-baselines.tex
\section{Details on Baselines}
\label{sec:baslines}
We provide additional details on how we implemented the baselines.

\parahead{ComplementMe} We re-implemented ComplentMe~\cite{ComplementMe} in PyTorch. We mostly used the original settings of ComplementMe, with the following exceptions:
\begin{packed_itemize}
\item
We set the maximum threshold for the standard deviation of the Gaussian Mixture model to $50$ instead of $0.05$, since we found that the standard deviation of all Gaussians saturate at the original threshold very quickly.
\item
ComplementMe sampled random triplets originally, we instead sample only the semi-hard triplets i.e. triplets that give a non-zero training loss.
\item
In the paper, ComplementMe suggests that the placement networks do not share weights with the retrieval/embedding networks. This is not the case in their official implementation. We followed the description in the paper.
\item
We removed all BatchNorm layers from the PointNet backbone since we observed that including them hurts the evaluation performance.
\end{packed_itemize}
We train ComplementMe until convergence.
\vspace{1em}

\parahead{StructureNet} We use the pre-trained models provided by StructureNet~\cite{StructureNet}, which are trained on the same split as what we use in the paper. Do note that StructureNet uses a different data filtering strategy than ours, so the training set will differ slightly.
We encode every part in the test set using the pre-trained part encoder, with each part centered and normalized in the same way as they would be if used to train StructureNet.
We then use the provided evaluation script to randomly sample outputs. Instead of decoding child-level latent code to point clouds, we directly retrieve the test set part that is the closest in the latent space, and then apply the predicted transformations to the retrived part.

%% file: supp/6-results.tex
\section{More Results}
\label{sec:results}
\begin{figure*}[t!]
    \vspace{-3em}
    \centering
    \setlength{\tabcolsep}{1pt}
    \begin{tabular}{cccccccccc}
        \multirow{3}{*}{\raisebox{-4em}{\rotatebox{90}{Ours}}} &
        \includegraphics[width=.105\linewidth]{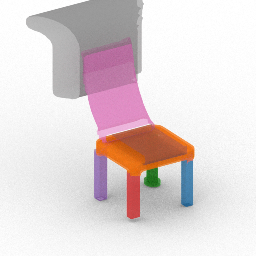} &
        \includegraphics[width=.105\linewidth]{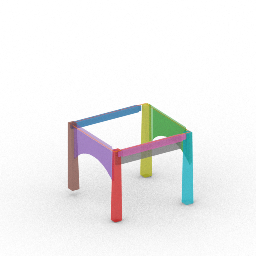} &
        \includegraphics[width=.105\linewidth]{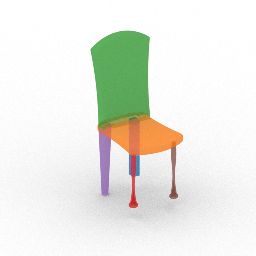} &
        \includegraphics[width=.105\linewidth]{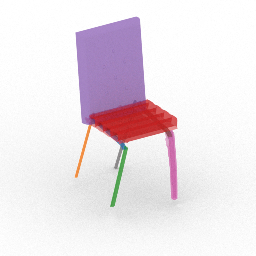} &
        \includegraphics[width=.105\linewidth]{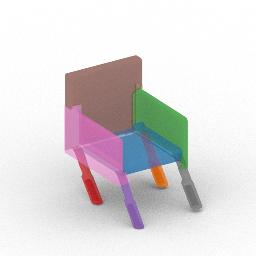} &
        \includegraphics[width=.105\linewidth]{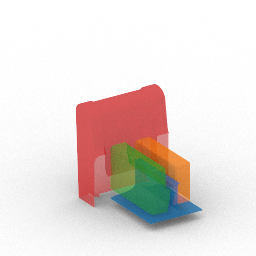} &
        \includegraphics[width=.105\linewidth]{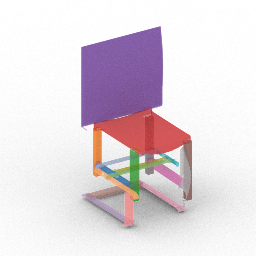} &
        \includegraphics[width=.105\linewidth]{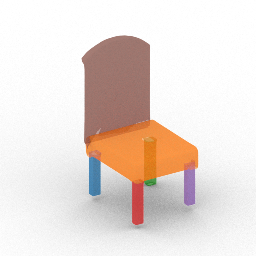} &
        \includegraphics[width=.105\linewidth]{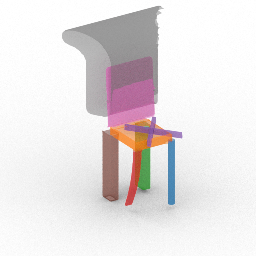}
        \\
        &
        \includegraphics[width=.105\linewidth]{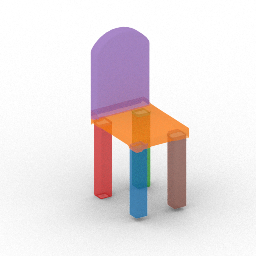} &
        \includegraphics[width=.105\linewidth]{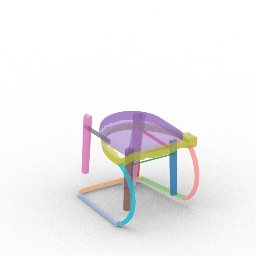} &
        \includegraphics[width=.105\linewidth]{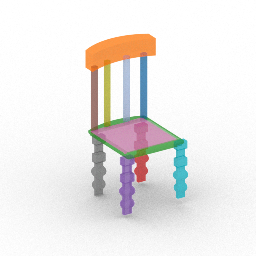} &
        \includegraphics[width=.105\linewidth]{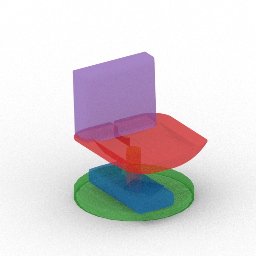} &
        \includegraphics[width=.105\linewidth]{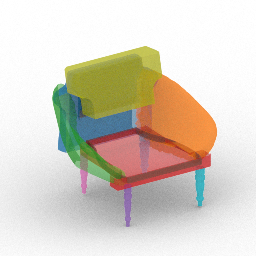} &
        \includegraphics[width=.105\linewidth]{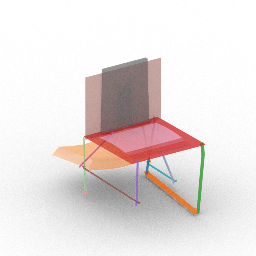} &
        \includegraphics[width=.105\linewidth]{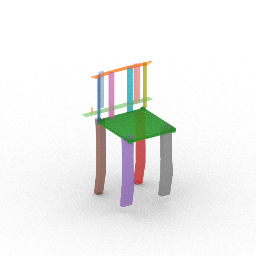} &
        \includegraphics[width=.105\linewidth]{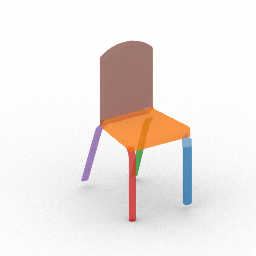} &
        \includegraphics[width=.105\linewidth]{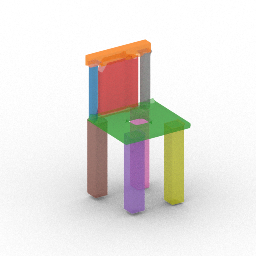}
        \\
        &
        \includegraphics[width=.105\linewidth]{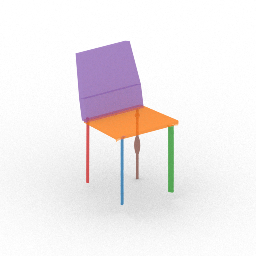} &
        \includegraphics[width=.105\linewidth]{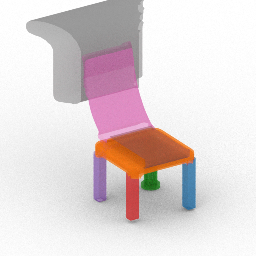} &
        \includegraphics[width=.105\linewidth]{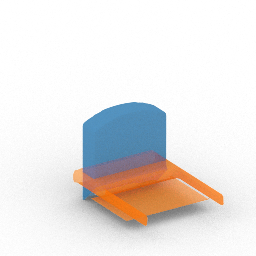} &
        \includegraphics[width=.105\linewidth]{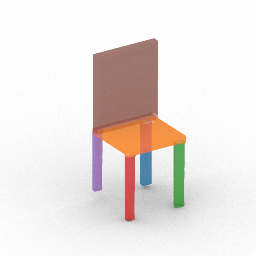} &
        \includegraphics[width=.105\linewidth]{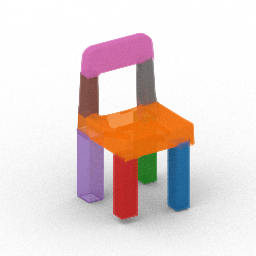} &
        \includegraphics[width=.105\linewidth]{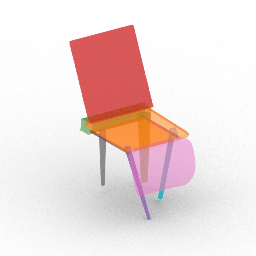} &
        \includegraphics[width=.105\linewidth]{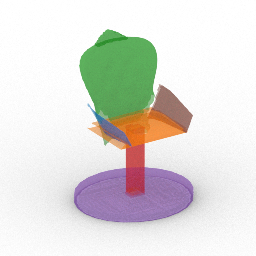} &
        \includegraphics[width=.105\linewidth]{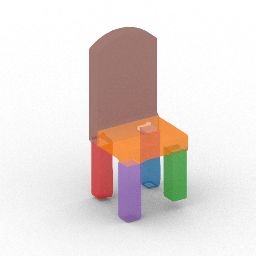} &
        \includegraphics[width=.105\linewidth]{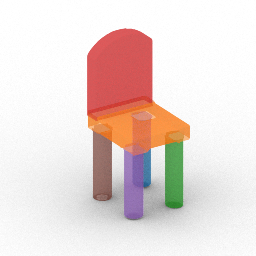}
        \\ \midrule
        \multirow{3}{*}{\raisebox{-2em}{\rotatebox{90}{ComplementMe(w/sym)}}} &
        \includegraphics[width=.105\linewidth]{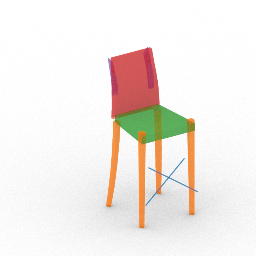} &
        \includegraphics[width=.105\linewidth]{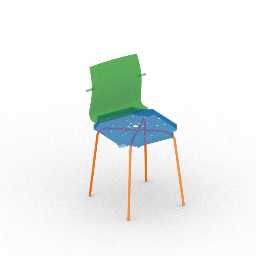} &
        \includegraphics[width=.105\linewidth]{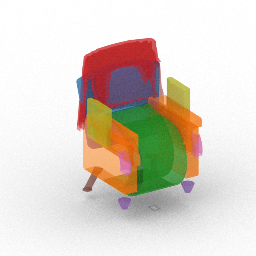} &
        \includegraphics[width=.105\linewidth]{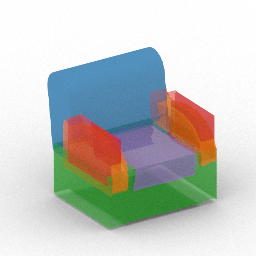} &
        \includegraphics[width=.105\linewidth]{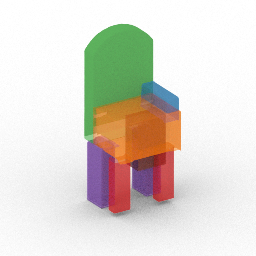} &
        \includegraphics[width=.105\linewidth]{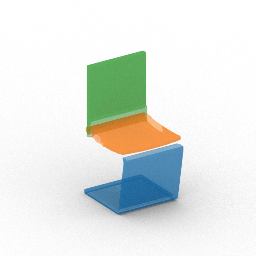} &
        \includegraphics[width=.105\linewidth]{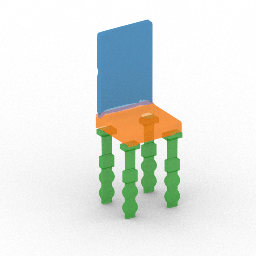} &
        \includegraphics[width=.105\linewidth]{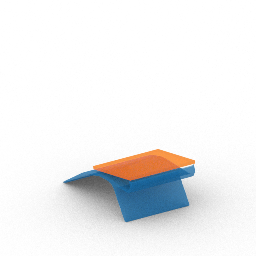} &
        \includegraphics[width=.105\linewidth]{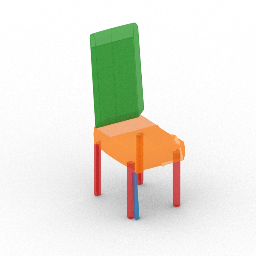}
        \\
        &
        \includegraphics[width=.105\linewidth]{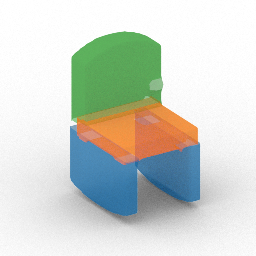} &
        \includegraphics[width=.105\linewidth]{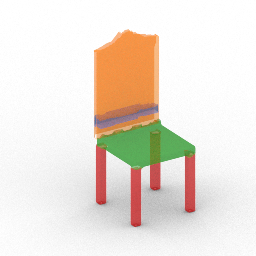} &
        \includegraphics[width=.105\linewidth]{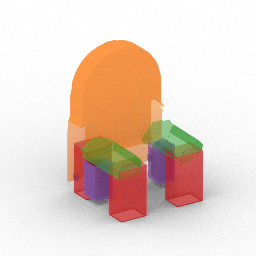} &
        \includegraphics[width=.105\linewidth]{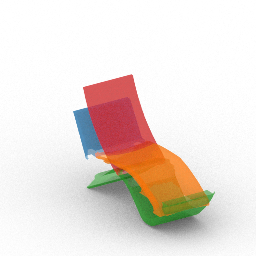} &
        \includegraphics[width=.105\linewidth]{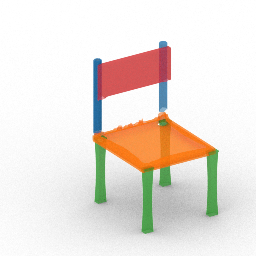} &
        \includegraphics[width=.105\linewidth]{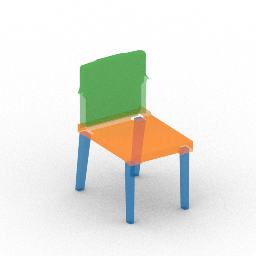} &
        \includegraphics[width=.105\linewidth]{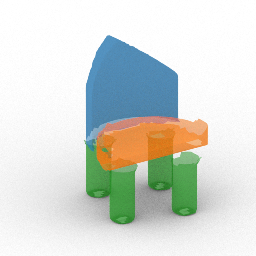} &
        \includegraphics[width=.105\linewidth]{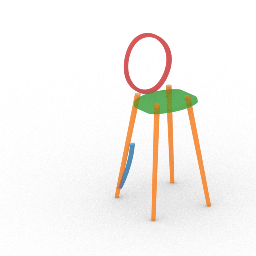} &
        \includegraphics[width=.105\linewidth]{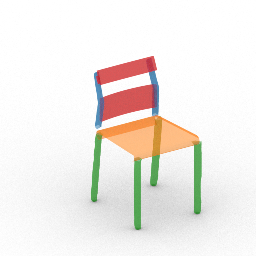}
        \\
        &
        \includegraphics[width=.105\linewidth]{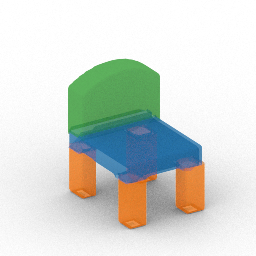} &
        \includegraphics[width=.105\linewidth]{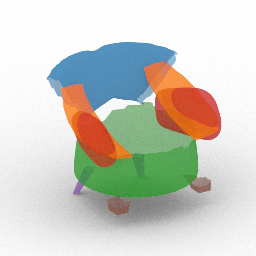} &
        \includegraphics[width=.105\linewidth]{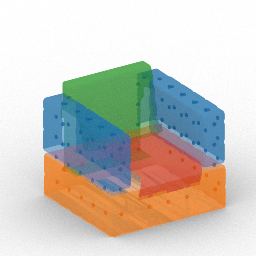} &
        \includegraphics[width=.105\linewidth]{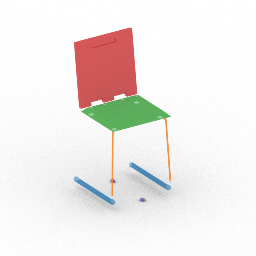} &
        \includegraphics[width=.105\linewidth]{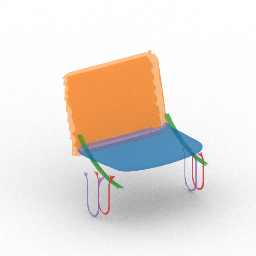} &
        \includegraphics[width=.105\linewidth]{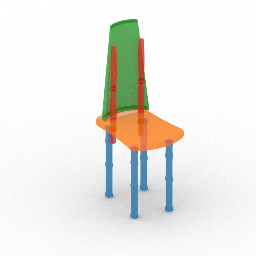} &
        \includegraphics[width=.105\linewidth]{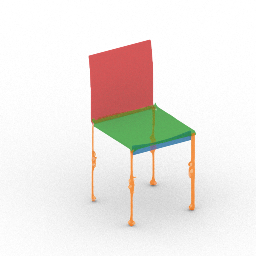} &
        \includegraphics[width=.105\linewidth]{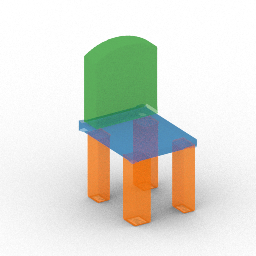} &
        \includegraphics[width=.105\linewidth]{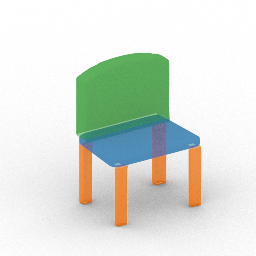}
        \\ \midrule
        \multirow{3}{*}{\raisebox{1em}{\rotatebox{90}{ComplementMe}}} &
        \includegraphics[width=.105\linewidth]{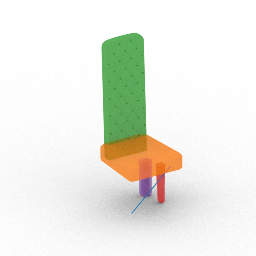} &
        \includegraphics[width=.105\linewidth]{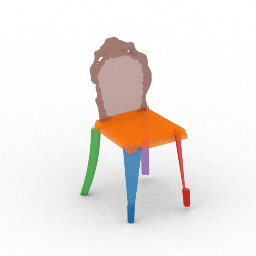} &
        \includegraphics[width=.105\linewidth]{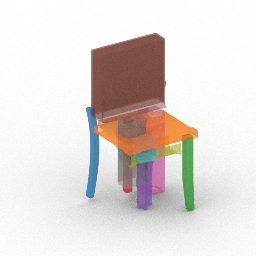} &
        \includegraphics[width=.105\linewidth]{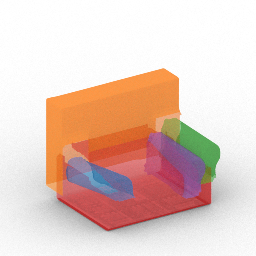} &
        \includegraphics[width=.105\linewidth]{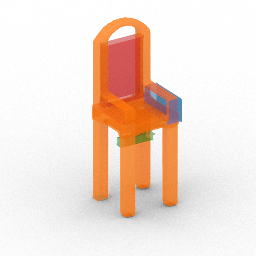} &
        \includegraphics[width=.105\linewidth]{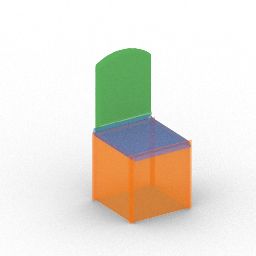} &
        \includegraphics[width=.105\linewidth]{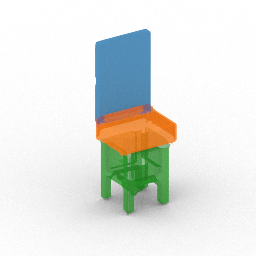} &
        \includegraphics[width=.105\linewidth]{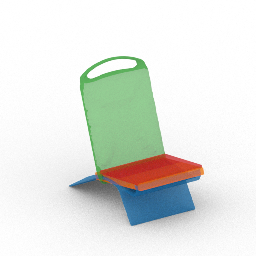} &
        \includegraphics[width=.105\linewidth]{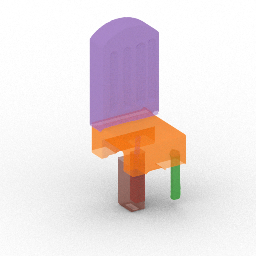}
        \\
        &
        \includegraphics[width=.105\linewidth]{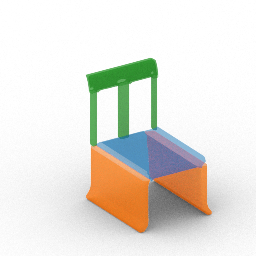} &
        \includegraphics[width=.105\linewidth]{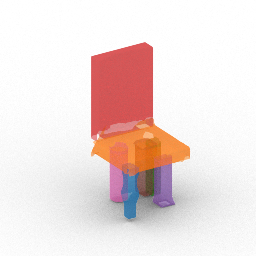} &
        \includegraphics[width=.105\linewidth]{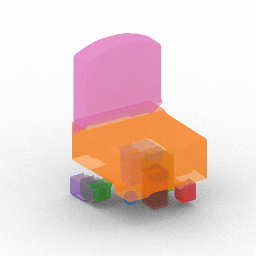} &
        \includegraphics[width=.105\linewidth]{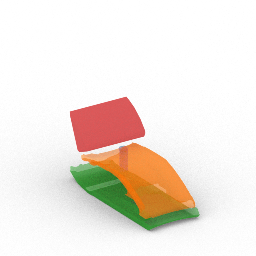} &
        \includegraphics[width=.105\linewidth]{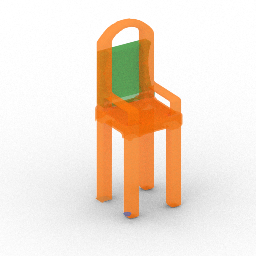} &
        \includegraphics[width=.105\linewidth]{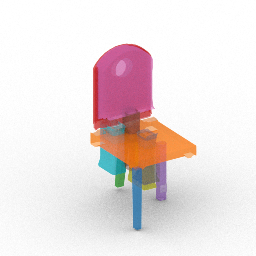} &
        \includegraphics[width=.105\linewidth]{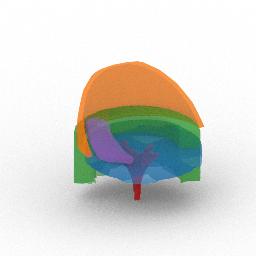} &
        \includegraphics[width=.105\linewidth]{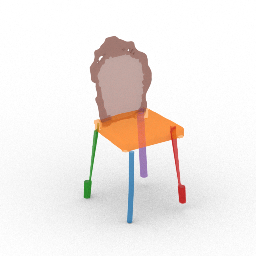} &
        \includegraphics[width=.105\linewidth]{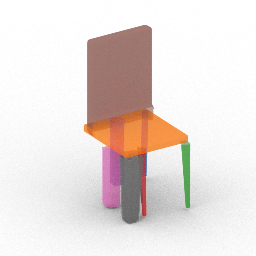}
        \\
        &
        \includegraphics[width=.105\linewidth]{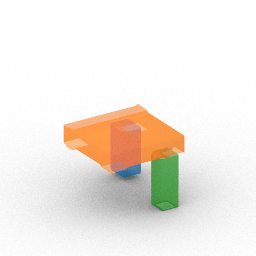} &
        \includegraphics[width=.105\linewidth]{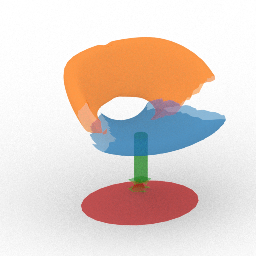} &
        \includegraphics[width=.105\linewidth]{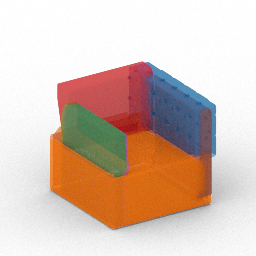} &
        \includegraphics[width=.105\linewidth]{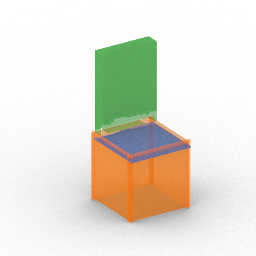} &
        \includegraphics[width=.105\linewidth]{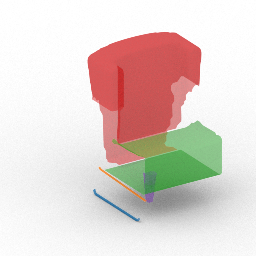} &
        \includegraphics[width=.105\linewidth]{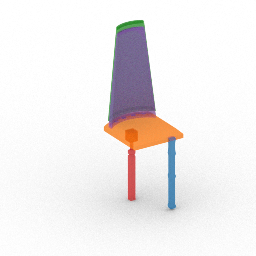} &
        \includegraphics[width=.105\linewidth]{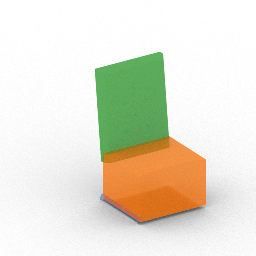} &
        \includegraphics[width=.105\linewidth]{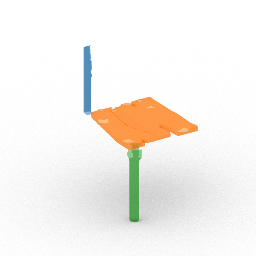} &
        \includegraphics[width=.105\linewidth]{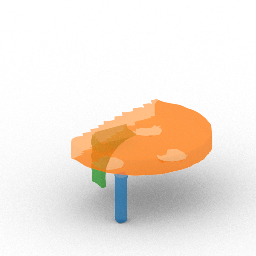}
        \\ \midrule
        \multirow{3}{*}{\raisebox{-4em}{\rotatebox{90}{StructureNet}}} &
        \includegraphics[width=.105\linewidth]{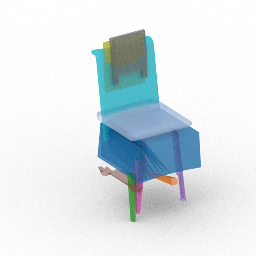} &
        \includegraphics[width=.105\linewidth]{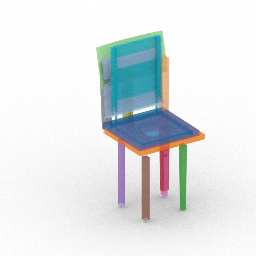} &
        \includegraphics[width=.105\linewidth]{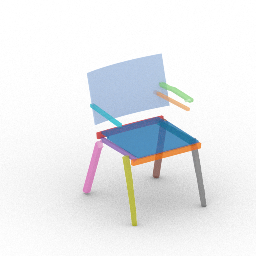} &
        \includegraphics[width=.105\linewidth]{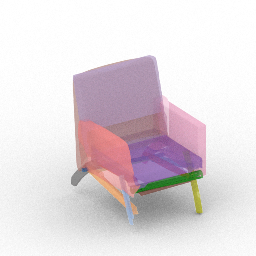} &
        \includegraphics[width=.105\linewidth]{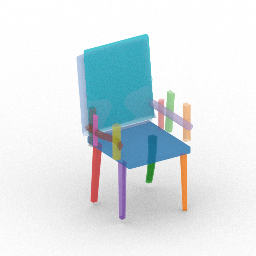} &
        \includegraphics[width=.105\linewidth]{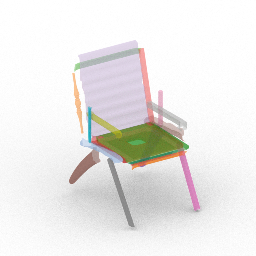} &
        \includegraphics[width=.105\linewidth]{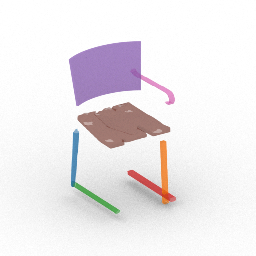} &
        \includegraphics[width=.105\linewidth]{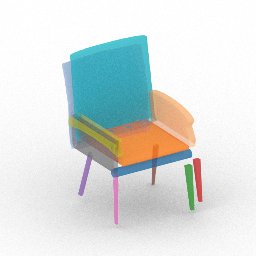} &
        \includegraphics[width=.105\linewidth]{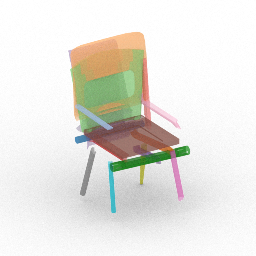}
        \\
        &
        \includegraphics[width=.105\linewidth]{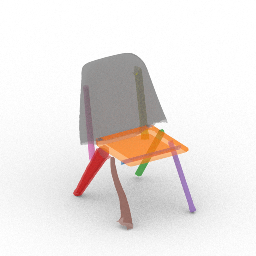} &
        \includegraphics[width=.105\linewidth]{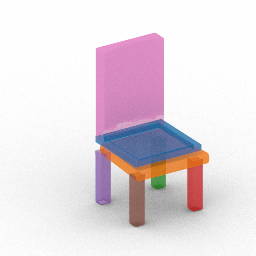} &
        \includegraphics[width=.105\linewidth]{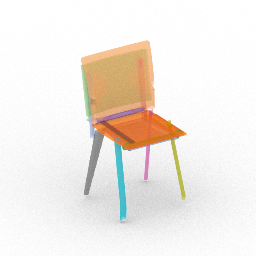} &
        \includegraphics[width=.105\linewidth]{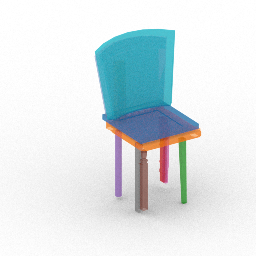} &
        \includegraphics[width=.105\linewidth]{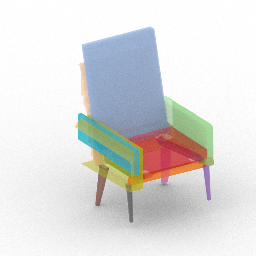} &
        \includegraphics[width=.105\linewidth]{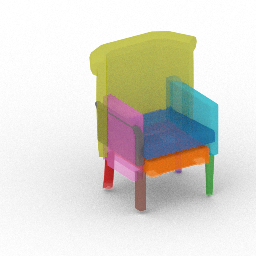} &
        \includegraphics[width=.105\linewidth]{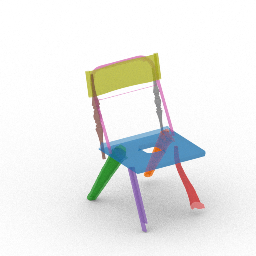} &
        \includegraphics[width=.105\linewidth]{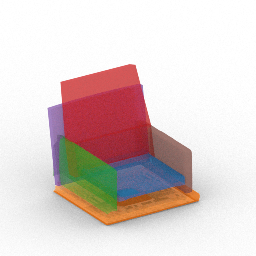} &
        \includegraphics[width=.105\linewidth]{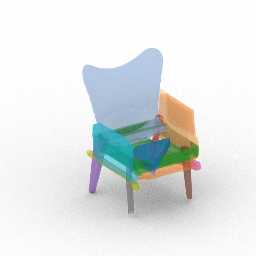}
        \\
        &
        \includegraphics[width=.105\linewidth]{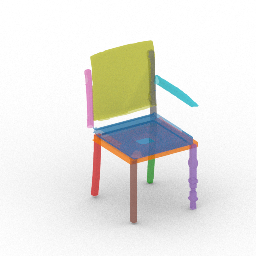} &
        \includegraphics[width=.105\linewidth]{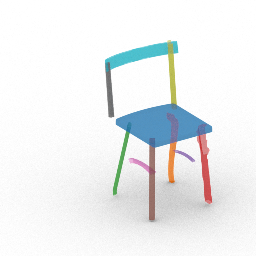} &
        \includegraphics[width=.105\linewidth]{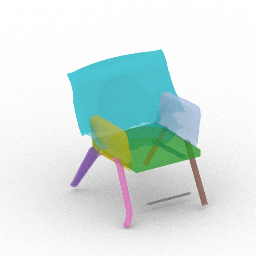} &
        \includegraphics[width=.105\linewidth]{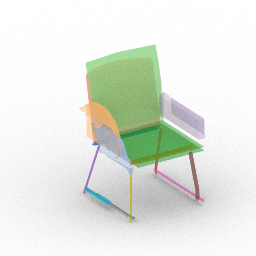} &
        \includegraphics[width=.105\linewidth]{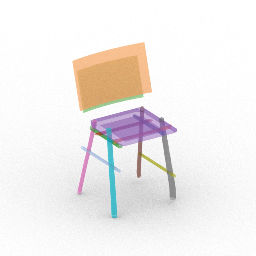} &
        \includegraphics[width=.105\linewidth]{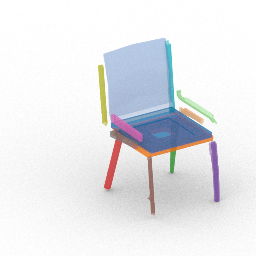} &
        \includegraphics[width=.105\linewidth]{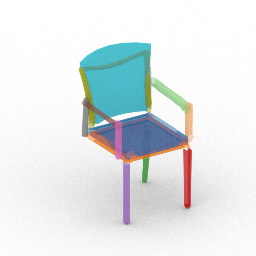} &
        \includegraphics[width=.105\linewidth]{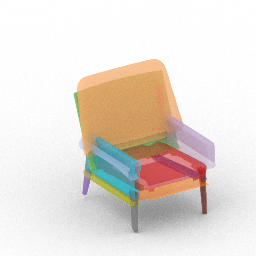} &
        \includegraphics[width=.105\linewidth]{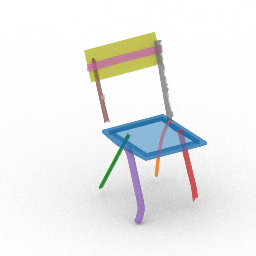}
        \\
    \end{tabular}
    \vspace{-1.5em}
    \caption{Chair Unconditional Samples}
    \label{fig:chair}
\end{figure*}

\begin{figure*}[t!]
    \vspace{-3em}
    \centering
    \setlength{\tabcolsep}{1pt}
    \begin{tabular}{cccccccccc}
        \multirow{3}{*}{\raisebox{-4em}{\rotatebox{90}{Ours}}} &
        \includegraphics[width=.105\linewidth]{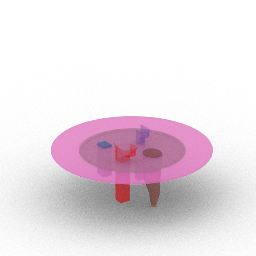} &
        \includegraphics[width=.105\linewidth]{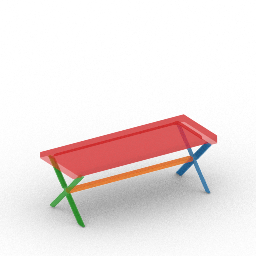} &
        \includegraphics[width=.105\linewidth]{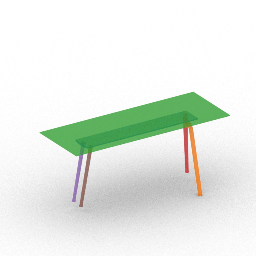} &
        \includegraphics[width=.105\linewidth]{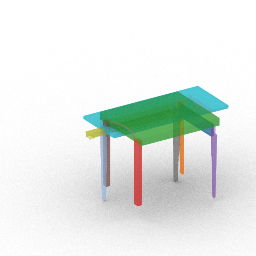} &
        \includegraphics[width=.105\linewidth]{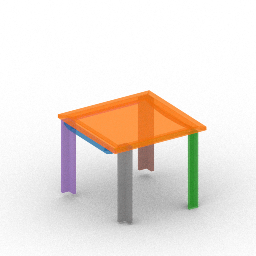} &
        \includegraphics[width=.105\linewidth]{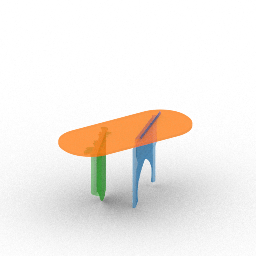} &
        \includegraphics[width=.105\linewidth]{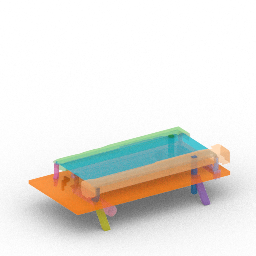} &
        \includegraphics[width=.105\linewidth]{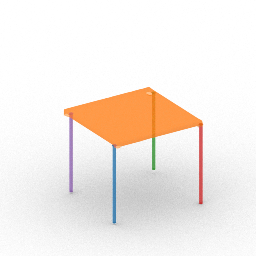} &
        \includegraphics[width=.105\linewidth]{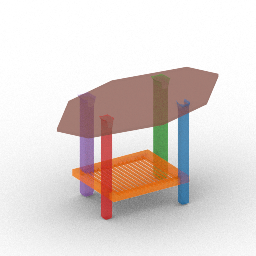}
        \\
        &
        \includegraphics[width=.105\linewidth]{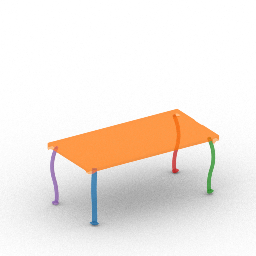} &
        \includegraphics[width=.105\linewidth]{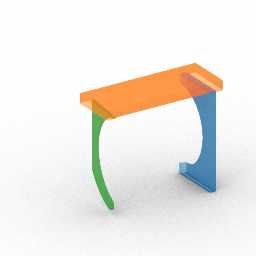} &
        \includegraphics[width=.105\linewidth]{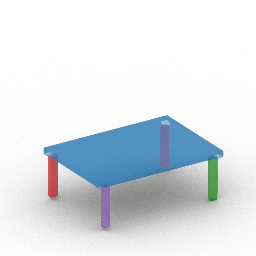} &
        \includegraphics[width=.105\linewidth]{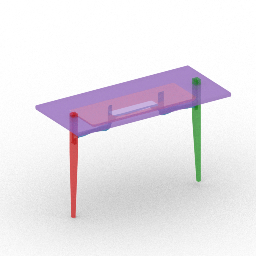} &
        \includegraphics[width=.105\linewidth]{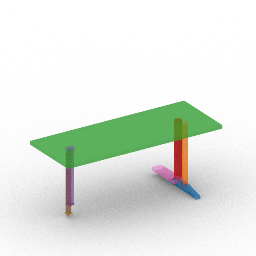} &
        \includegraphics[width=.105\linewidth]{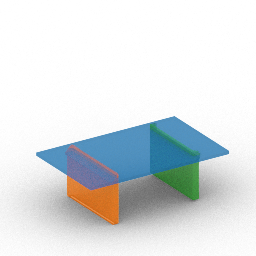} &
        \includegraphics[width=.105\linewidth]{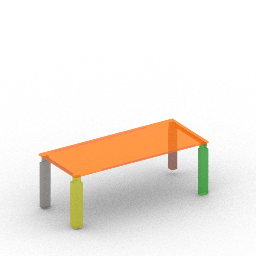} &
        \includegraphics[width=.105\linewidth]{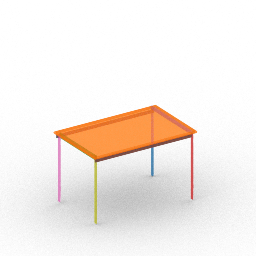} &
        \includegraphics[width=.105\linewidth]{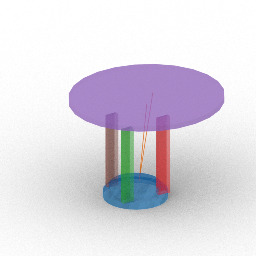}
        \\
        &
        \includegraphics[width=.105\linewidth]{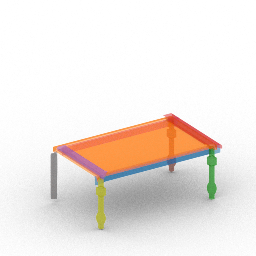} &
        \includegraphics[width=.105\linewidth]{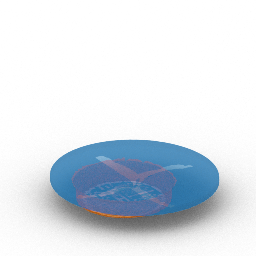} &
        \includegraphics[width=.105\linewidth]{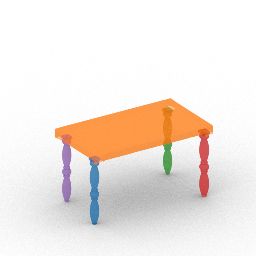} &
        \includegraphics[width=.105\linewidth]{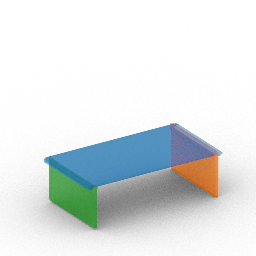} &
        \includegraphics[width=.105\linewidth]{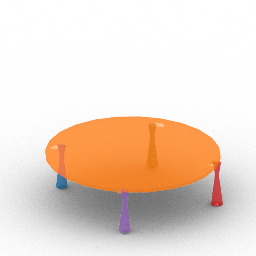} &
        \includegraphics[width=.105\linewidth]{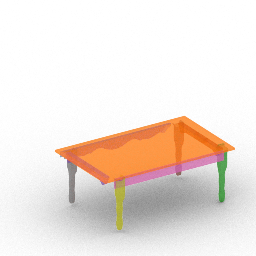} &
        \includegraphics[width=.105\linewidth]{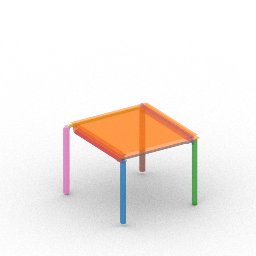} &
        \includegraphics[width=.105\linewidth]{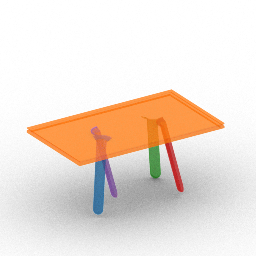} &
        \includegraphics[width=.105\linewidth]{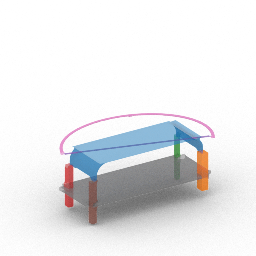}
        \\ \midrule
        \multirow{3}{*}{\raisebox{8em}{\rotatebox{90}{ComplementMe (w/sym)}}} &
        \includegraphics[width=.105\linewidth]{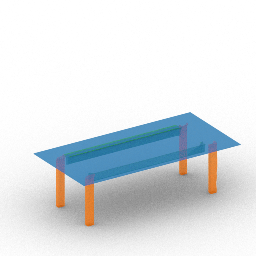} &
        \includegraphics[width=.105\linewidth]{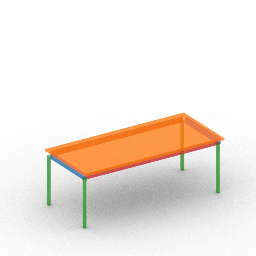} &
        \includegraphics[width=.105\linewidth]{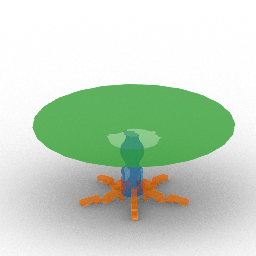} &
        \includegraphics[width=.105\linewidth]{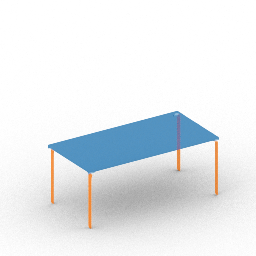} &
        \includegraphics[width=.105\linewidth]{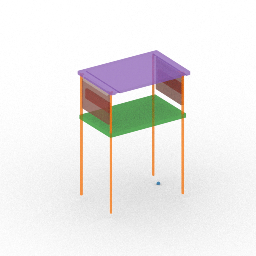} &
        \includegraphics[width=.105\linewidth]{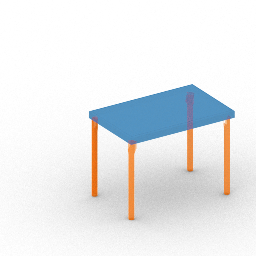} &
        \includegraphics[width=.105\linewidth]{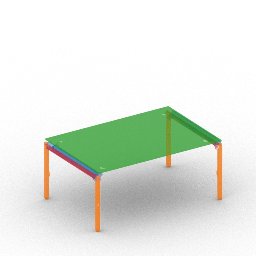} &
        \includegraphics[width=.105\linewidth]{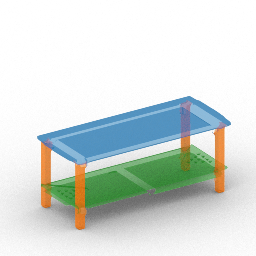} &
        \includegraphics[width=.105\linewidth]{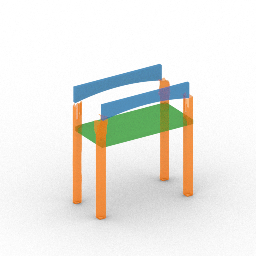}
        \\
        &
        \includegraphics[width=.105\linewidth]{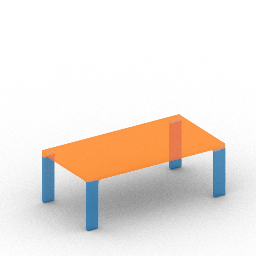} &
        \includegraphics[width=.105\linewidth]{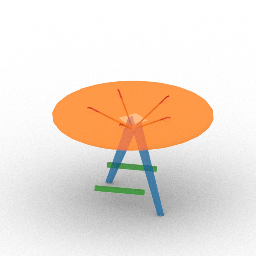} &
        \includegraphics[width=.105\linewidth]{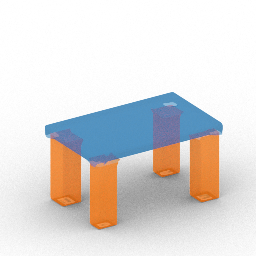} &
        \includegraphics[width=.105\linewidth]{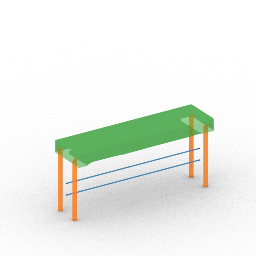} &
        \includegraphics[width=.105\linewidth]{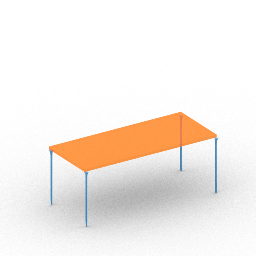} &
        \includegraphics[width=.105\linewidth]{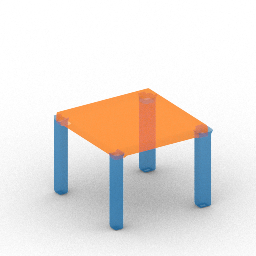} &
        \includegraphics[width=.105\linewidth]{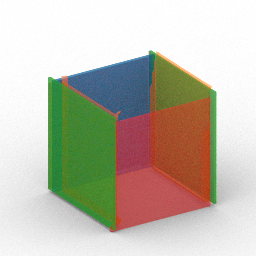} &
        \includegraphics[width=.105\linewidth]{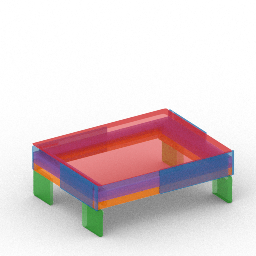} &
        \includegraphics[width=.105\linewidth]{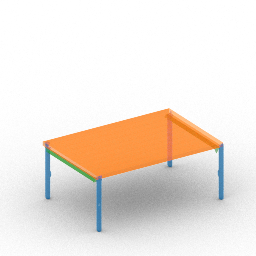}
        \\
        &
        \includegraphics[width=.105\linewidth]{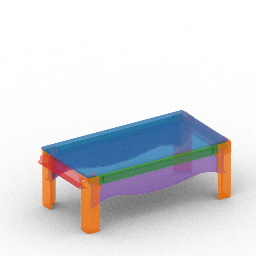} &
        \includegraphics[width=.105\linewidth]{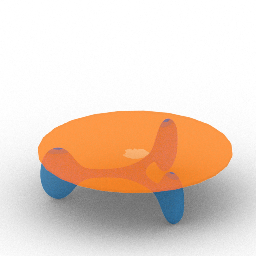} &
        \includegraphics[width=.105\linewidth]{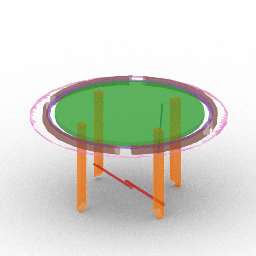} &
        \includegraphics[width=.105\linewidth]{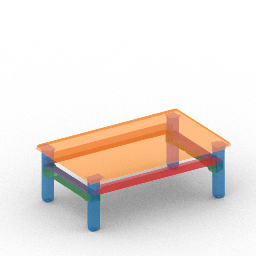} &
        \includegraphics[width=.105\linewidth]{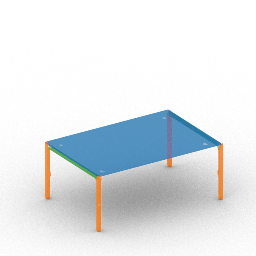} &
        \includegraphics[width=.105\linewidth]{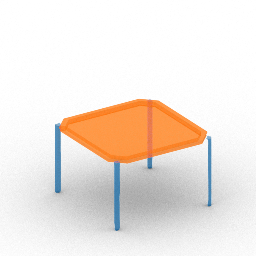} &
        \includegraphics[width=.105\linewidth]{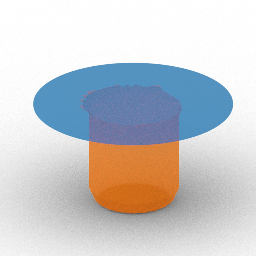} &
        \includegraphics[width=.105\linewidth]{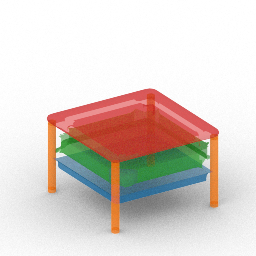} &
        \includegraphics[width=.105\linewidth]{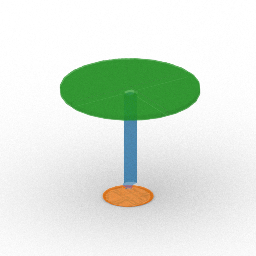}
        \\ \midrule
        \multirow{3}{*}{\raisebox{10em}{\rotatebox{90}{ComplementMe}}} &
        \includegraphics[width=.105\linewidth]{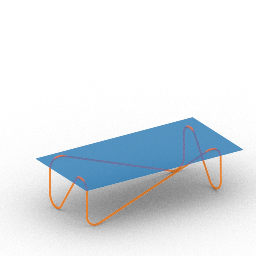} &
        \includegraphics[width=.105\linewidth]{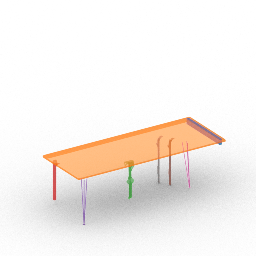} &
        \includegraphics[width=.105\linewidth]{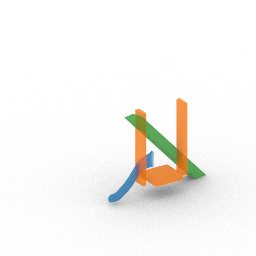} &
        \includegraphics[width=.105\linewidth]{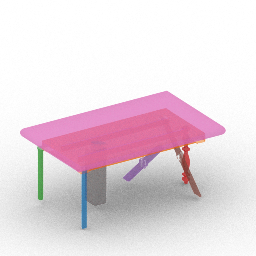} &
        \includegraphics[width=.105\linewidth]{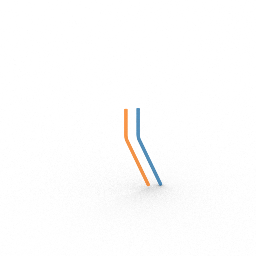} &
        \includegraphics[width=.105\linewidth]{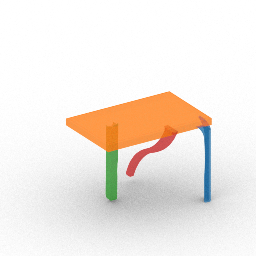} &
        \includegraphics[width=.105\linewidth]{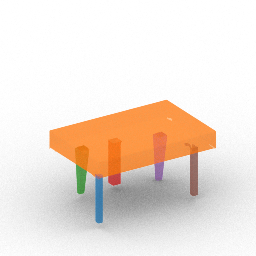} &
        \includegraphics[width=.105\linewidth]{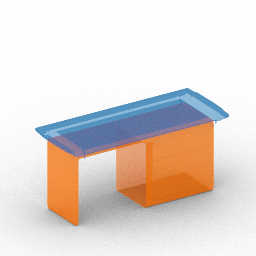} &
        \includegraphics[width=.105\linewidth]{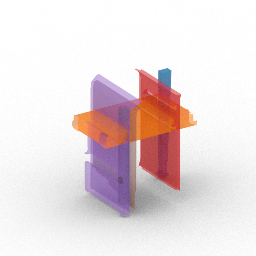}
        \\
        &
        \includegraphics[width=.105\linewidth]{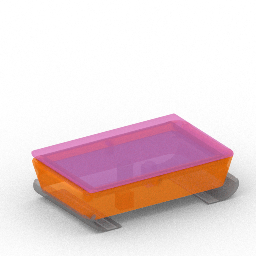} &
        \includegraphics[width=.105\linewidth]{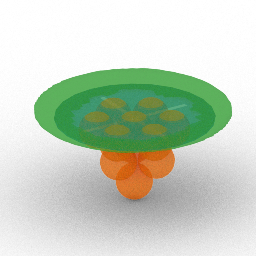} &
        \includegraphics[width=.105\linewidth]{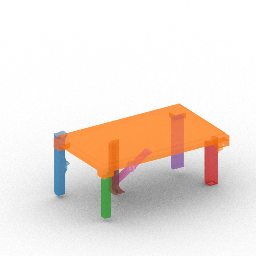} &
        \includegraphics[width=.105\linewidth]{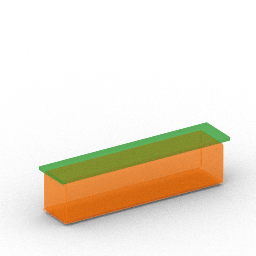} &
        \includegraphics[width=.105\linewidth]{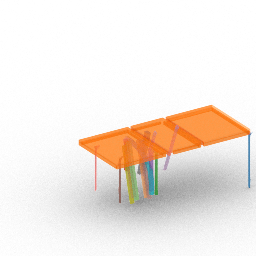} &
        \includegraphics[width=.105\linewidth]{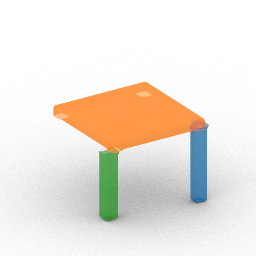} &
        \includegraphics[width=.105\linewidth]{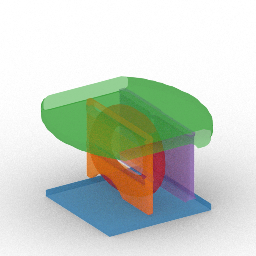} &
        \includegraphics[width=.105\linewidth]{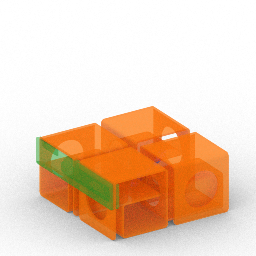} &
        \includegraphics[width=.105\linewidth]{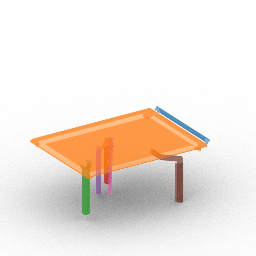}
        \\
        &
        \includegraphics[width=.105\linewidth]{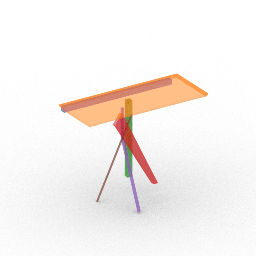} &
        \includegraphics[width=.105\linewidth]{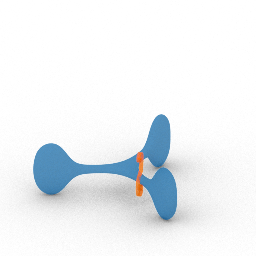} &
        \includegraphics[width=.105\linewidth]{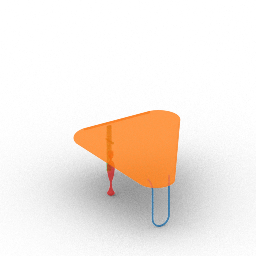} &
        \includegraphics[width=.105\linewidth]{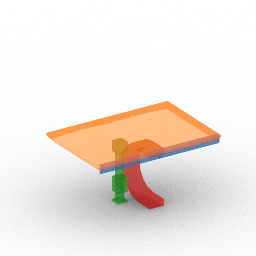} &
        \includegraphics[width=.105\linewidth]{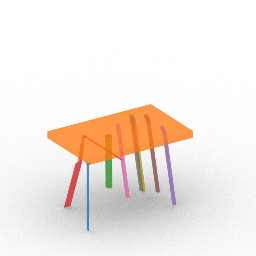} &
        \includegraphics[width=.105\linewidth]{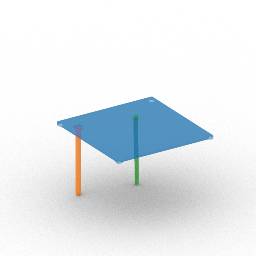} &
        \includegraphics[width=.105\linewidth]{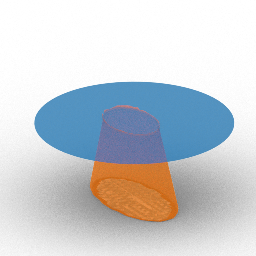} &
        \includegraphics[width=.105\linewidth]{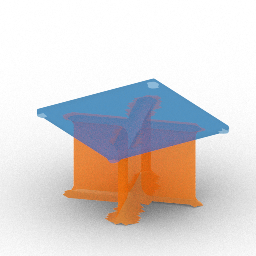} &
        \includegraphics[width=.105\linewidth]{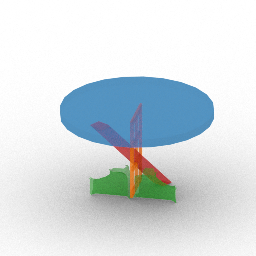}
        \\ \midrule
        \multirow{3}{*}{\raisebox{8em}{\rotatebox{90}{StructureNet}}} &
        \includegraphics[width=.105\linewidth]{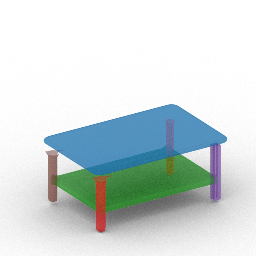} &
        \includegraphics[width=.105\linewidth]{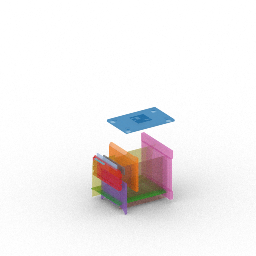} &
        \includegraphics[width=.105\linewidth]{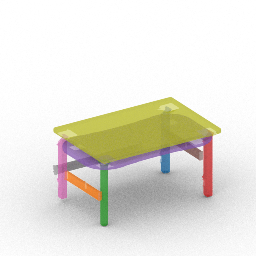} &
        \includegraphics[width=.105\linewidth]{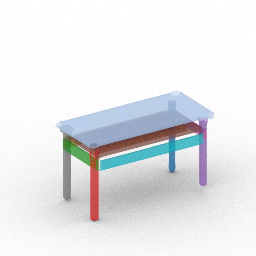} &
        \includegraphics[width=.105\linewidth]{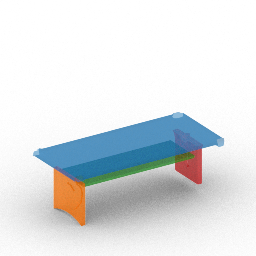} &
        \includegraphics[width=.105\linewidth]{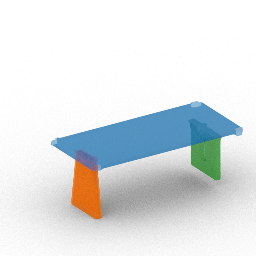} &
        \includegraphics[width=.105\linewidth]{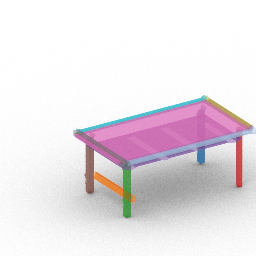} &
        \includegraphics[width=.105\linewidth]{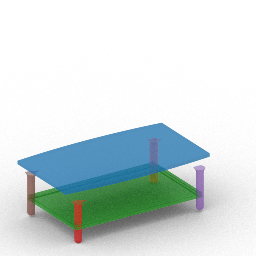} &
        \includegraphics[width=.105\linewidth]{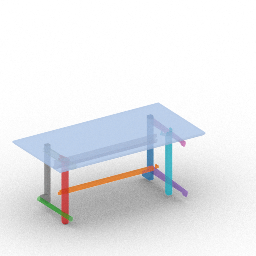}
        \\
        &
        \includegraphics[width=.105\linewidth]{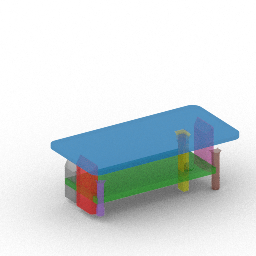} &
        \includegraphics[width=.105\linewidth]{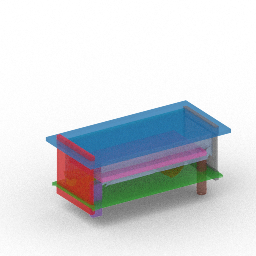} &
        \includegraphics[width=.105\linewidth]{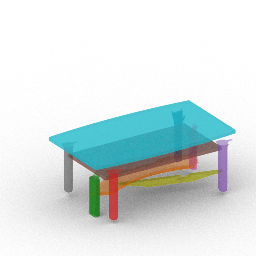} &
        \includegraphics[width=.105\linewidth]{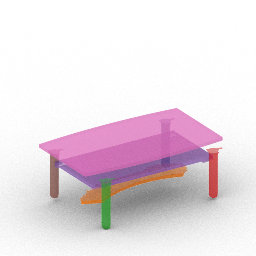} &
        \includegraphics[width=.105\linewidth]{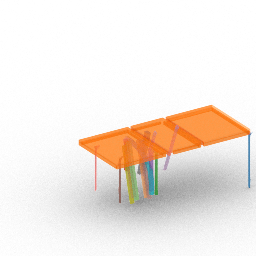} &
        \includegraphics[width=.105\linewidth]{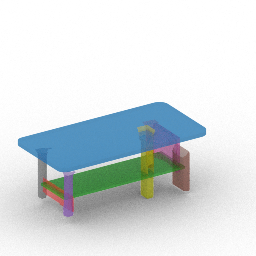} &
        \includegraphics[width=.105\linewidth]{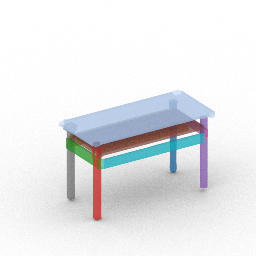} &
        \includegraphics[width=.105\linewidth]{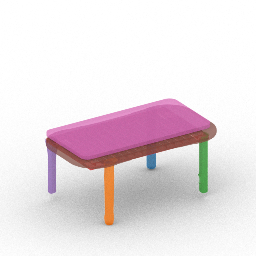} &
        \includegraphics[width=.105\linewidth]{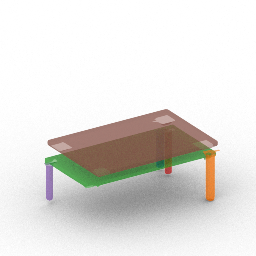}
        \\
        &
        \includegraphics[width=.105\linewidth]{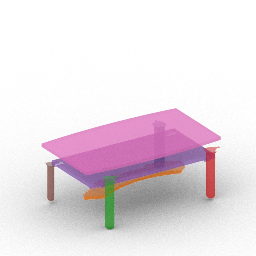} &
        \includegraphics[width=.105\linewidth]{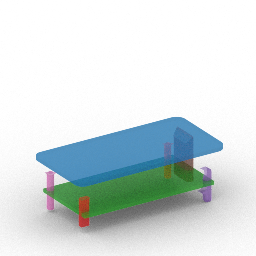} &
        \includegraphics[width=.105\linewidth]{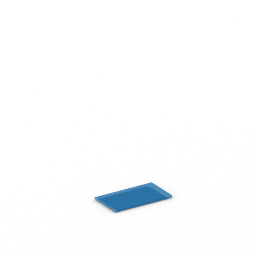} &
        \includegraphics[width=.105\linewidth]{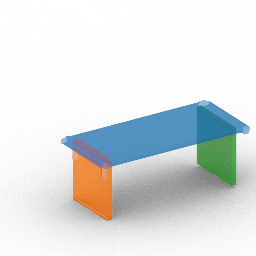} &
        \includegraphics[width=.105\linewidth]{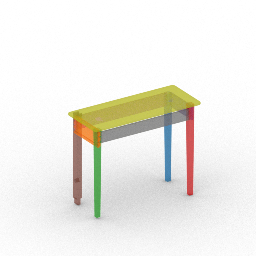} &
        \includegraphics[width=.105\linewidth]{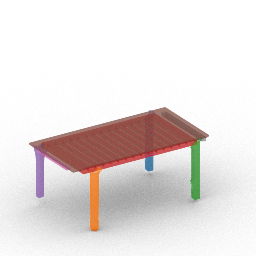} &
        \includegraphics[width=.105\linewidth]{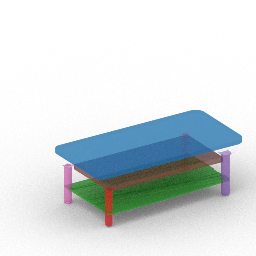} &
        \includegraphics[width=.105\linewidth]{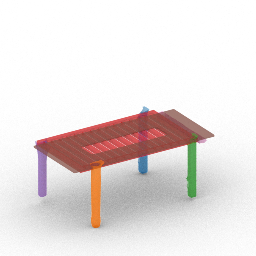} &
        \includegraphics[width=.105\linewidth]{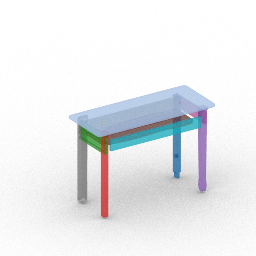}
        \\
    \end{tabular}
    \vspace{-1.5em}
    \caption{Table Unconditional Samples}
    \label{fig:table}
\end{figure*}

\begin{figure*}[t!]
    \centering
    \setlength{\tabcolsep}{1pt}
    \begin{tabular}{cccccccccc}
        \multirow{3}{*}{\raisebox{-4em}{\rotatebox{90}{Ours}}} &
        \includegraphics[width=.105\linewidth]{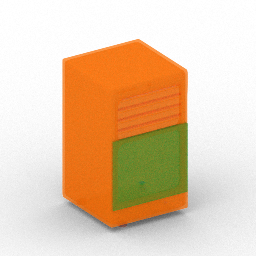} &
        \includegraphics[width=.105\linewidth]{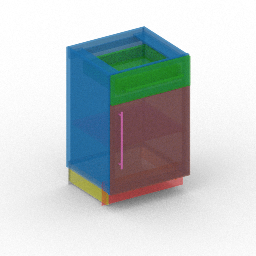} &
        \includegraphics[width=.105\linewidth]{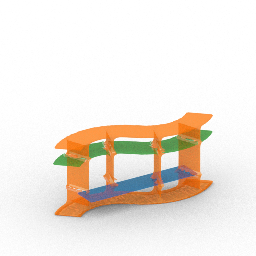} &
        \includegraphics[width=.105\linewidth]{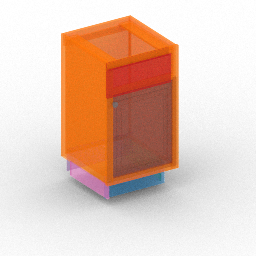} &
        \includegraphics[width=.105\linewidth]{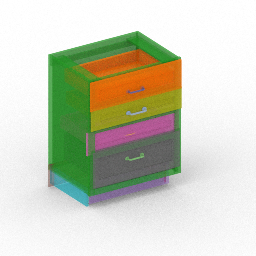} &
        \includegraphics[width=.105\linewidth]{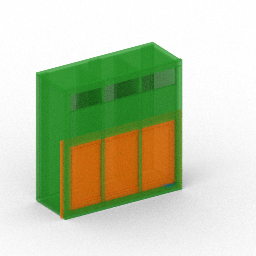} &
        \includegraphics[width=.105\linewidth]{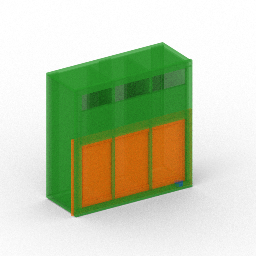} &
        \includegraphics[width=.105\linewidth]{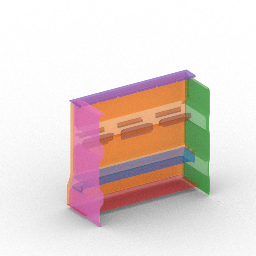} &
        \includegraphics[width=.105\linewidth]{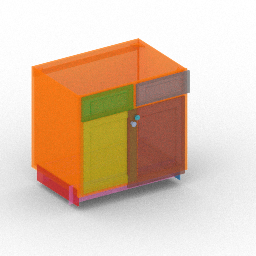}
        \\
        &
        \includegraphics[width=.105\linewidth]{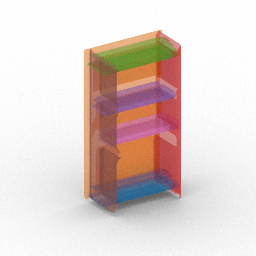} &
        \includegraphics[width=.105\linewidth]{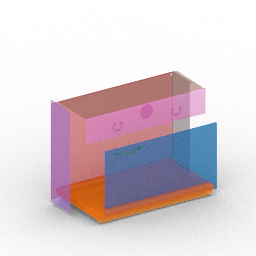} &
        \includegraphics[width=.105\linewidth]{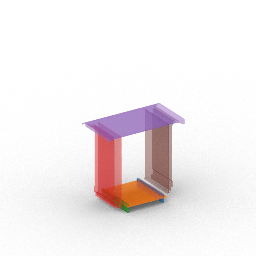} &
        \includegraphics[width=.105\linewidth]{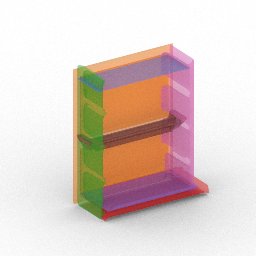} &
        \includegraphics[width=.105\linewidth]{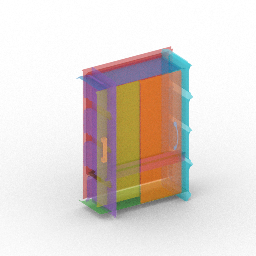} &
        \includegraphics[width=.105\linewidth]{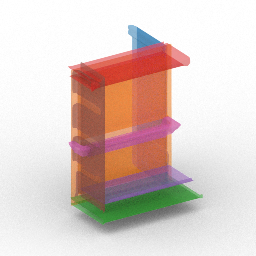} &
        \includegraphics[width=.105\linewidth]{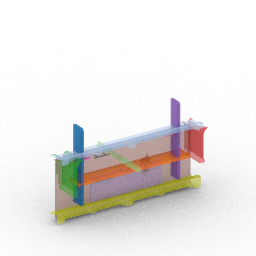} &
        \includegraphics[width=.105\linewidth]{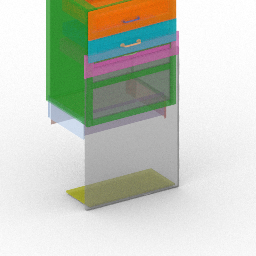} &
        \includegraphics[width=.105\linewidth]{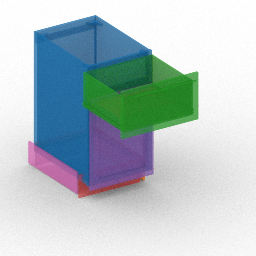}
        \\
        &
        \includegraphics[width=.105\linewidth]{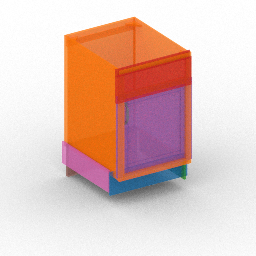} &
        \includegraphics[width=.105\linewidth]{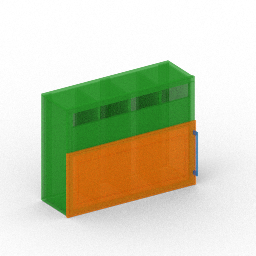} &
        \includegraphics[width=.105\linewidth]{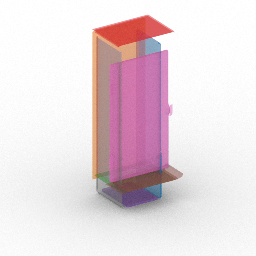} &
        \includegraphics[width=.105\linewidth]{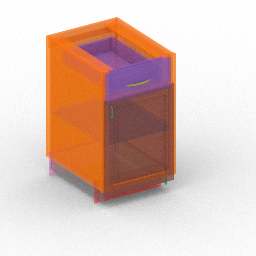} &
        \includegraphics[width=.105\linewidth]{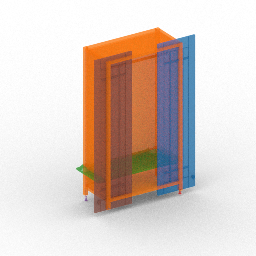} &
        \includegraphics[width=.105\linewidth]{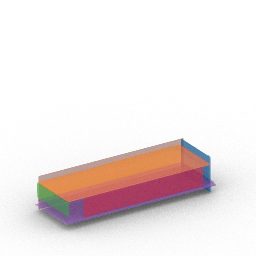} &
        \includegraphics[width=.105\linewidth]{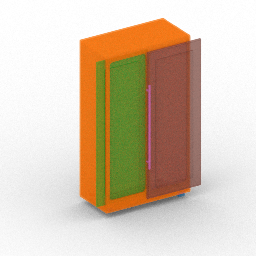} &
        \includegraphics[width=.105\linewidth]{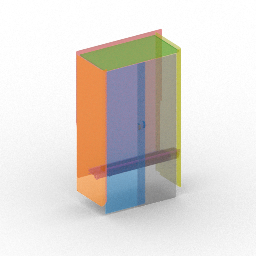} &
        \includegraphics[width=.105\linewidth]{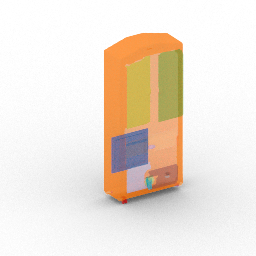}
        \\ \midrule
        \multirow{3}{*}{\raisebox{8em}{\rotatebox{90}{ComplementMe (w/sym)}}} &
        \includegraphics[width=.105\linewidth]{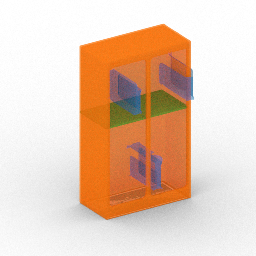} &
        \includegraphics[width=.105\linewidth]{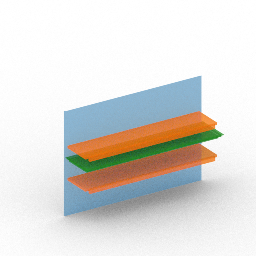} &
        \includegraphics[width=.105\linewidth]{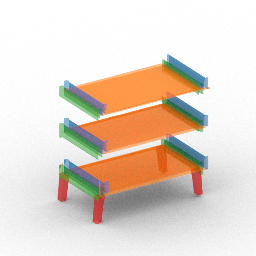} &
        \includegraphics[width=.105\linewidth]{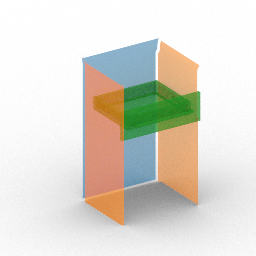} &
        \includegraphics[width=.105\linewidth]{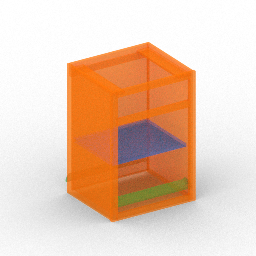} &
        \includegraphics[width=.105\linewidth]{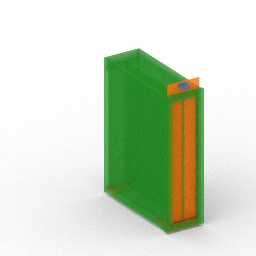} &
        \includegraphics[width=.105\linewidth]{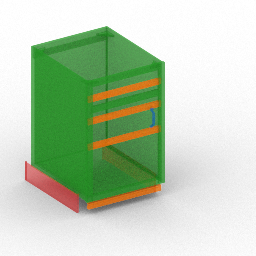} &
        \includegraphics[width=.105\linewidth]{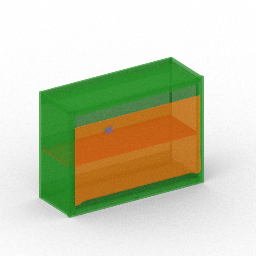} &
        \includegraphics[width=.105\linewidth]{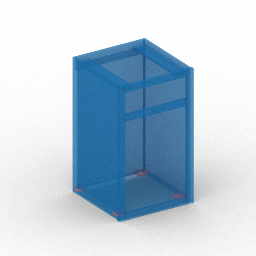}
        \\
        &
        \includegraphics[width=.105\linewidth]{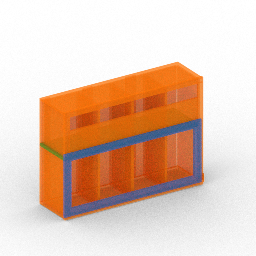} &
        \includegraphics[width=.105\linewidth]{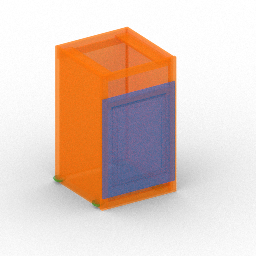} &
        \includegraphics[width=.105\linewidth]{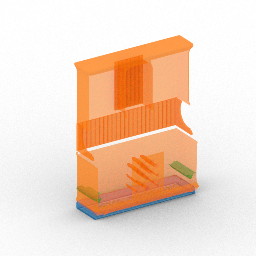} &
        \includegraphics[width=.105\linewidth]{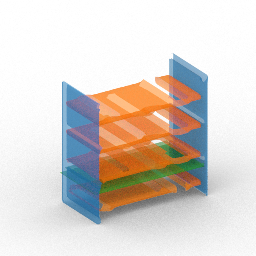} &
        \includegraphics[width=.105\linewidth]{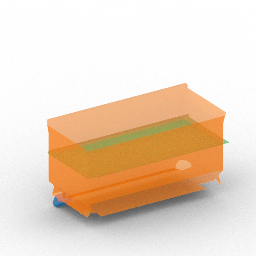} &
        \includegraphics[width=.105\linewidth]{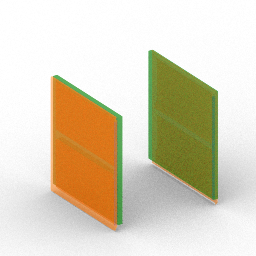} &
        \includegraphics[width=.105\linewidth]{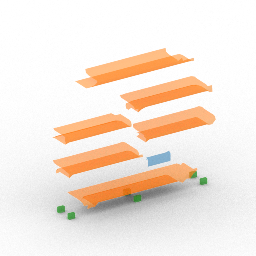} &
        \includegraphics[width=.105\linewidth]{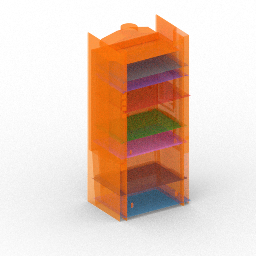} &
        \includegraphics[width=.105\linewidth]{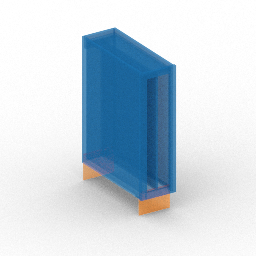}
        \\
        &
        \includegraphics[width=.105\linewidth]{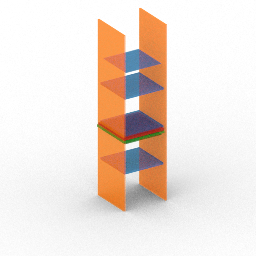} &
        \includegraphics[width=.105\linewidth]{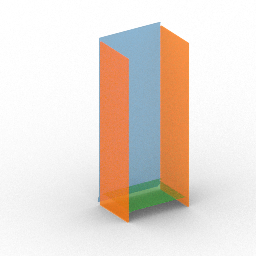} &
        \includegraphics[width=.105\linewidth]{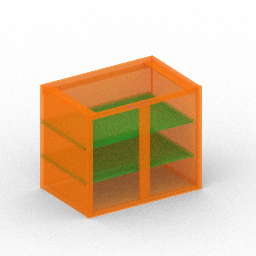} &
        \includegraphics[width=.105\linewidth]{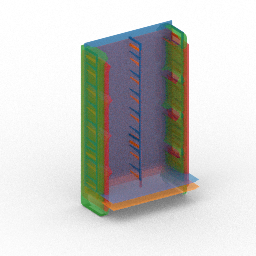} &
        \includegraphics[width=.105\linewidth]{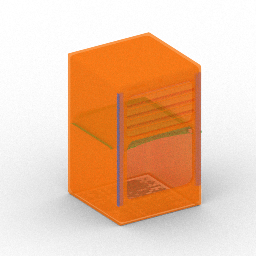} &
        \includegraphics[width=.105\linewidth]{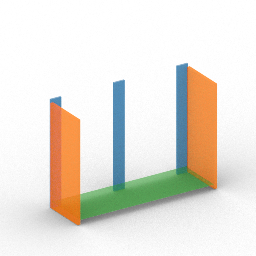} &
        \includegraphics[width=.105\linewidth]{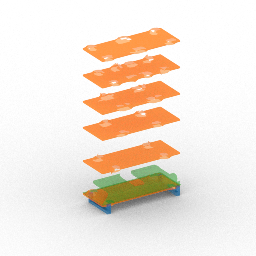} &
        \includegraphics[width=.105\linewidth]{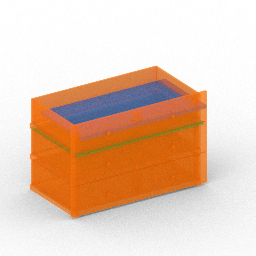} &
        \includegraphics[width=.105\linewidth]{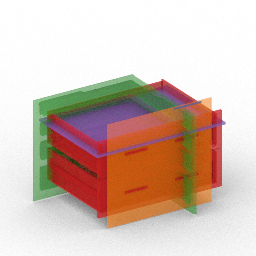}
        \\ \midrule
        \multirow{3}{*}{\raisebox{10em}{\rotatebox{90}{ComplementMe}}} &
        \includegraphics[width=.105\linewidth]{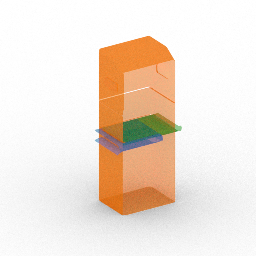} &
        \includegraphics[width=.105\linewidth]{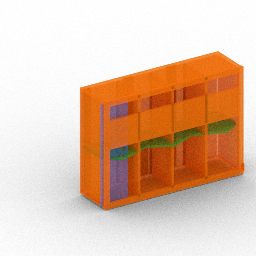} &
        \includegraphics[width=.105\linewidth]{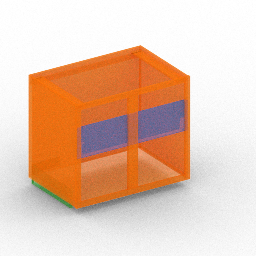} &
        \includegraphics[width=.105\linewidth]{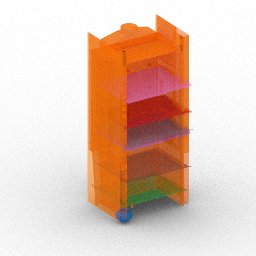} &
        \includegraphics[width=.105\linewidth]{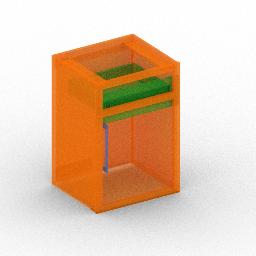} &
        \includegraphics[width=.105\linewidth]{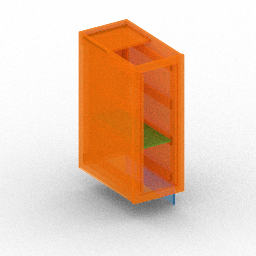} &
        \includegraphics[width=.105\linewidth]{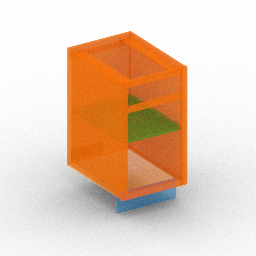} &
        \includegraphics[width=.105\linewidth]{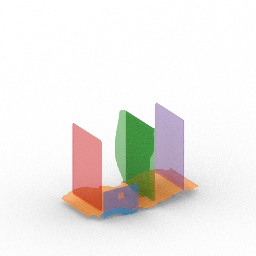} &
        \includegraphics[width=.105\linewidth]{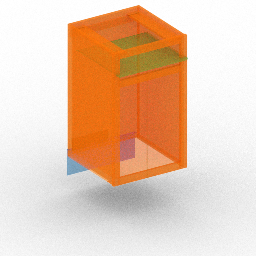}
        \\
        &
        \includegraphics[width=.105\linewidth]{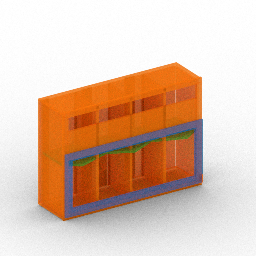} &
        \includegraphics[width=.105\linewidth]{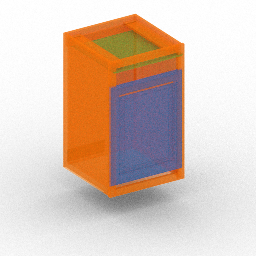} &
        \includegraphics[width=.105\linewidth]{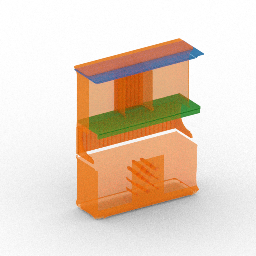} &
        \includegraphics[width=.105\linewidth]{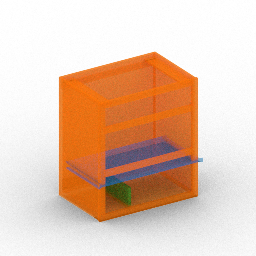} &
        \includegraphics[width=.105\linewidth]{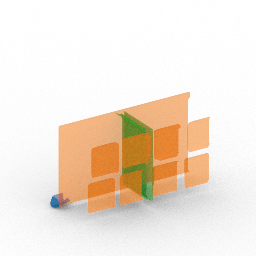} &
        \includegraphics[width=.105\linewidth]{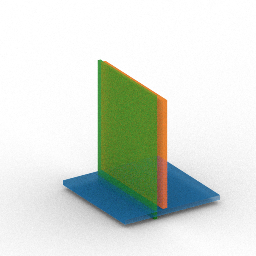} &
        \includegraphics[width=.105\linewidth]{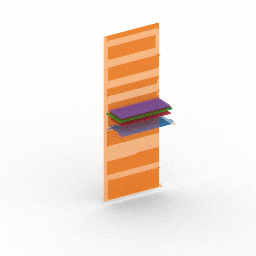} &
        \includegraphics[width=.105\linewidth]{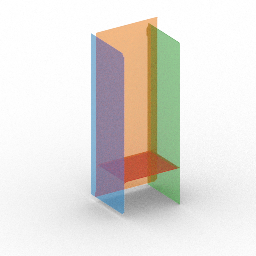} &
        \includegraphics[width=.105\linewidth]{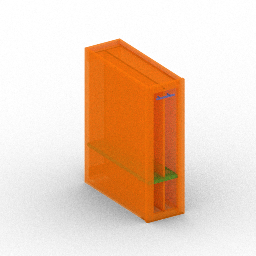}
        \\
        &
        \includegraphics[width=.105\linewidth]{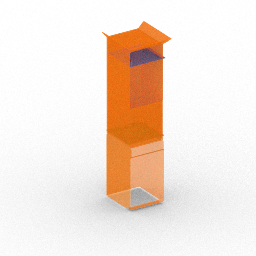} &
        \includegraphics[width=.105\linewidth]{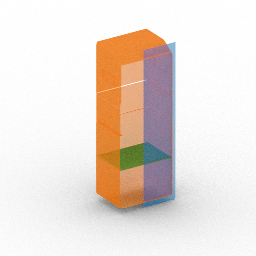} &
        \includegraphics[width=.105\linewidth]{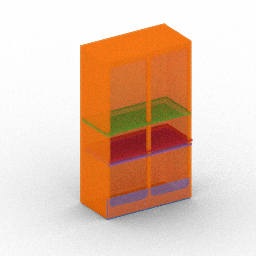} &
        \includegraphics[width=.105\linewidth]{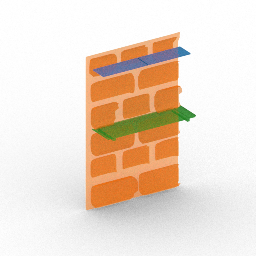} &
        \includegraphics[width=.105\linewidth]{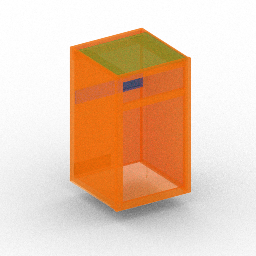} &
        \includegraphics[width=.105\linewidth]{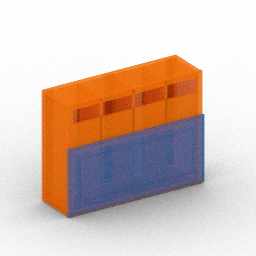} &
        \includegraphics[width=.105\linewidth]{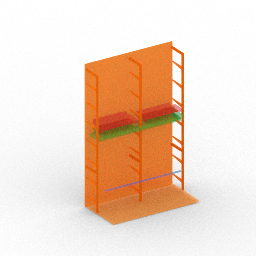} &
        \includegraphics[width=.105\linewidth]{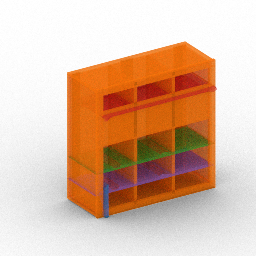} &
        \includegraphics[width=.105\linewidth]{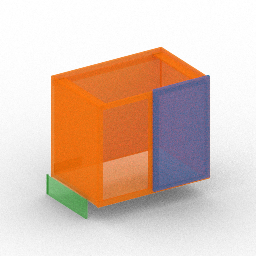}
        \\ \midrule
        \multirow{3}{*}{\raisebox{8em}{\rotatebox{90}{StructureNet}}} &
        \includegraphics[width=.105\linewidth]{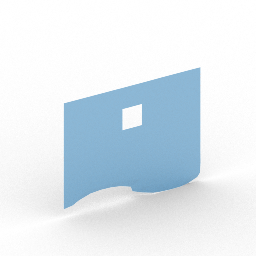} &
        \includegraphics[width=.105\linewidth]{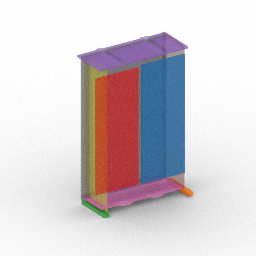} &
        \includegraphics[width=.105\linewidth]{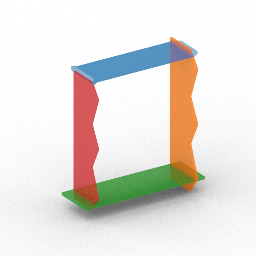} &
        \includegraphics[width=.105\linewidth]{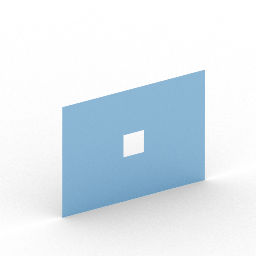} &
        \includegraphics[width=.105\linewidth]{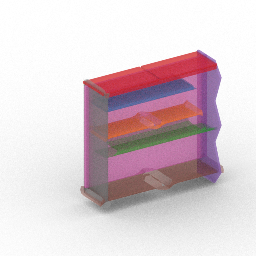} &
        \includegraphics[width=.105\linewidth]{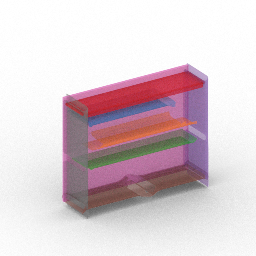} &
        \includegraphics[width=.105\linewidth]{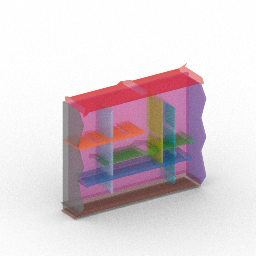} &
        \includegraphics[width=.105\linewidth]{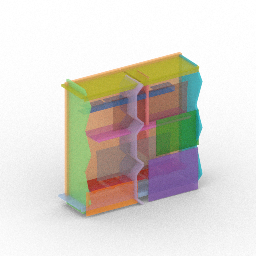} &
        \includegraphics[width=.105\linewidth]{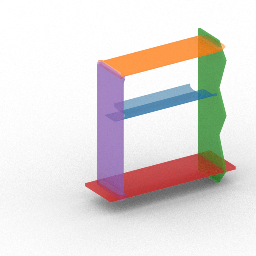}
        \\
        &
        \includegraphics[width=.105\linewidth]{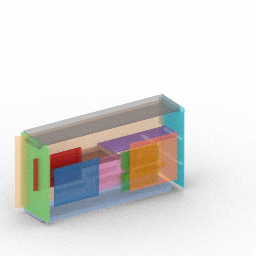} &
        \includegraphics[width=.105\linewidth]{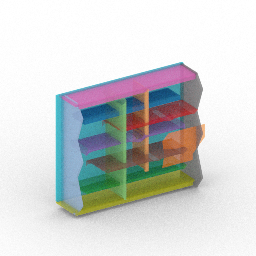} &
        \includegraphics[width=.105\linewidth]{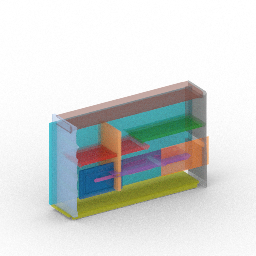} &
        \includegraphics[width=.105\linewidth]{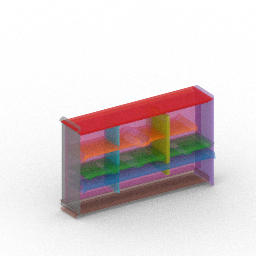} &
        \includegraphics[width=.105\linewidth]{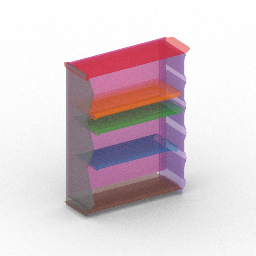} &
        \includegraphics[width=.105\linewidth]{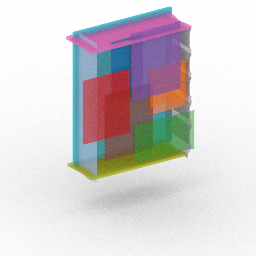} &
        \includegraphics[width=.105\linewidth]{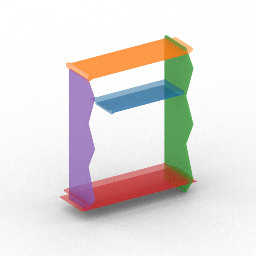} &
        \includegraphics[width=.105\linewidth]{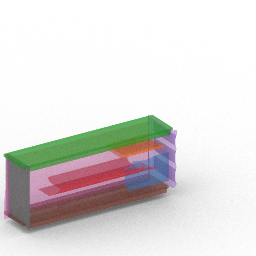} &
        \includegraphics[width=.105\linewidth]{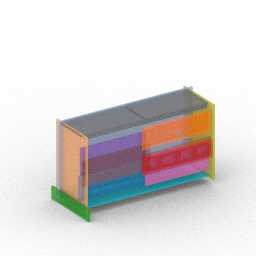}
        \\
        &
        \includegraphics[width=.105\linewidth]{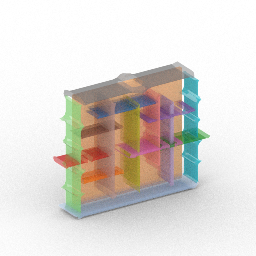} &
        \includegraphics[width=.105\linewidth]{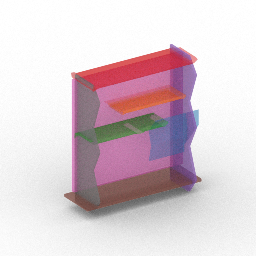} &
        \includegraphics[width=.105\linewidth]{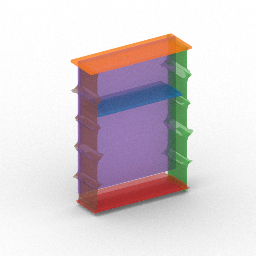} &
        \includegraphics[width=.105\linewidth]{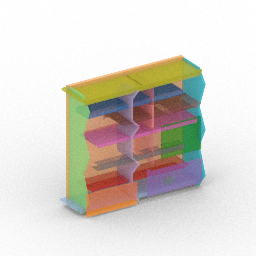} &
        \includegraphics[width=.105\linewidth]{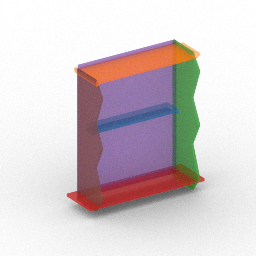} &
        \includegraphics[width=.105\linewidth]{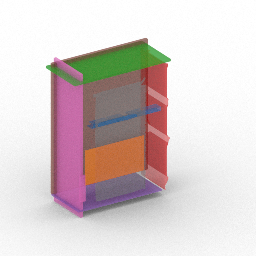} &
        \includegraphics[width=.105\linewidth]{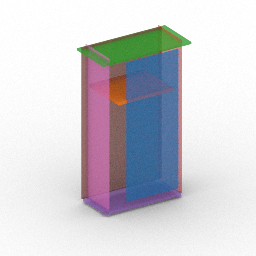} &
        \includegraphics[width=.105\linewidth]{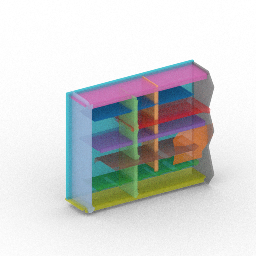} &
        \includegraphics[width=.105\linewidth]{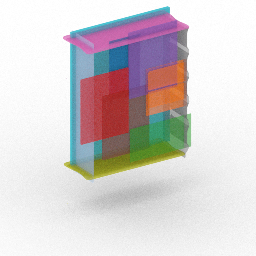}
        \\
    \end{tabular}
    \caption{Storage Unconditional Samples}
    \label{fig:storage}
\end{figure*}

\begin{figure*}[t!]
    \centering
    \setlength{\tabcolsep}{1pt}
    \begin{tabular}{cccccccccc}
        \multirow{3}{*}{\raisebox{-4em}{\rotatebox{90}{Ours}}} &
        \includegraphics[width=.105\linewidth]{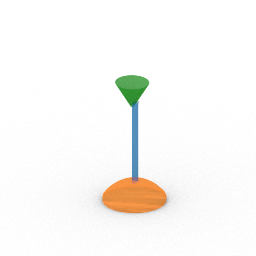} &
        \includegraphics[width=.105\linewidth]{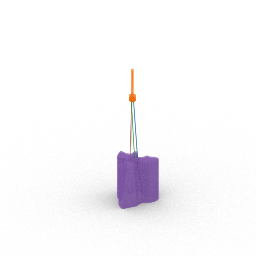} &
        \includegraphics[width=.105\linewidth]{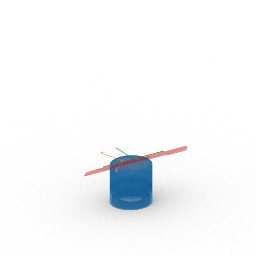} &
        \includegraphics[width=.105\linewidth]{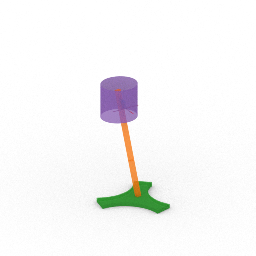} &
        \includegraphics[width=.105\linewidth]{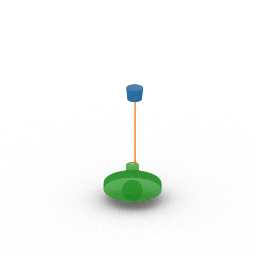} &
        \includegraphics[width=.105\linewidth]{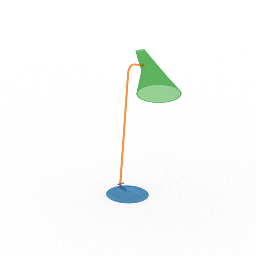} &
        \includegraphics[width=.105\linewidth]{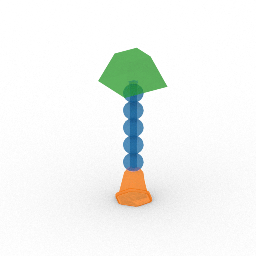} &
        \includegraphics[width=.105\linewidth]{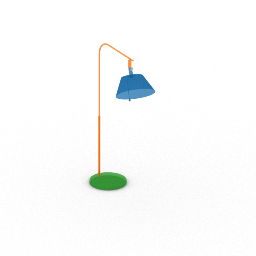} &
        \includegraphics[width=.105\linewidth]{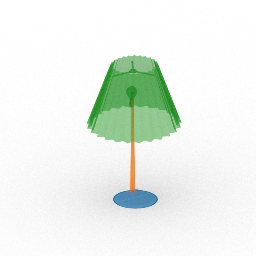}
        \\
        &
        \includegraphics[width=.105\linewidth]{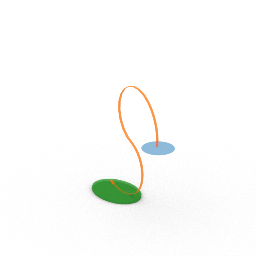} &
        \includegraphics[width=.105\linewidth]{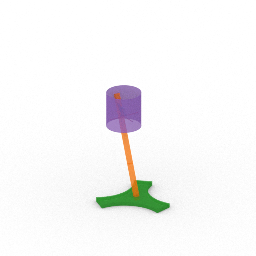} &
        \includegraphics[width=.105\linewidth]{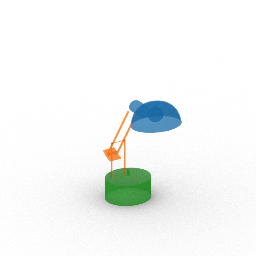} &
        \includegraphics[width=.105\linewidth]{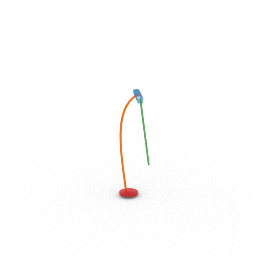} &
        \includegraphics[width=.105\linewidth]{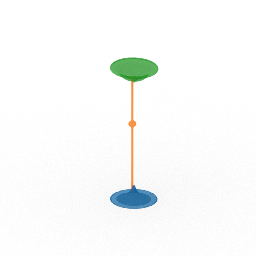} &
        \includegraphics[width=.105\linewidth]{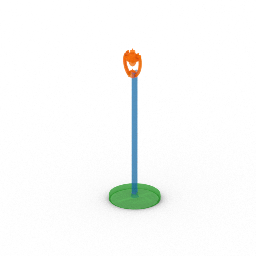} &
        \includegraphics[width=.105\linewidth]{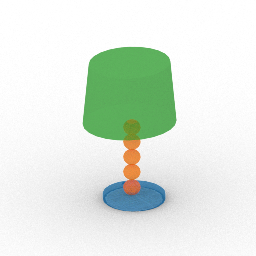} &
        \includegraphics[width=.105\linewidth]{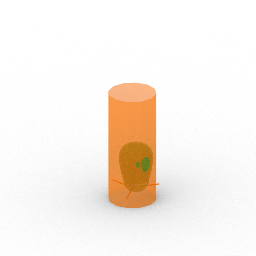} &
        \includegraphics[width=.105\linewidth]{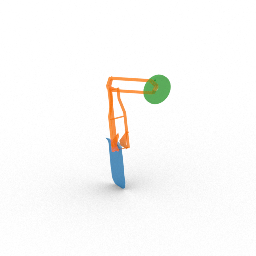}
        \\
        &
        \includegraphics[width=.105\linewidth]{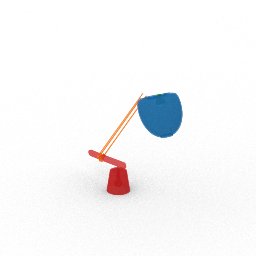} &
        \includegraphics[width=.105\linewidth]{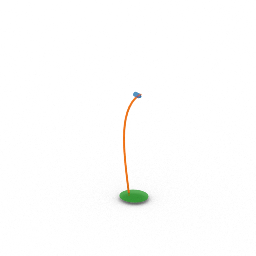} &
        \includegraphics[width=.105\linewidth]{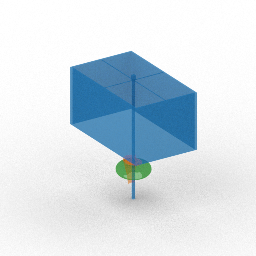} &
        \includegraphics[width=.105\linewidth]{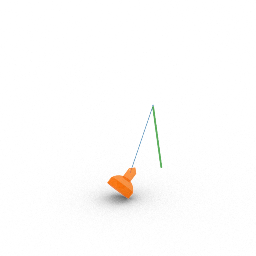} &
        \includegraphics[width=.105\linewidth]{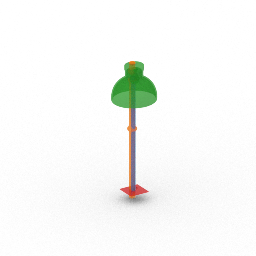} &
        \includegraphics[width=.105\linewidth]{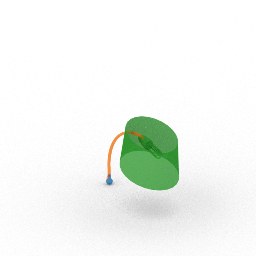} &
        \includegraphics[width=.105\linewidth]{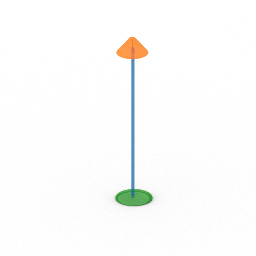} &
        \includegraphics[width=.105\linewidth]{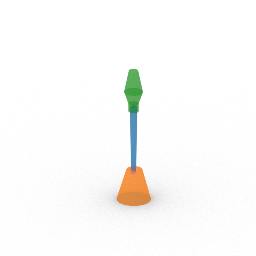} &
        \includegraphics[width=.105\linewidth]{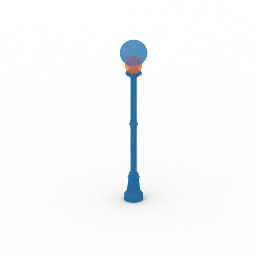}
        \\ \midrule
        \multirow{3}{*}{\raisebox{8em}{\rotatebox{90}{ComplementMe}}} &
        \includegraphics[width=.105\linewidth]{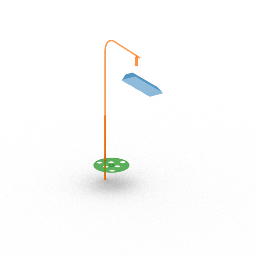} &
        \includegraphics[width=.105\linewidth]{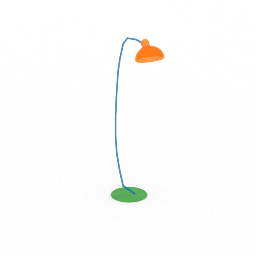} &
        \includegraphics[width=.105\linewidth]{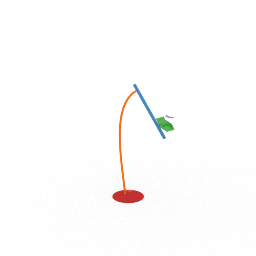} &
        \includegraphics[width=.105\linewidth]{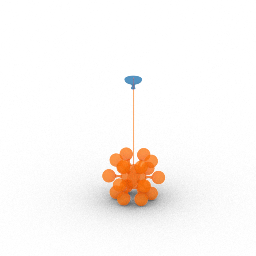} &
        \includegraphics[width=.105\linewidth]{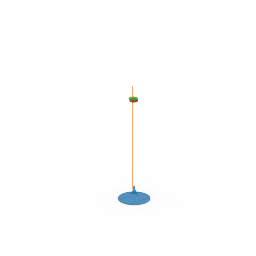} &
        \includegraphics[width=.105\linewidth]{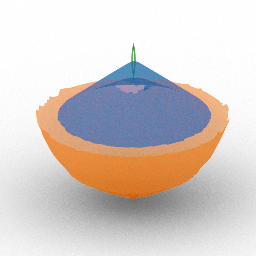} &
        \includegraphics[width=.105\linewidth]{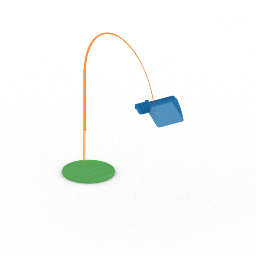} &
        \includegraphics[width=.105\linewidth]{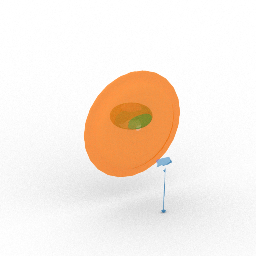} &
        \includegraphics[width=.105\linewidth]{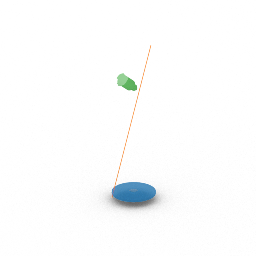}
        \\
        &
        \includegraphics[width=.105\linewidth]{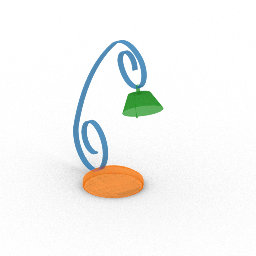} &
        \includegraphics[width=.105\linewidth]{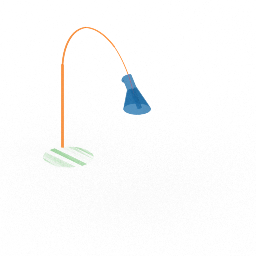} &
        \includegraphics[width=.105\linewidth]{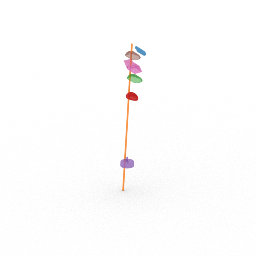} &
        \includegraphics[width=.105\linewidth]{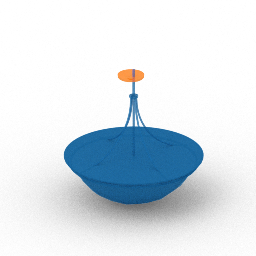} &
        \includegraphics[width=.105\linewidth]{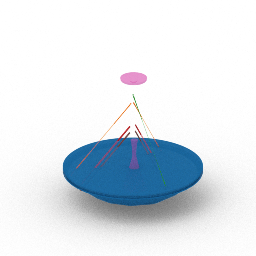} &
        \includegraphics[width=.105\linewidth]{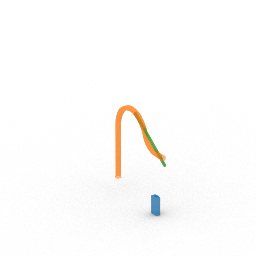} &
        \includegraphics[width=.105\linewidth]{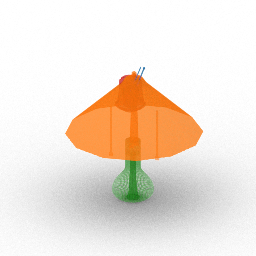} &
        \includegraphics[width=.105\linewidth]{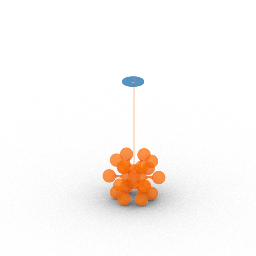} &
        \includegraphics[width=.105\linewidth]{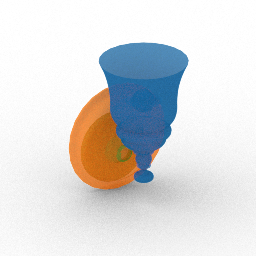}
        \\
        &
        \includegraphics[width=.105\linewidth]{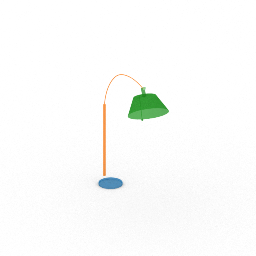} &
        \includegraphics[width=.105\linewidth]{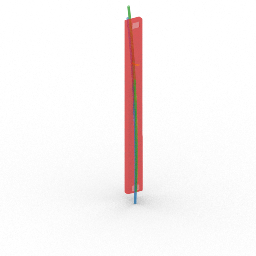} &
        \includegraphics[width=.105\linewidth]{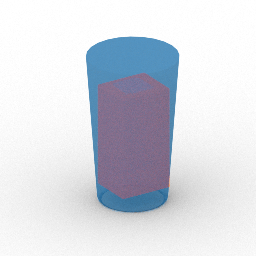} &
        \includegraphics[width=.105\linewidth]{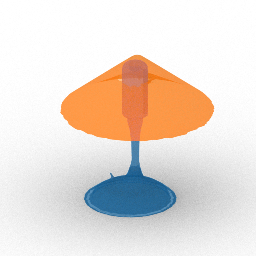} &
        \includegraphics[width=.105\linewidth]{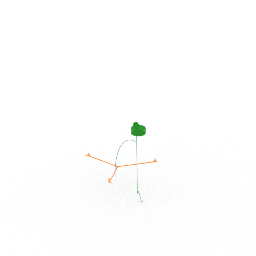} &
        \includegraphics[width=.105\linewidth]{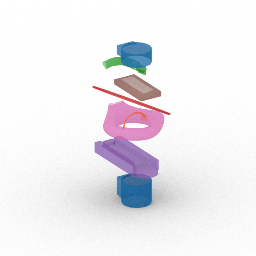} &
        \includegraphics[width=.105\linewidth]{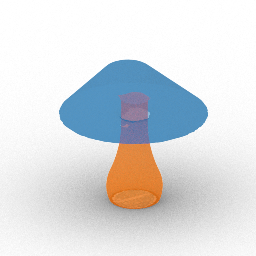} &
        \includegraphics[width=.105\linewidth]{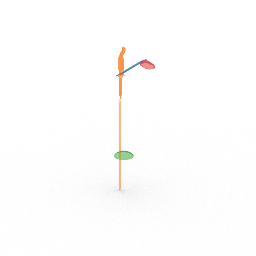} &
        \includegraphics[width=.105\linewidth]{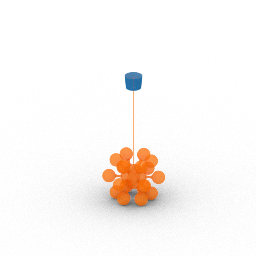}
    \end{tabular}
    \caption{Lamp Unconditional Samples}
    \label{fig:lamp}
\end{figure*}

We show random samples of our method and the baselines on all four categories in Figure~\ref{fig:chair},~\ref{fig:table},~\ref{fig:storage} and~\ref{fig:lamp}. 

For the methods that are autoregressive (ours and ComplementMe), the color of the parts correlate with the order in which they are inserted. We use the Tableu 10 color palette\footnote{https://public.tableau.com/views/TableauColors/ColorPaletteswithRGBValues} for the first 10 parts, and add the remaining 10 colors in the Tableu 20 color palette for shapes more than 10 parts: the blue part is used for initialization, and the subsequent parts inserted are colored orange, green, red, purple, etc. respectively.

Overall, the quality and physical plausibility of the generated shapes correlate well with the quantitative metrics.
ComplementMe benefits considerably from grouping parts by symmetry, as it simplifies the task of predicting global poses of shapes significantly.
When parts are not grouped by symmetry, it often fails to predict the right pose of parts, and sometimes is not able to complete a shape at all.
It also produces a lot of incorrect and incomplete storages, even with the help of symmetry.

StructureNet is usually able to generate shapes that are plausible, though often with a few missing parts. However, it has the tendency to generate only a subset of shape types. This is most apparent for table and storage, when it generates mostly square tables and storages with open shelves.
Large gaps sometimes exist between the individual parts, leading to problems with physical plausibility. Note that this problem is not caused by us retrieving parts directly using the latent code --- the box version of StructureNet has similar issues (see the evaluation of ShapeAssembly~\cite{jones2020shapeAssembly}).

\begin{figure*}[t!]
    \centering
    \setlength{\tabcolsep}{1pt}
    \begin{tabular}{ccccc}
        \includegraphics[width=.15\linewidth]{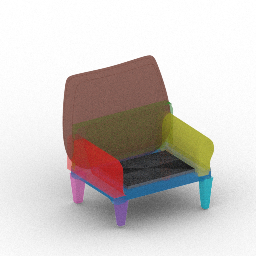} &
        \includegraphics[width=.15\linewidth]{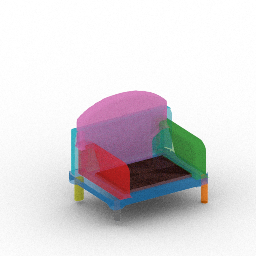} &
        \includegraphics[width=.15\linewidth]{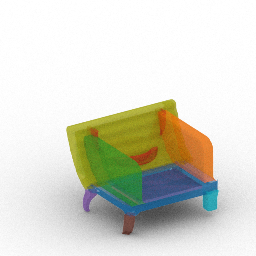} &
        \includegraphics[width=.15\linewidth]{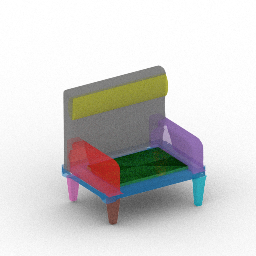} &
        \includegraphics[width=.15\linewidth]{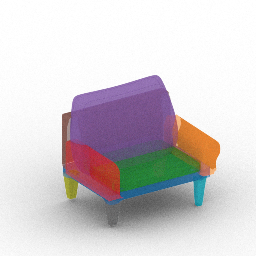}
        \\
        \includegraphics[width=.15\linewidth]{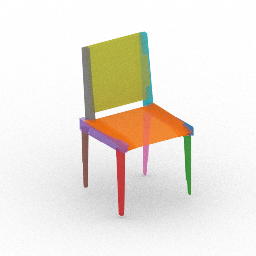} &
        \includegraphics[width=.15\linewidth]{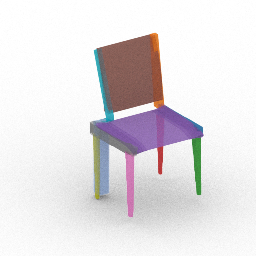} &
        \includegraphics[width=.15\linewidth]{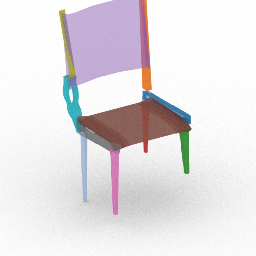} &
        \includegraphics[width=.15\linewidth]{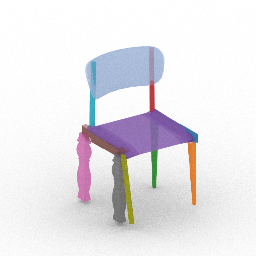} &
        \includegraphics[width=.15\linewidth]{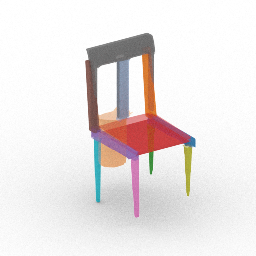}
        \\
        \includegraphics[width=.15\linewidth]{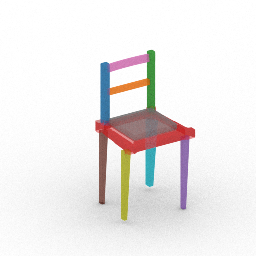} &
        \includegraphics[width=.15\linewidth]{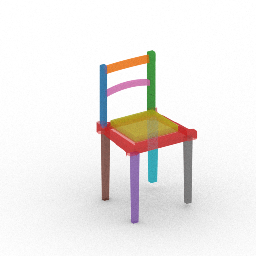} &
        \includegraphics[width=.15\linewidth]{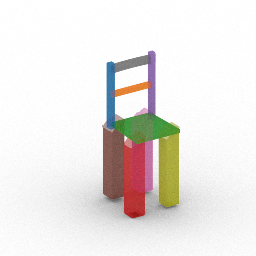} &
        \includegraphics[width=.15\linewidth]{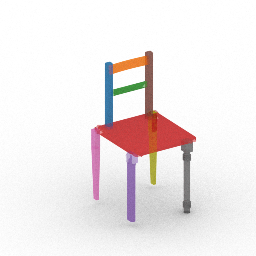} &
        \includegraphics[width=.15\linewidth]{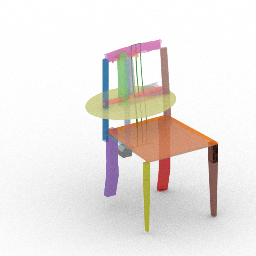}
        \\
        \includegraphics[width=.15\linewidth]{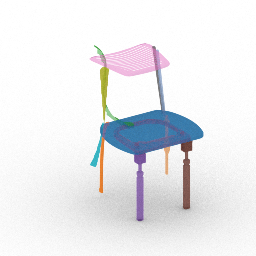} &
        \includegraphics[width=.15\linewidth]{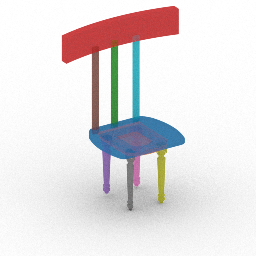} &
        \includegraphics[width=.15\linewidth]{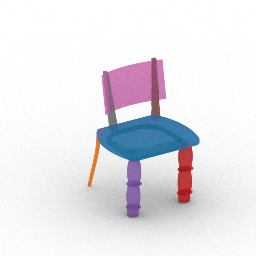} &
        \includegraphics[width=.15\linewidth]{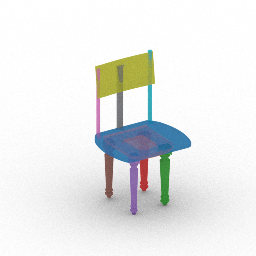} &
        \includegraphics[width=.15\linewidth]{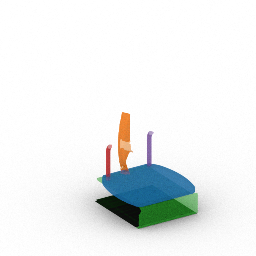}
        \\
    \end{tabular}
    \caption{Multiple output per initialization, achieved by sampling the neural networks randomly instead of doing MAP inference. Each row uses a different part as initialization.}
    \label{fig:multiple}
\end{figure*}

The behavior of our method is more polarizing: it generates a lot of high quality shapes; however, some other generated shapes are totally incorrect. The high quality shapes fare better than the baselines in terms of quality, physical plausibility, and diversity to some extent. 
The incorrect shapes exhibit a wide range of failure mode, which we hypothesize can be traced back to a few incorrect steps in the autoregressive generation process. Reducing the chance of these incorrect steps, and identifying them when they happen, is an important future direction to take in order to further improve the quality of the generated shapes.
Our method also has the tendency to generate simpler shapes when sampling randomly. This is not caused by the neural networks learning biased distributions, but caused by the higher failure rate for more complex structure during autoregressive sampling. We also notice a few repeated shapes, especially for storages. This can be addressed by sampling the neural network modules randomly, as opposed to doing MAP inference.
In figure~\ref{fig:multiple}, we show examples of random sampling: our method is able to produce multiple output per initialization (blue).

\begin{figure*}[t!]
    \vspace{-3em}
    \centering
    \setlength{\tabcolsep}{1pt}
    \begin{tabular}{cccccccccc}
        \multirow{3}{*}{\raisebox{-4em}{\rotatebox{90}{Chair}}} &
        \includegraphics[width=.105\linewidth]{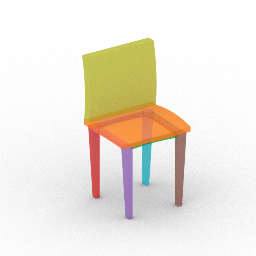} &
        \includegraphics[width=.105\linewidth]{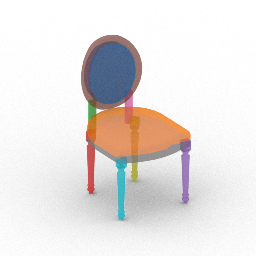} &
        \includegraphics[width=.105\linewidth]{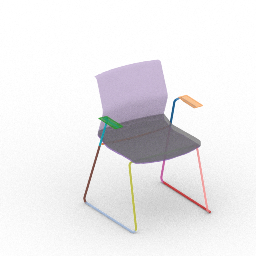} &
        \includegraphics[width=.105\linewidth]{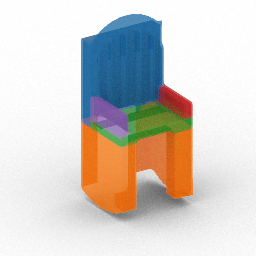} &
        \includegraphics[width=.105\linewidth]{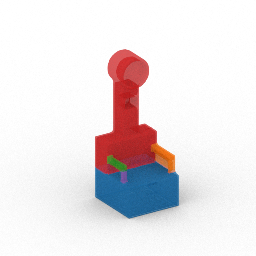} &
        \includegraphics[width=.105\linewidth]{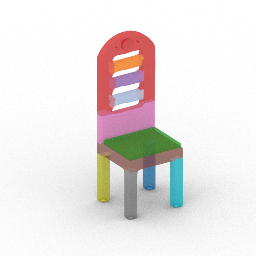} &
        \includegraphics[width=.105\linewidth]{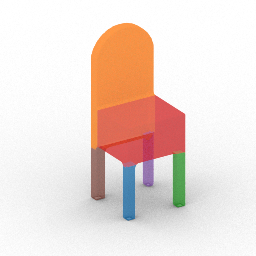} &
        \includegraphics[width=.105\linewidth]{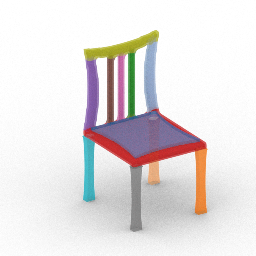} &
        \includegraphics[width=.105\linewidth]{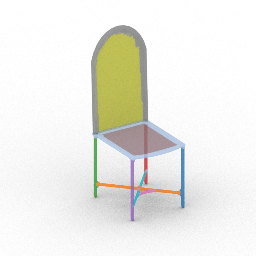}
        \\
        &
        \includegraphics[width=.105\linewidth]{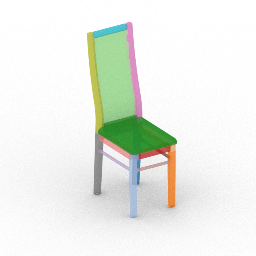} &
        \includegraphics[width=.105\linewidth]{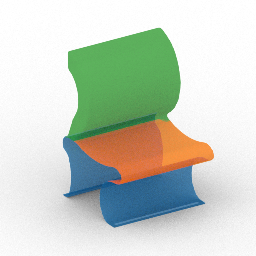} &
        \includegraphics[width=.105\linewidth]{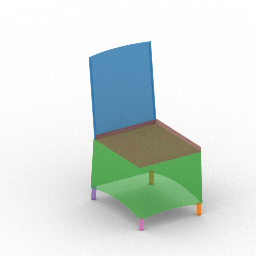} &
        \includegraphics[width=.105\linewidth]{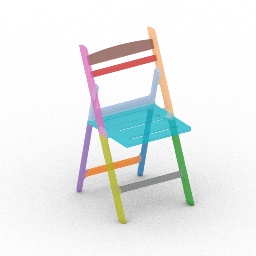} &
        \includegraphics[width=.105\linewidth]{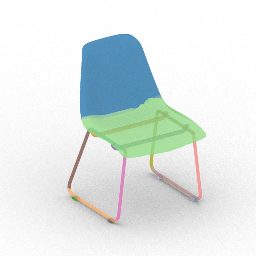} &
        \includegraphics[width=.105\linewidth]{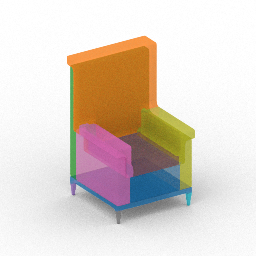} &
        \includegraphics[width=.105\linewidth]{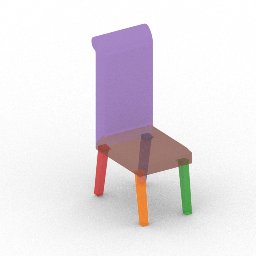} &
        \includegraphics[width=.105\linewidth]{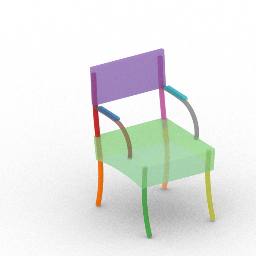} &
        \includegraphics[width=.105\linewidth]{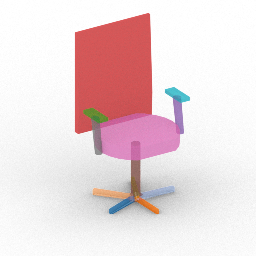}
        \\
        &
        \includegraphics[width=.105\linewidth]{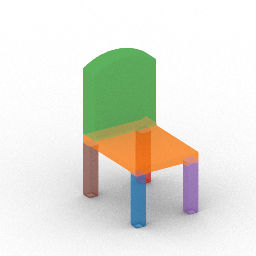} &
        \includegraphics[width=.105\linewidth]{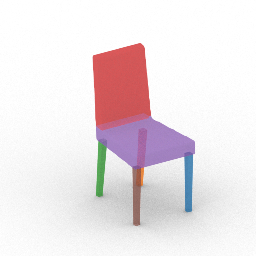} &
        \includegraphics[width=.105\linewidth]{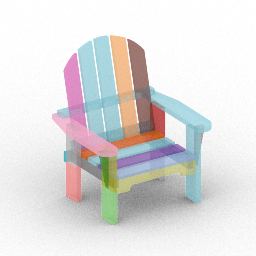} &
        \includegraphics[width=.105\linewidth]{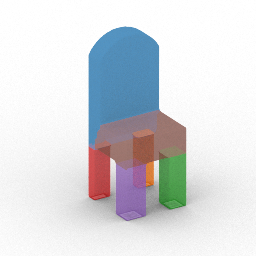} &
        \includegraphics[width=.105\linewidth]{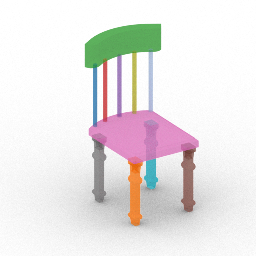} &
        \includegraphics[width=.105\linewidth]{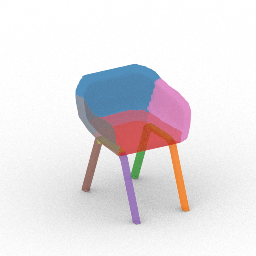} &
        \includegraphics[width=.105\linewidth]{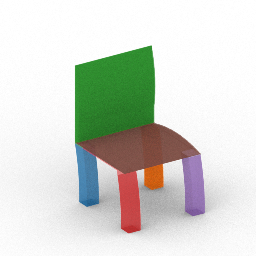} &
        \includegraphics[width=.105\linewidth]{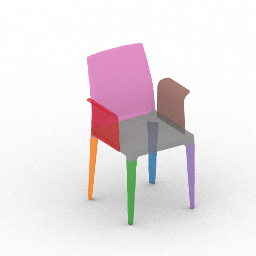} &
        \includegraphics[width=.105\linewidth]{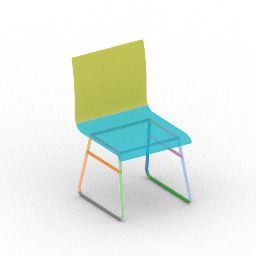}
        \\ \midrule
        \multirow{3}{*}{\raisebox{-4em}{\rotatebox{90}{Table}}} &
        \includegraphics[width=.105\linewidth]{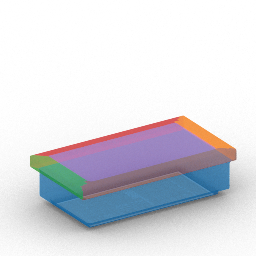} &
        \includegraphics[width=.105\linewidth]{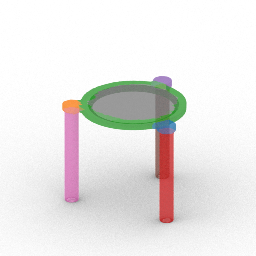} &
        \includegraphics[width=.105\linewidth]{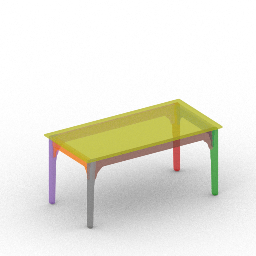} &
        \includegraphics[width=.105\linewidth]{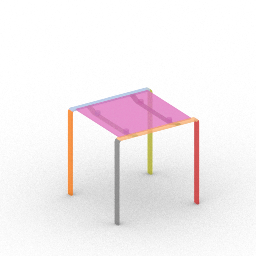} &
        \includegraphics[width=.105\linewidth]{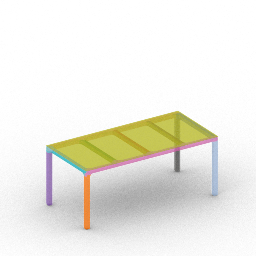} &
        \includegraphics[width=.105\linewidth]{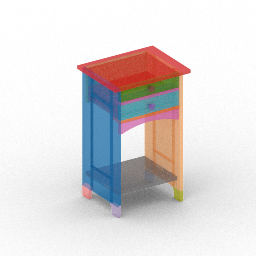} &
        \includegraphics[width=.105\linewidth]{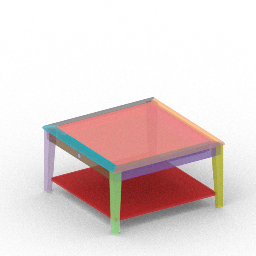} &
        \includegraphics[width=.105\linewidth]{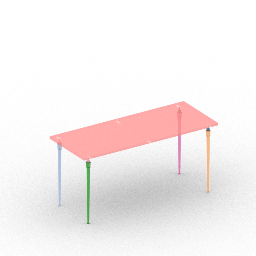} &
        \includegraphics[width=.105\linewidth]{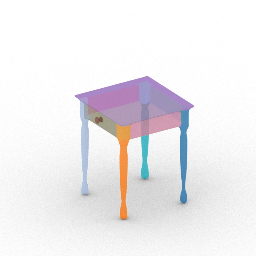}
        \\
        &
        \includegraphics[width=.105\linewidth]{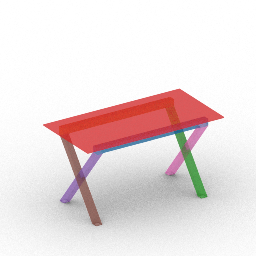} &
        \includegraphics[width=.105\linewidth]{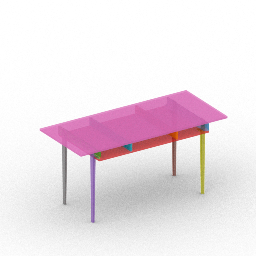} &
        \includegraphics[width=.105\linewidth]{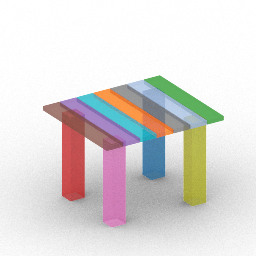} &
        \includegraphics[width=.105\linewidth]{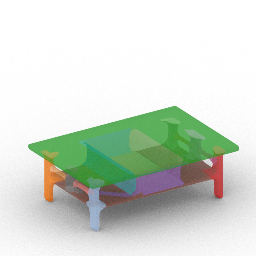} &
        \includegraphics[width=.105\linewidth]{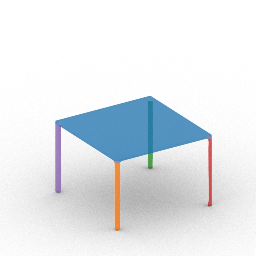} &
        \includegraphics[width=.105\linewidth]{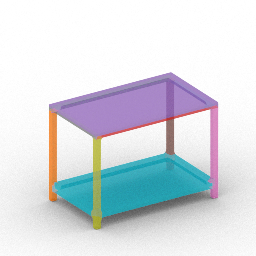} &
        \includegraphics[width=.105\linewidth]{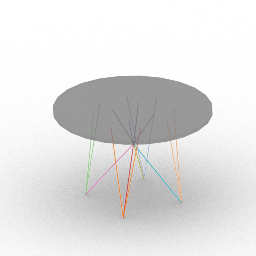} &
        \includegraphics[width=.105\linewidth]{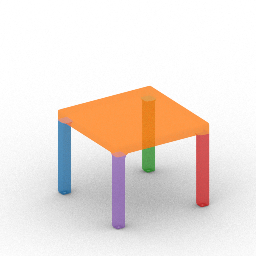} &
        \includegraphics[width=.105\linewidth]{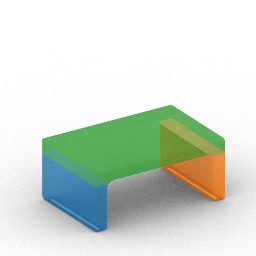}
        \\
        &
        \includegraphics[width=.105\linewidth]{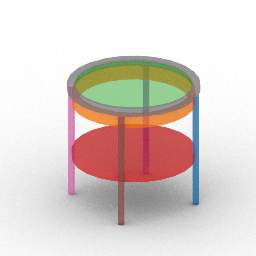} &
        \includegraphics[width=.105\linewidth]{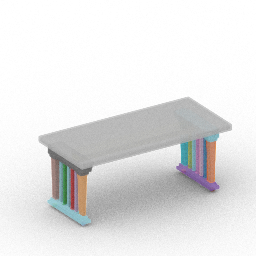} &
        \includegraphics[width=.105\linewidth]{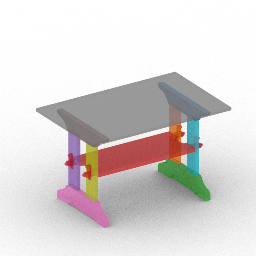} &
        \includegraphics[width=.105\linewidth]{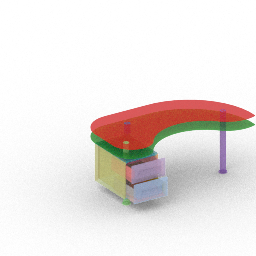} &
        \includegraphics[width=.105\linewidth]{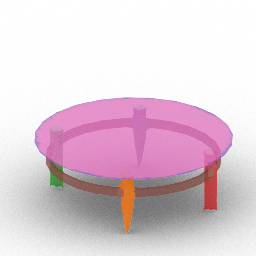} &
        \includegraphics[width=.105\linewidth]{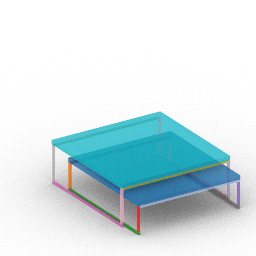} &
        \includegraphics[width=.105\linewidth]{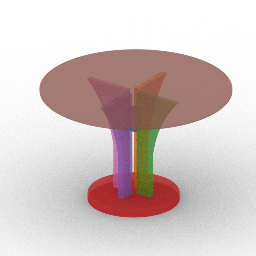} &
        \includegraphics[width=.105\linewidth]{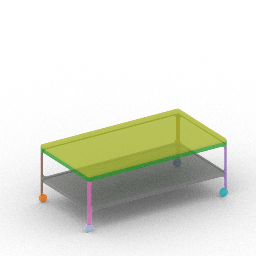} &
        \includegraphics[width=.105\linewidth]{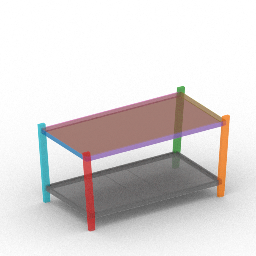}
        \\ \midrule
        \multirow{3}{*}{\raisebox{-4em}{\rotatebox{90}{Storage}}} &
        \includegraphics[width=.105\linewidth]{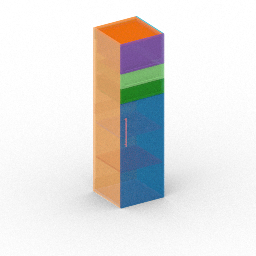} &
        \includegraphics[width=.105\linewidth]{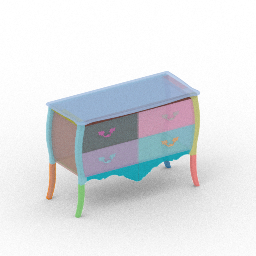} &
        \includegraphics[width=.105\linewidth]{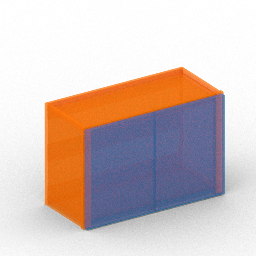} &
        \includegraphics[width=.105\linewidth]{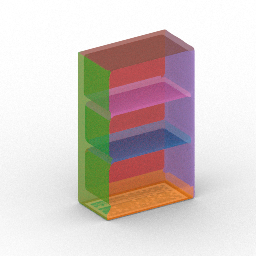} &
        \includegraphics[width=.105\linewidth]{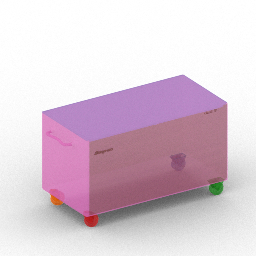} &
        \includegraphics[width=.105\linewidth]{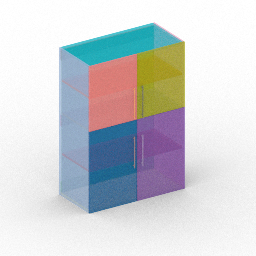} &
        \includegraphics[width=.105\linewidth]{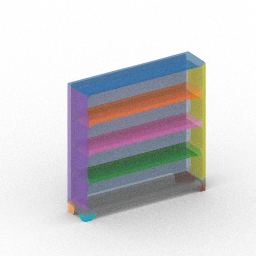} &
        \includegraphics[width=.105\linewidth]{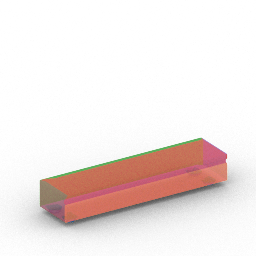} &
        \includegraphics[width=.105\linewidth]{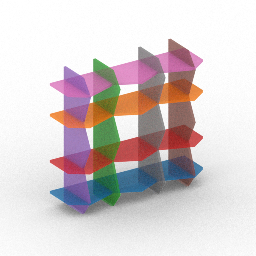}
        \\
        &
        \includegraphics[width=.105\linewidth]{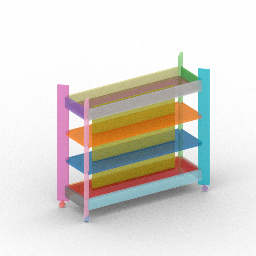} &
        \includegraphics[width=.105\linewidth]{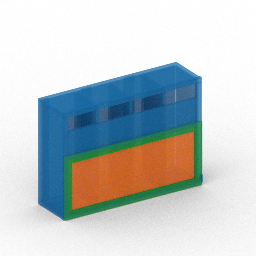} &
        \includegraphics[width=.105\linewidth]{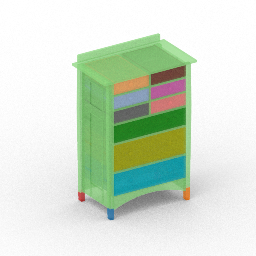} &
        \includegraphics[width=.105\linewidth]{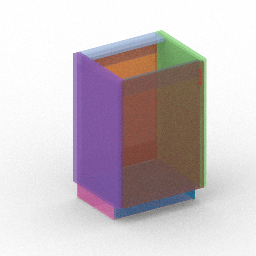} &
        \includegraphics[width=.105\linewidth]{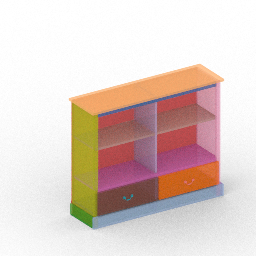} &
        \includegraphics[width=.105\linewidth]{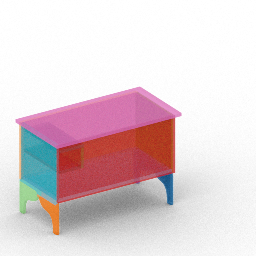} &
        \includegraphics[width=.105\linewidth]{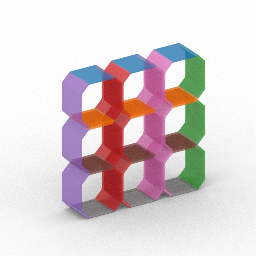} &
        \includegraphics[width=.105\linewidth]{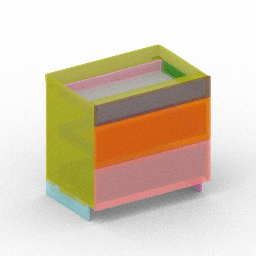} &
        \includegraphics[width=.105\linewidth]{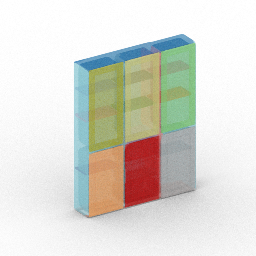}
        \\
        &
        \includegraphics[width=.105\linewidth]{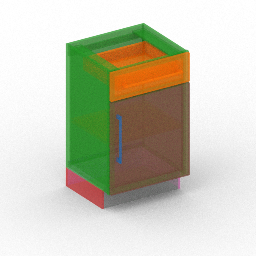} &
        \includegraphics[width=.105\linewidth]{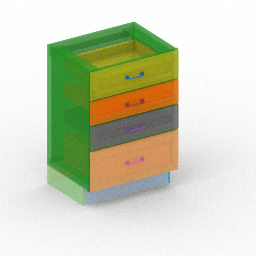} &
        \includegraphics[width=.105\linewidth]{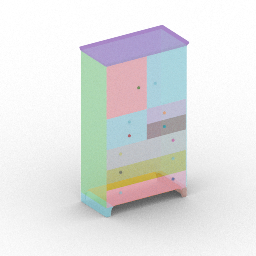} &
        \includegraphics[width=.105\linewidth]{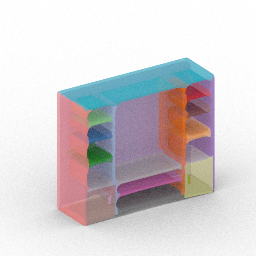} &
        \includegraphics[width=.105\linewidth]{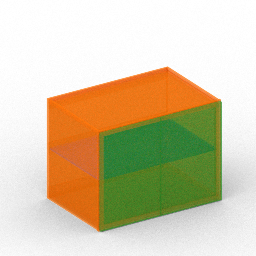} &
        \includegraphics[width=.105\linewidth]{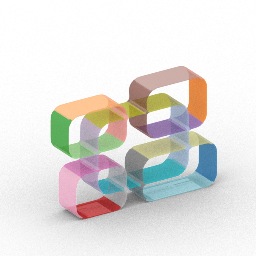} &
        \includegraphics[width=.105\linewidth]{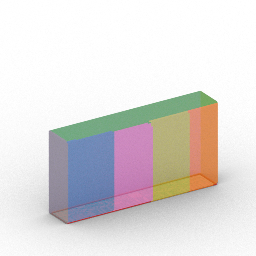} &
        \includegraphics[width=.105\linewidth]{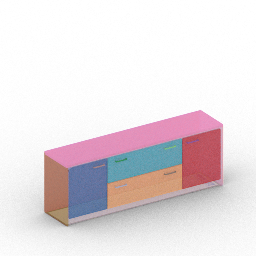} &
        \includegraphics[width=.105\linewidth]{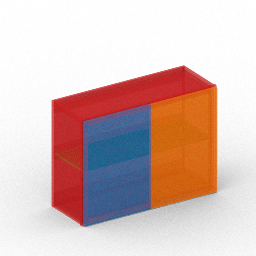}
        \\ \midrule
        \multirow{3}{*}{\raisebox{-4em}{\rotatebox{90}{Lamp}}} &
        \includegraphics[width=.105\linewidth]{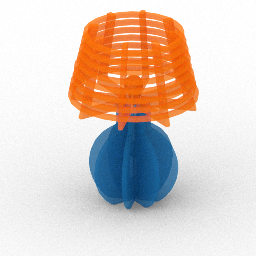} &
        \includegraphics[width=.105\linewidth]{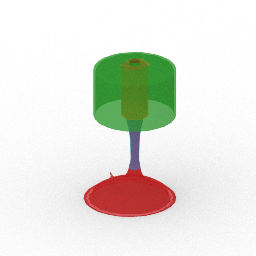} &
        \includegraphics[width=.105\linewidth]{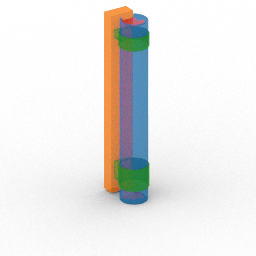} &
        \includegraphics[width=.105\linewidth]{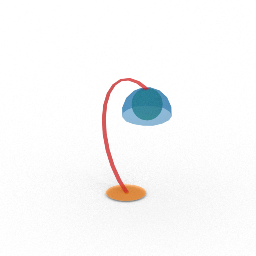} &
        \includegraphics[width=.105\linewidth]{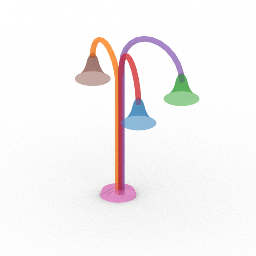} &
        \includegraphics[width=.105\linewidth]{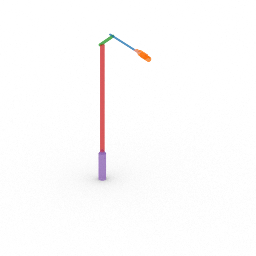} &
        \includegraphics[width=.105\linewidth]{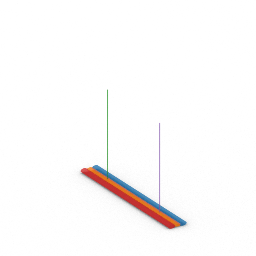} &
        \includegraphics[width=.105\linewidth]{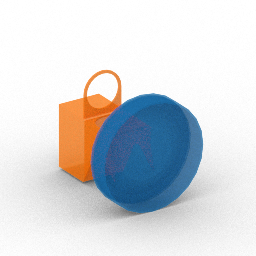} &
        \includegraphics[width=.105\linewidth]{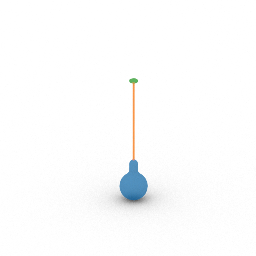}
        \\
        &
        \includegraphics[width=.105\linewidth]{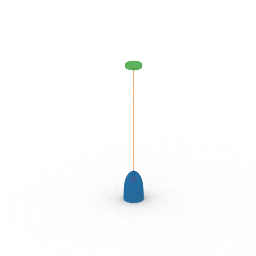} &
        \includegraphics[width=.105\linewidth]{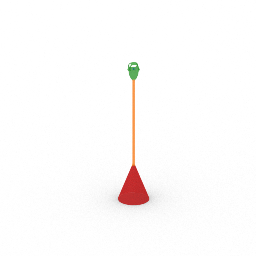} &
        \includegraphics[width=.105\linewidth]{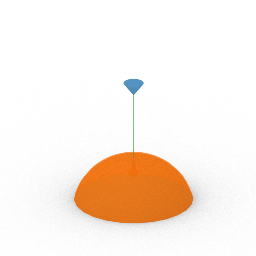} &
        \includegraphics[width=.105\linewidth]{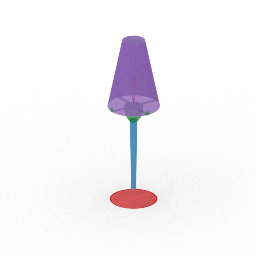} &
        \includegraphics[width=.105\linewidth]{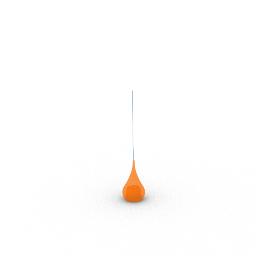} &
        \includegraphics[width=.105\linewidth]{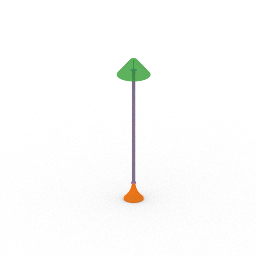} &
        \includegraphics[width=.105\linewidth]{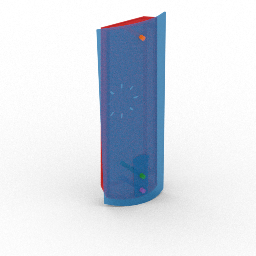} &
        \includegraphics[width=.105\linewidth]{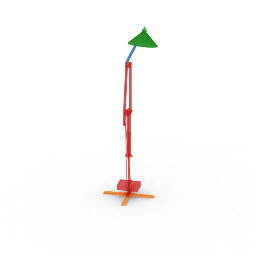} &
        \includegraphics[width=.105\linewidth]{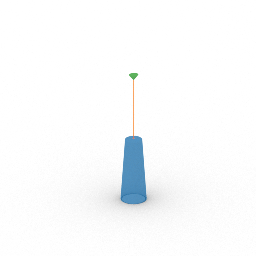}
        \\
        &
        \includegraphics[width=.105\linewidth]{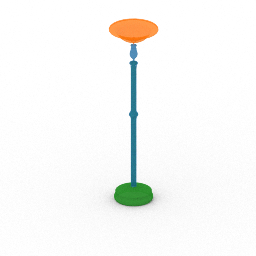} &
        \includegraphics[width=.105\linewidth]{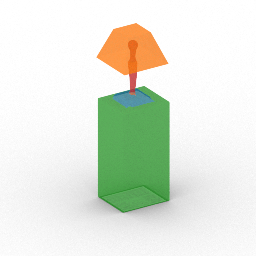} &
        \includegraphics[width=.105\linewidth]{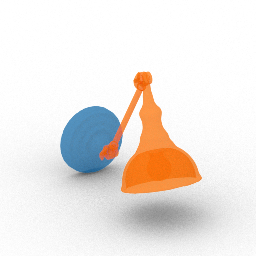} &
        \includegraphics[width=.105\linewidth]{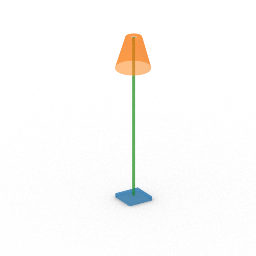} &
        \includegraphics[width=.105\linewidth]{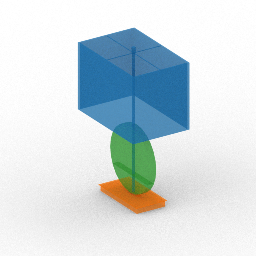} &
        \includegraphics[width=.105\linewidth]{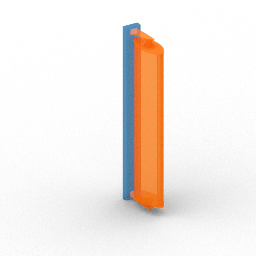} &
        \includegraphics[width=.105\linewidth]{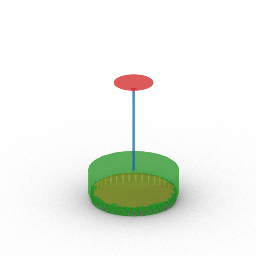} &
        \includegraphics[width=.105\linewidth]{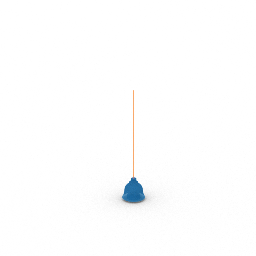} &
        \includegraphics[width=.105\linewidth]{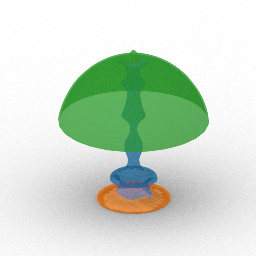}
        \\
    \end{tabular}
    \vspace{-1.5em}
    \caption{Dataset Unconditional Samples}
    \label{fig:dataset}
\end{figure*}

Finally, we show random samples from drawn the test set in Figure~\ref{fig:dataset}.
We note that none of the methods are able to generate shapes that are close to the dataset in terms of quality and diversity. This is especially the case for shapes with unique and complex structures: they are harder to learn, and there is often not enough training data for them. Learning these structures correctly and efficiently remains an open problem.

